\documentclass[12pt]{article}

 \usepackage[utf8]{inputenc}
 \usepackage[russian,english]{babel}
\usepackage{amsmath}
\usepackage{amsfonts}
 \usepackage[dvipsnames]{xcolor}

\def\lb{\label}
\def\be{\begin{equation}}
\def\ee{\end{equation}}
\def\ad{\rm ad}
\def\qed{\rule{5pt}{5pt}}

\begin{document}


\vspace{1cm}

\begin{center}
{\huge \bf Vogel universality and beyond}
\end{center}
\vspace{1cm}

\begin{center}
{\large \bf  A.P.Isaev${}^{a,b,c}$}
\end{center}

\vspace{0.2cm}

\begin{center}
{${}^a$ \it
Bogoliubov  Laboratory of Theoretical Physics,\\
Joint Institute for Nuclear Research,
141980, Dubna}\vspace{0.1cm} \\
{${}^b$ \it
St.Petersburg Department of
 Steklov Mathematical Institute of RAS,\\
Fontanka 27, 191023, St. Petersburg,  Russia}\vspace{0.1cm} \\
{${}^c$ \it Faculty of Physics,
Lomonosov Moscow State University, Russia}\vspace{0.3cm}

{\tt isaevap@theor.jinr.ru}
\end{center}
\vspace{3cm}

\begin{abstract}\noindent
 For simple Lie algebras we construct characteristic identities for split (polarized) Casimir operators in the representations $T \otimes Y_n$
  and $T \otimes Y_n'$,
  where $T$ is defining
  (minimal fundamental for exceptional Lie algebras) representation,
  $Y_n$  are Cartan $n$-powers
 of the adjoint representations $\ad = Y_1$, and $Y_n'$ are
 special representations arising in the Clebsch-Gordan
 decomposition of the totally symmetric part of $\ad^{\otimes n}$.
  By means of these characteristic identities, we derive
  (for all simple Lie algebras, except $\mathfrak{e}_8$)
  explicit formulas for invariant projectors onto
  irreducible subrepresentations appearing in the decomposition of
  $T \otimes Y_n$. These projectors and characteristic identities are written in the universal form for
   all simple Lie algebras (except $\mathfrak{e}_8$)
   in terms of the Vogel parameters. Universal formulas for the
 dimensions of Casimir subrepresentations arising in
 decompositions of $T \otimes Y_n$ are found. An example
 of evaluating of universal colour (group) factors for an infinite
 set of special Feynman ladder diagrams in
 non-Abelian gauge theories is given.
\end{abstract}

\newpage

\section{Introduction}

It was observed by P. Vogel \cite{Vogel},
in his investigation of the Vassiliev theory of
knot invariants,
that for all simple Lie algebras $\mathfrak{g}$ the symmetric
$Sym(\mathfrak{g}^{\otimes 2})$ and anti-symmetric
$\wedge(\mathfrak{g}^{\otimes 2})$ parts of the square
of the adjoint representation $\ad=\mathfrak{g}$
can be decomposed into subrepresentations in a uniform way.
 More precisely, we have \cite{Vogel}
 \be
 \lb{gsq}
Sym(\mathfrak{g}^{\otimes 2}) = 1+ Y_2 + Y_2' + Y_2'' \; , \;\;\;\;
\wedge(\mathfrak{g}^{\otimes 2}) = Y_1 + X_2 \; ,
 \ee
where $Y_2$, $Y_2'$, $Y_2''$, $X_2$ and
$Y_1 = \ad= \mathfrak{g}$  are
 Casimir subspaces \cite{LanMan} in $\mathfrak{g}^{\otimes 2}$
(Casimir representations of a simple Lie algebra $\mathfrak{g}$).
It was shown in \cite{Vogel} that the dimensions of all constituents
in (\ref{gsq}) are the rational homogeneous
functions of three universal
Vogel parameters $\alpha$, $\beta$, $\gamma$.
We note that the representation $Y_2$ in (\ref{gsq})
 is the second Cartan power of the adjoint representation
$\ad = \mathfrak{g}=Y_1$,
while dimensions of the representations
$Y_2'$ and $Y_2''$
are respectively obtained from $\dim Y_2$
by exchanging of the parameters $\alpha \leftrightarrow \beta$ and
$\alpha \leftrightarrow \gamma$. P. Vogel
then went on in \cite{Vogel} to find Casimir eigenspaces
in the higher tensor product $\mathfrak{g}^{\otimes 3}$.
For the exceptional simple Lie algebras, P.Deligne \cite{Del}
and A.M.~Cohen and R.~de Man \cite{CohMan}
showed that decompositions of the tensor powers
$\mathfrak{g}^{\otimes k}$ exhibit universal behavior
up to degree $k=4$. For all simple Lie algebras,
the uniform structure of $\mathfrak{g}^{\otimes 4}$
 was 
 investigated in \cite{IKMkr}
 by using the split Casimir operator technique. With the help
 of the same technique, a universal description
 of the representations arising in the decomposition
  of the antisymmetrized part of $\mathfrak{g}^{\otimes 5}$
  was also obtained in \cite{IsaKri2}. The set
  of universal Casimir eigenvalues
  for $\mathfrak{g}^{\otimes 5}$ was recently suggested in \cite{RMcr2}.

It is clear that
the Cartan $k$-th power $Y_k$ of the adjoint representations
 will always arise in the
decomposition of the totally
symmetric part of $\mathfrak{g}^{\otimes k}$.
The general universal formula for the dimensions
of the representations $Y_k = Y_k(\alpha)$
is known \cite{LanMan} (see also \cite{RMcr0}
 for its $q$-deformation)
 \be
 \lb{dimYk}
 \dim Y_k(\alpha) =
 \frac{\bigl(1-(2k-1)\hat{\alpha}\bigr)}{k! \, \hat{\alpha}^k \,
 \bigl(1- (k-1)\hat{\alpha}\bigr)}
 \prod_{\ell=0}^{k-1}
 \frac{(\ell \, \hat{\alpha}-1) (\hat{\beta}+ \ell \hat{\alpha} -1)
 (\hat{\gamma}+ \ell \hat{\alpha} -1)}{(\hat{\beta}- \ell \hat{\alpha})
 (\hat{\gamma}- \ell \hat{\alpha})}  \; ,
 \ee
 and in particular we have
 \be
 \lb{dimY1}
 \dim Y_1 = \dim \mathfrak{g} =
 \frac{(\hat{\alpha}-1)(\hat{\beta}-1)(\hat{\gamma}-1)}{
 \hat{\alpha}\, \hat{\beta}\, \hat{\gamma}}  \; .
 \ee
 Our homogeneous parameters $\hat{\alpha},\hat{\beta},\hat{\gamma}$
 are related to the standard Vogel
parameters $\alpha$, $\beta$, $\gamma$ by
 \be
 \lb{hompar}
 \hat{\alpha} = \frac{\alpha}{2 {\sf t}} \; , \;\;\;\;
 \hat{\beta}= \frac{\beta}{2{\sf t}} \; , \;\;\;\;
 \hat{\gamma}= \frac{\gamma}{2{\sf t}} \; ,  \;\;\;\;\;\;\;
 ({\sf t}:=\alpha+\beta+\gamma)  \; ,
 \ee
 so that
 $\hat{\alpha}+\hat{\beta}+\hat{\gamma}=1/2$.
 We give their values in Table {\sf \ref{tab1}} below
 (for the standard choice of the Vogel variables
 $\alpha$, $\beta$ and $\gamma$, see, e.g.,
 \cite{LanMan},\cite{MSeVes},\cite{IsKriv}).
The dimensions of the representations
$Y_k' = Y_k(\beta)$ and $Y_k'' = Y_k(\gamma)$
(which are also presented in the decomposition
of the totally symmetric part of $\mathfrak{g}^{\otimes k}$)
are  respectively obtained from $\dim Y_k(\alpha)$ given in (\ref{dimYk})
by permuting the parameters $\alpha \leftrightarrow \beta$ and
$\alpha \leftrightarrow \gamma$.

\vspace{0.1cm}

One of the motivations that also leads to the Vogel
universal description of Lie algebras is related to the calculation of group (colour) factors of Feynman diagrams in non-Abelian gauge theories (see works
\cite{Cvit2}, \cite{Cvit} that preceded the works
of Vogel and Deligne).
Evaluation of colour factors in
non-Abelian gauge theories is an extremely important task for applications
(see e.g. \cite{Vermas} and references therein).
The long-standing problem of finding a minimal set of group variables, through which all color factors in gauge models can be expressed, apparently cannot be solved by directly searching for a minimal set of such group invariants. A promising approach involves 
writing group factors in a universal form using three Vogel parameters. Moreover, this approach also allows one to write the color factors of Feynman diagrams in a form applicable to any simple gauge group.
Apparently, all color factors for Feynman diagrams in
 pure gluodynamics (a non-Abelian gauge theory without
 matter fields) can be written in a universal form using only three Vogel parameters. For example, in our paper \cite{IsaPro3}, universal
 expressions were found for colour factors of
an infinite set of Feynman diagrams in gluodynamics. In \cite{IsaPro3},
we also proposed a universal analog of the $1/N$ expansion
that applies even to
theories with exceptional Lie gauge groups.

It is also necessary to note various further applications of
Vogel universality to knot/ Chern-Simons theory
\cite{MorSlep,KhudSlep,KhudMoSlep,BiMiMo},
in particular to the possibility of an universal description of
Wilson loop averages, in Chern-Simons theory, which are represented as polynomials of knots and links \cite{BiMir,MirMor1,MirMor2,Miron}.
The latter papers continue the investigations of
applications of
Vogel universality to Chern-Simons theory that
began with the papers \cite{MkrVes,Mkr00}.

The results of the present paper show that
 Vogel universality can be extended to the
 class of representations $\Lambda_i^{(n)}|_{i=1,2,...}$
 that appear in the decomposition of tensor products
 $T \otimes Y_n$ of the defining (fundamental) representation
 $T \equiv \Box$
  and representations $Y_n$ that
include adjoint representation $Y_1=\ad$. The representations
$\Lambda_i^{(n)}$ do not appear in the decomposition of $\ad^{\otimes k}$.
The results obtained give hope that universal expressions for color factors
 (uniform for all simple gauge groups)
 can also be obtained for gauge theories that include matter fields (fermions and bosons) transforming according to the fundamental representations of the gauge groups. This could also probably be applied to constructing
 more than just adjoint knot invariants essential
 for the Vogel universality of Chern-Simons theory.

 \vspace{0.1cm}

The paper is organized as follows. In Subsection 2.1 of Section 2,
according to our approach
developed in \cite{IsKriv,IKP,IKMkr,Isa1},
we briefly describe a set of
representations that appear
in decompositions of the tensor products
 $\mathfrak{g}^{\otimes k}|_{k=2,3,...}$
for simple Lie algebras
$\mathfrak{g}=\mathfrak{sl}_N,\mathfrak{so}_N,
\mathfrak{sp}_{N=2r}$. The description of
such $\mathfrak{sl}_N$ and
$\mathfrak{so}_N$ representations was also given in recent papers
\cite{RMcr1,RMcr1a,RMcr2}.
In Subsection 2.2 of Section 2, we
recall the notion of $k$-split Casimir operators for simple
Lie algebras $\mathfrak{g}$
 and give explicit formulas needed to calculation of their eigenvalues.
 The main results of the paper are presented
  in Sections 3 and 4,
  where we consider possible generalizations of the standard Vogel universality that is only related to the study of tensor
 products $\mathfrak{g}^{\otimes k}$. In particular,
 for all simple Lie algebras $\mathfrak{g}$, we
 investigate the possibility
 of a universal
 prescription for decomposition of tensor products
 $\Box \otimes Y_n$ and $\Box \otimes Y_n'$, where
 $\Box$ denotes the defining
  (minimal fundamental for exceptional Lie algebras) representation
  of $\mathfrak{g}$. In Section 3, for all simple Lie algebras
  $\mathfrak{g}$ (except $\mathfrak{g}= \mathfrak{e}_8$),
  we find a universal characteristic identity (written
  in terms of the Vogel parameters) for the
  2-split Casimir operator in the representation $\Box \otimes Y_n$.
  Since, in the $\mathfrak{e}_8$ case, we
  have $\Box=\ad$, the decomposition of
  $\Box \otimes Y_n$ can be considered
  in the framework of the standard Vogel universality.
  In Section 4, by using the split Casimir operator
  technique, we deduce a universal form for projectors onto irreducible
   subrepresentations arising in the decomposition of $\Box \otimes Y_n$
   and find universal dimensions of these subrepresentations.
   At the end of this section, we discuss an example
 of evaluating universal colour (group) factors for an infinite
 set of special Feynman ladder diagrams in
 non-Abelian gauge theories.
   In Conclusion, we summarize the obtained results and
   outline some perspectives for further investigations.

\section{
Decomposition of $\ad^{\otimes k}$ for
 Lie algebras of classical series.
 Split Casimir operators for simple Lie
 algebras.\label{class0}}
 \setcounter{equation}0

\subsection{
Class of subrepresentations
in the decomposition of $\ad^{\otimes k}$ for
 Lie algebras $\mathfrak{sl}_N$
and $\mathfrak{so}_N$\label{class}}

Let ''$\ad$'' denote the adjoint representation of
 a Lie algebra $\mathfrak{g}$.
In this section,
following our approach
developed in \cite{IsPr1,IsKriv,IKP,IsPr2,IKMkr,IsaKri2,Isa1},
 we describe a set of 
subrepresentations  that appear
in decompositions of the tensor products
 $\ad^{\otimes k}$
of the  Lie algebras $\mathfrak{sl}_N$, $\mathfrak{so}_N$ and $\mathfrak{sp}_{N=2r}$.

Consider two irreducible
 representations (irreps) of $\mathfrak{sl}_N$ associated
with two Young diagrams (partitions) $\lambda=[\lambda_1,\lambda_2,...]$
and $\mu=[\mu_1,\mu_2,...]$. From these two irreps one can
form, by using the known procedure \cite{GrossT}, a new
irrep of $\mathfrak{sl}_N$ which is called ''composite''
representation of $\lambda$ and $\mu$ \cite{Koik}, \cite{GrossT}
(see also \cite{MMHopf} and references therein).
 The notion of the composite
 representations allows us
  to describe
   all irreps  of $\mathfrak{sl}_N$
 that appear in the decompositions of tensor products
 $\ad^{\otimes k}$.
 \newtheorem{pro2}{Proposition}[section]
 \begin{pro2}
 \label{test2}  (\cite{RMcr1}, see also \cite{Isa1}).
  For the Lie
 algebra $\mathfrak{sl}(N)$ and sufficiently large $N$, all
 irreps, which appear
 in the decomposition of ad$^{\otimes k}$,
 are associated with the Young diagrams $\Lambda$,
 with the $r+s \leq 2k$ columns,
 for which the transposed Young diagrams
 have the form\footnote{Here, in the
 square brackets, the heights of the
 columns of the diagram $\Lambda$ are indicated.}
 \be
 \lb{lambda2}
 \Lambda^{\sf T} = [N-\lambda_r', ...,N-\lambda_2',
 N-\lambda_1',\; \mu_1',\mu_2',...\mu_s'] \; ,
 \ee
 where ${\sf T}$ denotes the transposition of the
 Young diagrams $\Lambda$, the
 numbers $\lambda_1'  \geq ... \geq \lambda_r'$
 and $\mu_1'  \geq ... \geq \mu_s'$
 are the heights of the columns of two partitions
 $\lambda$ and $\mu$
 ($\lambda \vdash m$ and $\mu \vdash m$ of the integer
 $m \in [0,k]$).
 Thus, the Young diagram $\Lambda$ is defined
  by two Young diagrams $\lambda$ and $\mu$ so that
 $$
 \lambda^{\sf T} = [\lambda_1',\lambda_2',...\lambda_r'] \; , \;\;\;\;
  \mu^{\sf T} = [\mu_1',\mu_2',...\mu_s'] \; , \;\;\;\;
  |\lambda|=|\mu|=m   \; ,
 $$
 where we denote the number of cells in the Young
 diagram $\lambda $ as $|\lambda|$.
 The number of cells in $\Lambda$ is equal to
 $|\Lambda|=r \cdot N$. The irrep
 associated with the diagram
 $$
 \overline{\Lambda}^{\, \sf T} = [N-\mu_s', ...,N-\mu_2',
 N-\mu_1', \; \lambda_1',\lambda_2',...\lambda_r']
 \; ,
 $$
 is contravariant (dual) to the irrep associated
 with (\ref{lambda2}) and $|\overline{\Lambda}|=s \cdot N$.
 \end{pro2}
 {\bf Proof.}
 The adjoint representation (ad) of
 the Lie algebra $\mathfrak{sl}_N$ acts in the space of
 second rank traceless tensors with components
 $\psi^a_{\; b}$ ($\psi^a_{\; a}=0$),
 where the upper $a$ and lower $b$ indices
correspond to the defining $[1]$ and co-defining $[1^{N-1}]$
fundamental representations of $\mathfrak{sl}_N$. Then the representation
ad$^{\otimes k}$ of $\mathfrak{sl}_N$ acts in the space of tensors
with components $\psi^{a_1...a_k}_{\;\;\; b_1...b_k}$,
which are decomposed into invariant subspaces
$V_{\mu,\lambda}$ of traceless tensors
$\psi^{a_1...a_m}_{\;\;  b_1...b_m}$ $(m \leq k)$,
where the upper $\{a_1,...,a_m \}$ and lower
$\{b_1,...,b_m \}$ indices are symmetrized
according to the Young diagrams $\mu \vdash m$
and $\lambda \vdash m$:
$$
 (P_{(\mu)}\, \psi \, P_{(\lambda)})_{\;\; b_1...b_m}^{a_1...a_m} =
 P^{\; a_1...a_m}_{(\mu)\, a_1'...a_m'}\;
 \psi^{a_1'...a_m'}_{\;\;  b_1'...b_m'}\;
  P^{\; b_1'...b_m'}_{(\lambda) \, b_1...b_m}
 \; \in \; V_{\mu,\lambda} \; .
$$
Here $P_{(\lambda)}$ denotes
 the Young projector associated with the Young diagram $\lambda$.
Since any lower index $b_i$
is related to the antisymmetrization of
$N-1$ the upper indices $c_\ell|_{\ell =1,...,N-1}$ by means
of the antisymmetric tensor
$\varepsilon_{b_i c_1...c_{_{N-1}}}\psi^{c_1...c_{_{N-1}}...}_{\;\; ...}$, the invariant
subspaces $V_{\mu,\lambda}$ are associated
with the Young diagram $\Lambda$ presented in (\ref{lambda2}).
It is clear that irreducible tensors
$P_{(\lambda)}\, \psi \, P_{(\mu)}$
are contravariant to the irreducible tensors
$P_{(\mu)}\, \psi \, P_{(\lambda)} \in V_{\mu,\lambda}$
 and therefore
 the space $V_{\lambda,\mu}$ is dual
 to the space $V_{\mu,\lambda}$. \hfill \qed

 \vspace{0.1cm}


 \newtheorem{not2}{Remark\itshape}[section]
\begin{not2}\label{not2}
Any irrep $T$ of $\mathfrak{sl}_N$, which acts in the space
of the symmetrized traceless tensors 
$\psi^{a_1...a_m}_{\;\;\; b_1...b_n} \in V_{\mu,\lambda}$,
is the composite irrep of two irreps of $\mathfrak{sl}_N$
 with Young diagrams
$\mu  \vdash m ,\lambda \vdash n$, 
and $T$ is numerated
by a pair of diagrams\footnote{In the paper \cite{GrossT},
the composite representation $T=(\mu,\lambda)$ is denoted as  $\overline{\lambda}\mu$.} $(\mu,\lambda)$.
 The representations $(\mu,\lambda)$
  and $(\lambda,\mu)$ are
 contravariant (dual) to each other and this obviously means that
 $\dim(\mu,\lambda)=\dim(\lambda,\mu)$.
 The centralizer of the diagonal action of $\mathfrak{sl}_N$ on the mixed tensor space $V^{\otimes m} \otimes V^{*\otimes n}$,
 where $V=V_{[1],[\emptyset]}$, is the walled Brauer algebra $B_{m,n}$
(see, e.g., \cite{OgBul} and references therein).
 Proposition {\bf \ref{test2}}
 states that we can numerate
 irreducible subrepresen\-tations arising in the decomposition
 of ad$^{\otimes k}$ by the pairs $(\mu,\lambda)$ of Young diagrams
 $\mu \vdash m,\lambda \vdash n$ such that $m=n$
 ($|\mu|=|\lambda|$) and $|\mu|=m \leq k$. For the
$\mathfrak{sl}_N$ composite representation $T=(\mu,\lambda)$,
which is associated with the Young diagram
$\Lambda =[\Lambda_1,\Lambda_2,...,\Lambda_{N-1},\Lambda_N=0]$,
the partitions $\mu=[\mu_1,...,\mu_q]$
and $\lambda=[\lambda_1,...,\lambda_p]$ appear as
$$
(L_1,...,L_N) : = (\Lambda_{1}-L,...,\Lambda_{N-1}-L,-L)=
(\mu_1,...,\mu_q,\underbrace{0,...,0}_{N-p-q},
-\lambda_p,...,-\lambda_1) \; ,
$$
where $L=\Lambda_1-\mu_1=\lambda_1$, and the Dynkin labels are
 \be
 \lb{dynkin}
d_i =\Lambda_i-\Lambda_{i+1} = L_{i} - L_{i+1}
\; , \;\;\;\; (i=1,...,N-1) \; ,
 \ee
 such that
 \be
 \lb{dynkin2}
 (d_1,...,d_{N-1}) = (\mu_1-\mu_2,\, \mu_2-\mu_3,\,...,\,\mu_q,\,
 \underbrace{0,...,0}_{N-p-q-1},\,\lambda_p,\,\lambda_{p-1}-\lambda_p,\,
 ...,\, \lambda_{1}-\lambda_2)\; .
 \ee
 This parametrization of
 composite $\mathfrak{sl}_N$ representations $(\mu,\lambda)$
 was used in \cite{RMcr1a}, \cite{RMcr1} and \cite{RMcr2} to explore
 duality symmetry for Casimir eigenvalues
 and to find their universal expressions.
\end{not2}


An example of the notation used in Proposition {\bf \ref{test2}}
and in Remark {\bf \ref{not2}} is
 \be
  \lb{hidcol}
{\scriptsize \Lambda = \;\;
  N\!\! \left\{
\begin{array}{|c|c|c|c|c|c|c|}
\hline
\, & \;  & \;   & \; & \mu & \mu & \mu  \\ [0.1cm] \hline
\, &  \;  & \; & \;  & \mu  & \multicolumn{2}{c}{} \\ [0.1cm] \cline{1-5}
\vdots & \vdots  & \vdots  & \vdots &
\multicolumn{3}{c}{} \\ [0.1cm] \cline{1-4}
\, &  \; & \;  & \;  & \multicolumn{3}{c}{} \\ [0.1cm]
 \cline{1-4}
 \, &  \; & _\lambda  & _\lambda  & \multicolumn{3}{c}{} \\ [0.1cm]
 \cline{1-4}
 _\lambda &  _\lambda & _\lambda  & _\lambda  &
 \multicolumn{3}{c}{} \\ [0.1cm]
 \cline{1-4}
\end{array} 
 \right. }
\!\!\!\!\!\! \leftrightarrow \,
{\footnotesize
\left\{ \begin{array}{l}
\!\! \Lambda^{T} = [N_{-1}^2,N_{-2}^2,2,1^2] \;\;\; \leftrightarrow \;\;\;
(\mu,\lambda) = ([3,1],[4,2]) \, ; \\  [0.2cm]
\!\! \Lambda = [7,5,4^{N-4},2] , \;\; L=4 ,
\;\;
(L_1,...,L_N) = (3,1,0,...0,-2,-4),  \\  [0.2cm]
\!\! (d_1,...,d_{N-1}) = (2,1,0,...,0,2,2)\; .
\end{array} \right.  }
  \ee
Here $N_{-k}^\ell :=(N-k)^\ell$, the indices $\lambda$ and $\mu$ label cells
of the diagrams $\lambda$ and $\mu$
 and cells with indices
$\lambda$ are excluded from the diagram $\Lambda$.

    \vspace{0.2cm}


 \newtheorem{not2b}[not2]{Remark\itshape}
\begin{not2b}\label{not2b}
 The dimensions of the
 irreducible representations of $\mathfrak{sl}_N$
 associated to the Young diagram
 in (\ref{lambda2}) are evaluated by means of the
 useful formula
 \be
\lb{abc7}
\dim_{_{\mathfrak{sl}(N)}} (\Lambda) =
\prod_{i=1}^k \frac{(N+i-1)! }{(N-a_i +i-1)!(a_i+k-i)!} \prod_{\ell<j} (a_\ell-a_j+(j-\ell))  \; ,
\ee
where $\Lambda^{\sf T}=[a_1,a_2,a_3...,a_k]$ is
the Young diagram transposed to $\Lambda$,
i.e. $a_1,a_2,a_3...,a_k$ -- are the
 heights of the columns of $\Lambda$.
 Formula (\ref{abc7}) is derived from the
 Weyl hook formula.
 The quantum analog ($q$-deformation) of formula (\ref{abc7})
  is discussed in \cite{Isa2} (Remark 1 in Subsect. 4.3.6).
  With the help of (\ref{abc7}), for $|\lambda|=|\mu|$,
   one can prove that
  \be
  \lb{dim2T}
\left. \dim(\mu,\lambda) \right|_{N \to -N}
=\dim(\mu^{\sf T},\lambda^{\sf T}) \; ,
  \ee
 i.e. $\dim(\mu^{\sf T},\lambda^{\sf T})$ is obtained from
 $\dim(\mu,\lambda)$ by formal exchange $N \to -N$.
 Identity (\ref{dim2T}) is a special case of the general
 statement \cite{RMcr1a}
 on duality $N \leftrightarrow -N$
 for the composite representations $(\mu,\lambda)$:
 \be
  \lb{dim2Tb}
\left. \dim(\mu,\lambda) \right|_{N \to -N}
= (-1)^{|\lambda|+|\mu|} \, \dim(\lambda^{\sf T},\mu^{\sf T}) \; .
  \ee
 \end{not2b}

   \vspace{0.2cm}

  \newtheorem{not3}[not2]{Remark\itshape}
\begin{not3}\label{not3}
To construct
 decompositions of ad$^{\otimes k}$ into
 irreducible representations, we need to find
  the decomposition of the tensor products
  $\Lambda \times \ad \equiv (\lambda,\mu) \times ([1],[1])$,
  where a pair of diagrams $\Lambda=(\lambda,\mu)$
  characterizes an irrep appearing in the decomposition
  of ad$^{\otimes (k-1)}$.
  Since $\lambda \vdash m$ and $\mu \vdash m$
  are two Young diagrams associated with the symmetrizations
  of the lower and upper indices in the traceless tensors
  $\psi^{a_1...a_m}_{\;\; b_1...b_m}$ and $\ad=([1],[1])$
  associates with traceless tensors
  $\psi^{a}_{\;\; b}$, we need to extract invariant
  irreducible subspaces in the space of tensors
  $\psi^{a_1...a_m}_{\;\; b_1...b_m} \psi^{a}_{\;\; b}$.
  This can be done by making a special symmetrization of
  the upper $(a_1,...,a_m,a)$ and lower $(b_1,...,b_m,b)$ indices
  separately and by contracting one of the upper indices
   in $(a_1,...,a_m)$
  with $b$ and one of the lower indices in $(b_1,...,b_m)$
  with $a$. Thus, we arrive at the rule
  \be
  \lb{rule01}
  (\lambda,\mu) \times ([1],[1]) =
  (\lambda \times [1],\mu \times [1]) +
  (\sum_i \lambda_{i} \times [1],\mu) +
  (\lambda ,\sum_j  \mu_{j}\times [1]) +
  \sum_{i,j} (\lambda_{i},\mu_{j}),
  \ee
  where in the right-hand side
  we denote by tensor
  products $\lambda \times [1], \mu_{j}\times [1], ....$
  their decompositions in the irreducible
  components, while  $\lambda_{i}$ and $\mu_{j}$
  are the Young diagrams obtained by removing from
  diagrams $\lambda$ and $\mu$ one of their corner cells
  numbered respectively by $i$ and $j$.
  The special case of formula (\ref{rule01}), which is needed
  for calculation $([1],[1])^{\otimes k}$, is written as follows:
 \be
  \lb{rule02}
   \begin{array}{l}
  ([1]^{\otimes p},[1]^{\otimes p}) \times ([1],[1]) =   \\ [0.2cm]
 \quad \quad = ([1]^{\otimes (p+1)},[1]^{\otimes (p+1)}) +
  2 p \, ([1]^{\otimes p},[1]^{\otimes p}) +
 p^2 \, ([1]^{\otimes (p-1)},[1]^{\otimes (p-1)}) \, ,
 \end{array}
  \ee
  where we use the concise notation
  \be
  \lb{rule05}
 ([1]^{\otimes |\lambda|}, \; [1]^{\otimes |\mu|}) =
 (\sum_{\lambda \vdash |\lambda|}
 f_{(\lambda)} \; \lambda, \; \sum_{\mu \vdash |\mu|}f_{(\mu)} \; \mu)=
 \sum_{\lambda,\mu} f_{(\lambda)}\, f_{(\mu)}
 (\lambda, \; \mu) \; .
 \ee
 Here the sums are over all Young diagrams $\lambda$
 and $\mu$ with $|\lambda|$ and $|\mu|$ boxes, while
 multiplicities
 $f_{(\lambda)}$ are equal to the number of the standard
 Young tableaux of the shape $\lambda$; this number is given
 by the famous hook formula (see e.g. \cite{IsRub2},
 section 4.3.2).
 \end{not3}

  \noindent
  {\bf Examples 1.} According to the rules (\ref{rule01}) and
  (\ref{rule02}),
  the decomposition of $\ad^{\otimes 2}$
  (the case $p=1$ in (\ref{rule02}))  is
  \be
  \lb{ad2sl1}
  \begin{array}{c}
  ([1],[1])^{\otimes 2} = ([1]^{\otimes 2},[1]^{\otimes 2}) +
  2 \, ([1],[1]) + ([\emptyset],[\emptyset]) = \\ [0.2cm]
 = ([2]+[1^2],[2]+[1^2]) +
  2 \dot ([1],[1]) + ([\emptyset],[\emptyset])
 = \\ [0.2cm]
 = ([2],[2]) + \bigl(([2],[1^2]) + ([1^2],[2])\bigr)_{_{d}} +
 ([1^2],[1^2]) + 2 \dot ([1],[1]) + ([\emptyset],[\emptyset]) \; ,
  \end{array}
  \ee
  or in the standard Young diagram notation we have
   \be
  \lb{ad2sl2}
  \begin{array}{c}
  [2,1^{N-2}]^{\otimes 2} = [4,2^{N-2}] +
  \bigl([3^2,2^{N-3}]+[3,1^{N-3}]\bigr)_{_{d}}
  + [2^2,1^{N-4}]+2 [2,1^{N-2}] + [\emptyset] \; .
  \end{array}
  \ee
  The brackets $(...)_{_{d}}$ refer to the sum of
  two Young diagrams that are dual to each other.
  For a more complicated case $m=3$
  (we take in (\ref{rule01}) the partitions
  $\lambda \vdash 3$ and $\mu\vdash 3$), we have
  $$
  \begin{array}{l}
  ([3],[2,1]) \times ([1],[1]) = \\ [0.3cm]
  = ([3] \times [1],[2,1] \times [1]) +
  ([2] \times [1],[2,1]) + ([3] ,([2] + [1^2]) \times [1])
  + ([2] ,[2] + [1^2])= \\ [0.3cm]
  = \bigl([4]+[3,1] ,[3,1] + [2^2] + [2,1^2]\bigr) +
  \bigl(([3]+ [2,1]),[2,1]\bigr) \; + \\ [0.2cm]
  + \bigl([3] ,([3]+ 2 [2,1] + [1^3])\bigr) + ([2] ,[2])
   + ([2] ,[1^2]) = \\ [0.3cm]
  = \bigl([4],[3,1]\bigr) + \bigl([4],[2^2]\bigr)
   + \bigl([4],[2,1^2]\bigr) +
  \bigl([3,1] ,[3,1] + [2^2] + [2,1^2]\bigr) + \\ [0.2cm]
  + 3 \bigl([3],[2,1]\bigr) + \bigl([2,1],[2,1]\bigr)
  + \bigl([3] ,[3]\bigr) + \bigl([3],[1^3]\bigr)
   + ([2] ,[2]) + ([2] ,[1^2]) \; ,
  \end{array}
  $$
  or in the standard Young diagram notation
  $$
 \begin{array}{l}
 [5, 2^{N-3},1] \times [2, 1^{N-2}]_{s.d.}
 = [7,3^{N-3},2]  + [6, 2^{N-3}] + [6, 2^{N-4},1^2]
 + \\ [0.2cm]
 + [6,4, 3^{N-4}, 2]_{s.d.} + [5,3, 2^{N-4}]
+ [5, 3, 2^{N-5}, 1^2] + \\ [0.2cm]
+ 3[5, 2^{N-3},1]
 + [4, 3, 2^{N-4}, 1]_{s.d.} +  [6, 3^{N-2}]_{s.d.}
 + \\ [0.2cm]
 + [4, 1^{N-4}] + [4, 2^{N-2}]_{s.d.} + [3, 1^{N-3}] \; ,
\end{array}
  $$
  where $[\Lambda]_{s.d.}$ denotes a self-dual Young diagram
   which in terms of the composite notation $(\mu,\lambda)$
   corresponds to the choice $\mu=\lambda$.
  Thus, we have a remarkably effective method of the
  Clebsch-Gordan
  decomposition of the representations $\ad^{\otimes k}$,
  which is drastically simplifies the application
  of the Littlewood-Richardson rule in the case of the Lie algebra
  $\mathfrak{sl}_N$.

  \vspace{0.2cm}

 \noindent
 {\bf Examples 2.} According to Proposition
 {\bf \ref{test2}} and making use of the rules of
 Remark {\bf \ref{not3}},
 the decomposi\-tions of the representations ad$^{\otimes k}$
 (for $k=3,4,5$; the case $k=2$ is given in (\ref{ad2sl1})) are  written
 in terms of pairs of diagrams $(\lambda,\mu)$
  for the $\mathfrak{sl}_N$ algebra  as follows
 \be
 \lb{ad3sl}
  \begin{array}{c}
  ([1],[1])^{\otimes 3} = 2\, ([\emptyset],[\emptyset]) +
  9\, ([1],[1]) +6\, ([1]^{\otimes 2},[1]^{\otimes 2}) +
  ([1]^{\otimes 3},[1]^{\otimes 3}) =
  \\ [0.2cm]
 =  2\, ([\emptyset],[\emptyset]) +
  9\, ([1],[1]) +6\, \bigl( ([2], [2]) + ([1^2], [2])+
 ([2],[1^2])+ ([1^2],[1^2]) \bigr) +   \\ [0.3cm]
 + ([3],[3]) +([3],[1^3])+([1^3],[3]) + ([1^3],[1^3])
  + 4([2,1],[2,1]) +\\ [0.2cm]
 + 2 \bigl( ([2,1],[3])+  ([2,1],[1^3])+([3],[2,1])+
 ([1^3],[2,1])\bigr) \; ,
 \end{array}
 \ee
 {\small \be
 \lb{ad4sl}
  \begin{array}{c}
 ([1],[1])^{\otimes 4}  = 9([\emptyset],\cdot)
 + 44([1],\cdot)   + 42 ([1]^{\otimes 2},[1]^{\otimes 2})
  + 12 ([1]^{\otimes 3},[1]^{\otimes 3})
 + ([1]^{\otimes 4},[1]^{\otimes 4})
 =  \\ [0.3cm]
 = 9\, ([\emptyset],\cdot) +
  44\, ([1],\cdot) +42 \, ([2],\cdot)+42 \, ([1^2], [2])_{_{dual}}
  +42 \, ([1^2],\cdot)  +
  12([3],\cdot)  +  \\ [0.2cm] + 48([2,1],\cdot) + 12([1^3],\cdot)+
 24([2,1],[3])_{_{dual}} +  24([2,1],[1^3])_{_{dual}} +12([3],[1^3])_{_{dual}}  +
 \\ [0.2cm]
  + ([4],\cdot)+9([3,1],\cdot)+ 4([2^2],\cdot)
 +9([2,1^2],\cdot)+([1^4],\cdot) + 3([4],[3,1])_{_{dual}} +
  \\ [0.2cm]
\!\!\!\!\!\!\! + 2([4],[2^2])_{_{dual}}+ 3([4],[2,1^2])_{_{dual}}
 +([4],[1^4])_{_{dual}}
  + 6 ([3,1],[2^2])_{_{dual}} + 9 ([3,1],[2,1^2])_{_{dual}}
 + \\ [0.2cm] + 3([3,1],[1^4])_{_{dual}}
 + 6([2^2],[2,1^2])_{_{dual}} + 2([2^2],[1^4])_{_{dual}}
 +3([2,1^2],[1^4])_{_{dual}}   \; ,
 \end{array}
 \ee }
  \be
 \lb{ad5sl}
  \begin{array}{c}
 ([1],[1])^{\otimes 5} = 44 \, ([\emptyset],[\emptyset]) +
  265 \, ([1],[1]) + 320 \, ([1]^{\otimes 2},[1]^{\otimes 2}) +
  \\ [0.2cm]
  + 130 \, ([1]^{\otimes 3},[1]^{\otimes 3}) +
  20\, ([1]^{\otimes 4},[1]^{\otimes 4}) +
  ([1]^{\otimes 5},[1]^{\otimes 5}) \; .
 \end{array}
 \ee
 Here
 we use the concise notation
 $([1]^{\otimes k},[1]^{\otimes k})$,
which is defined in (\ref{rule05}), where
$$
\begin{array}{c}
[1]^{\otimes 2} = [2]+[1^2] , \;\;
[1]^{\otimes 3} = [3]+2 [2,1]+[1^3] , \;\;
[1]^{\otimes 4} = [4]+3[3,1]+ 2[2^2]
 +3[2,1^2]+[1^4], \\ [0.2cm]
 [1]^{\otimes 5} = [5]+4[4,1]+ 5[2^2,1]+ 6[3,1^2]
 +5[3,2] +4[2,1^3]+[1^5] , \;\;\; \dots ,
\end{array}
$$
and
$$
(\mu,\cdot):=(\mu,\mu) \; , \;\;\;\;\;\;
 (\mu,\lambda)_{_{dual}} := (\mu,\lambda)+(\lambda,\mu) \; .
$$

 The general statement is the following.
  \newtheorem{pro3}[pro2]{Proposition}
 \begin{pro3}\label{pro3}
 The representation ad$^{\otimes m}$
   for $\mathfrak{sl}_N$
 is decomposed into the sum of the irreps
 (written in terms of pairs of diagrams $(\lambda,\mu)$)
 as follows
 \be
 \lb{adksl}
 ([1],[1])^{\otimes m} = \sum_{k=0}^{m} a_k^{(m)} \;
 \, ([1]^{\otimes k},[1]^{\otimes k}) \; ,
 \ee
 where $([1]^{\otimes k},[1]^{\otimes k})|_{k=0}:=
 ([\emptyset],[\emptyset])$ and the coefficients $a_k^{(m)}$
 satisfy the recurrence relation
  \be
 \lb{adksl2}
 a_k^{(m+1)} =  a_{k-1}^{(m)} + 2k \, a_{k}^{(m)} +
 (k+1)^2 \, a_{k+1}^{(m)} \; , \;\;\;\;\;
 a_{k>m}^{(m)} = 0  =  a_{k<0}^{(m)} \; .
 \ee
 In particular, we have
 \be
 \lb{adksl3}
 \begin{array}{c}
 a_m^{(m)} = 1 \; , \;\;\;\;\;  a_0^{(m+1)} = a_1^{(m)}
\; , \;\;\;\;\;  a_{m-1}^{(m)} = m(m-1) \; , \\ [0.2cm]
a_{m-1}^{(m+1)} = \frac{1}{2}m(m^3+1)\; , \;\;\;\;\;
a_{m-1}^{(m+2)} = \frac{1}{6}m(m+1)(m+2)(m^3+2m-1)\; ,
\end{array}
 \ee
 and the coefficient $a_0^{(m)}$ is equal to the number
 of elements $\sigma$ of the symmetric group $S_m$
 such that $\sigma(i) \neq i$ for all $i=1,...,m$
 (number of permutations of $n$ elements
 with no fixed points);
 in other words, the coefficient $a_0^{(m)}$ is the number
 of the elements $\sigma \in S_m$
 which do not have cycles of length 1.
  \end{pro3}
  {\bf Proof.} Applying the rule (\ref{rule02}), we deduce
  the relation
  $$
  ([1],[1])^{\otimes (m+1)} = \sum_{k=0}^{m} a_k^{(m)} \;
 \Bigl( ([1]^{\otimes (k+1)},[1]^{\otimes (k+1)}) +
 2k \, ([1]^{\otimes k},[1]^{\otimes k}) +
  k^2 ([1]^{\otimes  (k-1)},[1]^{\otimes  (k-1)})
 \Bigr) \; ,
  $$
from which we immediately obtain (\ref{adksl2}). To
obtain the first two identities in (\ref{adksl3}), we
take $k=m+1$ and $k=0$ in (\ref{adksl2}) and
use the initial $a^{(1)}_1 =1$ and boundary $a_{k<0}^{(m)} =0$
conditions. If we take $k=m$ in (\ref{adksl2}) and
use $a^{(m)}_m =1$ with the boundary condition $a_{k>m}^{(m)} =0$,
we deduce the equation $a^{(m+1)}_m = a^{(m)}_{m-1} + 2m$ which
is solved as the third identity in (\ref{adksl3}). Then we take
$k=(m-1)$ and $k=(m-2)$, and using the previous results, we deduce
 the recurrence equations
$$
\begin{array}{c}
a^{(m+1)}_{m-1} = a^{(m)}_{m-2} + 2m(m-1)^2 + m^2 \; , \\ [0.2cm]
a^{(m+2)}_{m-1} = a^{(m+1)}_{m-2} + m(m-1)(m^3+1) + m^3(m+1)  \; ,
\end{array}
$$
 the solutions of which give two last expressions in (\ref{adksl3}).
One can continue this procedure to obtain all coefficients
$a^{(m+k)}_{m-1}$ for fixed $k>2$.
The representation  $([1],[1])^{\otimes m}$
acts in the space of tensors
$(\psi_1)^{a_1}_{b_1} \cdots (\psi_m)^{a_m}_{b_m}$, where
$\sum_{a_i}(\psi_i)^{a_i}_{b_i}|_{b_i=a_i}=0$ for all $i=1,...,m$.
Thus, to obtain the space of the singlet representation,
we need to contract all indices $a_i$ and $b_j$
with $i \neq j$. The number of all possible
 contractions of that type is equal to the number
 of elements $\sigma \in S_m$
 such that $\sigma(i) \neq i$ (for all $i=1,...,m$)
 and at the same time gives the multiplicity
$a^{(m)}_{0}$ of the trivial subrepresentation
$([\emptyset],[\emptyset])$ in $([1],[1])^{\otimes m}$.
\hfill \qed
\newtheorem{not3a}[not2]{Remark\itshape}
\begin{not3a}\label{not3a}
The sequence
of numbers $a_0^{(m)}$ ($m=0,1,2,...$) was first considered by
L.Euler in 1809 (see \cite{Sloane}
and references therein), who
also proved two recurrences
$$
a_0^{(m)} = (m-1) \bigl(a_0^{(m-1)} + a_0^{(m-2)} \bigr) \; , \;\;\;\;
a_0^{(m)} = m \, a_0^{(m-1)} + (-1)^m \; .
$$
These recurrences give $a_0^{(m)} =
m! \, \sum\limits_{k=0}^m (-1)^k/k!$. The first ten numbers $a_0^{(m)}|_{m=0,1,...,9}$ are
$1,0,1,2,9,44,265,1854,14833,133496$.
\end{not3a}

 \vspace{0.2cm}

 An analog of Proposition {\bf \ref{test2}}
 for the case of the $\mathfrak{so}(N)$ algebra is formulated as follows.
 \newtheorem{pro2a}[pro2]{Conjecture}
 \begin{pro2a}\lb{pro2a}
 For the Lie
 algebra $\mathfrak{so}(N)$ and sufficiently large $N$,
 all irreps, which appear
 in the decomposition of ad$^{\otimes k}$,
 are associated with the Young diagrams $\lambda \vdash 2m$
 with $m \leq k$ and
 with the number of columns not bigger than
 $k$. All these diagrams are obtained from
 the diagram $[k^2]$ by allowable shifting of boxes to the left,
 e.g. $[k^2] \to [k,k-1,1] \to [k,k-2,2] \to [k,k-2,1^2] \to .... $
 and allowable\footnote{Here the word allowable means
 that as a result of the cancelations we again obtain Young
 diagrams but with a reduced number of boxes.}
 cancelation of pairs of boxes in one row.
 \end{pro2a}

\vspace{0.2cm}

  \noindent
 {\bf Examples 3.} For the decompositions of
 the $\mathfrak{so}_N$ representations ad$^{\otimes 2}$,
  ad$^{\otimes 3}$ and ad$^{\otimes 4}$ we have
 (see e.g. \cite{IKP}, \cite{IKMkr})
  {\small  \be
 \lb{soad23}
 \begin{array}{c}
 [1^2] \otimes [1^2] =
 \emptyset + [1^2] + [2] + [2^2] + [2,1^2]+ [1^4] \; , \\ [0.3cm]
 [1^2]^{\otimes 3} = \emptyset + 6 \cdot [1^2] + 3 \cdot [2] +   3 \cdot [2^2] +
 6 \cdot [2,1^2]  + 3 \cdot  [1^4] +   \\ [0.2cm]
 + [1^6]+ 3 \cdot[3,1] +
 2 \, [2,1^4] + [2^3]  + [3^2]  + 3 \, [2^2,1^2]
 +  [3,1^3]  + 2 \cdot[3,2,1]  \; ,
  \end{array}
 \ee }
 {\small  \be
 \lb{ad4}
 \begin{array}{c}
 [1^2]^{\otimes 4} = 6 \cdot \emptyset + 120 \cdot [1^2]  +
 18 \cdot [2]    + 24 \cdot [2^2] +  39 \cdot [2,1^2]
 + 21 \cdot [1^4]  +   21 \cdot [3,1] +   \\ [0.2cm]
 + 6 \cdot [1^6]  + 18 \cdot [2,1^4] + 12 \cdot [2^3]
 + 6 \cdot [3^2]  + 24 \cdot [2^2,1^2]
  + 18 \cdot [3,1^3] + 24 \cdot  [3,2,1] +
  \\ [0.2cm]
   + [1^8] + 3 \cdot [2,1^6] + 3 \cdot [4]
 + 6 \cdot [4,1^2] + 6 \cdot [4,2]   + \\ [0.2cm]
 +  6 \cdot [2^2,1^4] + 3 \cdot [2^4] + 3 \cdot  [3,1^5]
 + 6 \cdot [2^3,1^2] + [4^2] + [4,1^4] + 8\cdot [3,2,1^3] +
  \\ [0.2cm]
  + 3\cdot [3^2,2] + 6 \cdot [3^2,1^2]  + 7 \cdot [3,2^2,1] +
  2\cdot [4,2^2]  +  3\cdot [4,2,1^2] +  3\cdot [4,3,1] \; .
 \end{array}
 \ee}
  Note that the decompositions
  in the right-hand sides
  of (\ref{soad23}) and (\ref{ad4}) are valid only for $\mathfrak{so}_N$ with a sufficiently large $N$.
 Indeed, the decomposition in (\ref{ad4})
is valid only for $N > 16$, since for $\mathfrak{so}_M$
 (where $M < 16$) the representation $[1^8]$ is equivalent
 to the representation $[1^{M-8}]$, while
 for $\mathfrak{so}_{16}$
 the representation $[1^8]$ splits into two
 subrepresentations
 (self-dual and anti-self-dual), etc.

  \vspace{0.2cm}


 \newtheorem{not4}[not2]{Remark\itshape}
\begin{not4}\label{not4}The dimensions of the
 irreducible tensor representations of $\mathfrak{so}_N$
 (associated with the Young diagrams $\Lambda$)
 are evaluated by means of the 
 convenient formula (cf. (\ref{abc7}))
  \be
\lb{dimak}
\begin{array}{rl}
\displaystyle
 {\rm dim}_{\mathfrak{so}_{_N}}(\Lambda)
&
\displaystyle
= \prod\limits_{i=1}^k  \frac{(N+2(i-1))!}{
(N-a_i+k-2+i)!(a_i+k-i)!} \; \cdot
\\ [0.5cm]
& \cdot \prod\limits_{\ell<j} (a_\ell-a_j + j-\ell))
(N-(a_\ell+a_j) + (\ell+j-2))) \; .
\end{array}
\ee
where $a_1,a_2,...,a_k$ -- are the heights of the columns of
the diagram $\Lambda$, i.e.
$\Lambda^{T} = [a_1,a_2,...,a_k]$
is the Young diagram transposed to $\Lambda$.
  A quantum analog of formula (\ref{dimak})
  (its $q$-deformation)
 was found in \cite{Isa2} (see Remark 1 in Subsect. 4.5.5).
 \end{not4}

 \vspace{0.2cm}

\subsection{Split Casimir operators and
universal description of Lie algebras}

 \subsubsection{$k$-split Casimir operators
 and their eigenvalues}

Let $X_a$ $(a=1,...,\dim(\mathfrak{g}))$
be the generators of the enveloping
 algebra ${\cal U}(\mathfrak{g})$ of the simple Lie algebra
  $\mathfrak{g}$ with the commutation
   relations $[X_a, \; X_b] = X_d \; X^d_{ab}$.
  We define the Cartan-Killing metric by the
  formula
   \be
 \lb{defMK}
 {\sf g}_{ab} = Tr({\rm ad} X_a
\cdot {\rm ad} X_b) = X^d_{ac} X^c_{ad} \; ,
\ee
and define the quadratic
Casimir operator
\be
 \lb{Qcas}
 c_{(2)} = {\sf g}^{ab} X_a X_b \; \in \;  {\cal U}(\mathfrak{g}) \; ,
 \ee
where ${\sf g}^{ab}$ is inverse to the Cartan-Killing metric
(\ref{defMK}). The element (\ref{Qcas}) is central in
${\cal U}(\mathfrak{g})$ and its image
in any irreducible representation $T$
has the form $T(c_{(2)})= c_{(2)}^{(T)} I_T$, where
$c_{(2)}^{(T)}$ is the eigenvalue
 of $c_{(2)}$ and $I_T$
 is the unit operator in the space of the irrep $T$. For any irrep $T$
 of the simple Lie algebra $\mathfrak{g}$ we have (cf. (\ref{defMK})
 \be
 \lb{casnorm1}
 {\rm Tr}\bigl( T(X_a) T(X_b) \bigr) = d^{(T)} \,  {\sf g}_{ab} \; ,
 \ee
  where the parameter $d^{(T)}$ characterizes the irrep $T$ and
 is related to the eigenvalue $c_{(2)}^{(T)}$:
 \be
 \lb{casnorm2}
 \begin{array}{c}
 {\rm Tr}(T(c_{(2)}))= {\sf g}^{ab} {\rm Tr}\bigl( T(X_a) T(X_b) \bigr) =
 {\sf g}^{ab}  {\sf g}_{ab} \, d^{(T)} = d^{(T)} \, \dim \mathfrak{g}
  \, , \\ [0.2cm]
  {\rm Tr}(T(c_{(2)}))= c_{(2)}^{(T)} {\rm Tr}(I_T)=
  c_{(2)}^{(T)} \, \dim T   \;\;\; \Rightarrow \;\;\;
 d^{(T)} = c_{(2)}^{(T)} \, \frac{\dim T}{\dim \mathfrak{g}} \; .
 \end{array}
 \ee
The definition of the quadratic Casimir operator (\ref{Qcas})
   fixes its value in the adjoint representation
 so that $c_{(2)}^{(\ad)}=1$. Indeed \cite{IsRub1},
in view of the definition
of the metric (\ref{defMK}), we have $ d^{(\ad)} =1$ and since
$\dim (\ad) = \dim \mathfrak{g}$, we obtain
 $c_{(2)}^{(\ad)}=1$ from (\ref{casnorm2}).

 To simplify the notation, we denote
  $$
  C \equiv c_{(2)} = {\sf g}^{ab} X_a X_b \; .
  $$
Then we introduce the $k$-split Casimir operator \cite{IKP}
 \be
 \lb{char03}
\hat{C}_{(1...k)} := \sum_{i <j}^k \hat{C}_{ij} =
\frac{1}{2} \bigl(\Delta^{k-1} (C) - \sum_{i=1}^k  C_i \bigr) \; \in \;
{\cal U}(\mathfrak{g})^{\otimes k} \; ,
 \ee
where
$$
\hat{C}_{ij} := {\sf g}^{ab} \bigl(I^{\otimes (i-1)} \otimes X_a \otimes
I^{\otimes (j-i-1)} \otimes X_b \otimes I^{\otimes (k -j)} \bigr)
\, , \;\;\;\;
C_\ell := I^{\otimes (\ell-1)} \otimes C \otimes I^{\otimes (k -\ell)}\, ,
$$
and $\Delta$ is the comultiplication in
${\cal U}(\mathfrak{g})$ (see e.g. \cite{IsRub1}, sect. 3.7.4):
$$
\Delta(X_a) = X_a \otimes I + I \otimes X_a \, , \;\;\;
\Delta^{k-1} := (\Delta \otimes I^{\otimes (k-2)}) \cdots
(\Delta \otimes I) \Delta \; .
$$

\vspace{0.2cm}

For the decomposition of the tensor product of
 $k$ irreducible representations $T^{(\Lambda_i)}$
 of the algebra $\mathfrak{g}$:
 \be
 \lb{tttt}
 T^{(\Lambda_1)} \otimes \cdots \otimes T^{(\Lambda_k)} =
 \sum_\Lambda T^{(\Lambda)}
 \ee
 ($\Lambda,\Lambda_i$ -- are the highest weights
 of the representations),
 we deduce \cite{IKP} (taking into account the definition of the $k$-split
 Casimir operator (\ref{char03}))
 a formula for the eigenvalues $\hat{c}_{(1...k)}^{(\Lambda)}$
  of the $k$-split Casimir
  operator
  $\hat{C}_{\Lambda_1,\dots,\Lambda_k} =
   \bigl(T^{(\Lambda_1)} \otimes \cdots \otimes T^{(\Lambda_k)}\bigr)
   \hat{C}_{(1...k)}$ which acts on the subspace
   $V_\Lambda \subset V_{\Lambda_1} \otimes \cdots \otimes V_{\Lambda_k}$ of
   the irrep $\Lambda$:
 \be
 \lb{gfad}
 \hat{c}_{(1...k)}^{(\Lambda)} = \frac{1}{2} \bigl(c_{(2)}^{(\Lambda)} -
 \sum_{i=1}^k c_{(2)}^{(\Lambda_i)} \bigr) \; .
 \ee
 Here $c_{(2)}^{(\Lambda)}, c_{(2)}^{(\Lambda_i)}$ are eigenvalues of the
  quadratic Casimir operator (\ref{Qcas})
  on the spaces $V_\Lambda,V_{\Lambda_i}$
  of irreps with the highest weights $\Lambda,\Lambda_i$.
  In the representation of $\mathfrak{g}$ with the highest
weight $\Lambda$, due to the well-known formula,
the quadratic Casimir operator $C$ has
the eigenvalue
  \be
  \lb{2casim}
 C_{(2)}(\Lambda) = (\Lambda, \, \Lambda + 2\rho) \; ,
 \ee
 where $(.,.)$ is a scalar product in the root space of $\mathfrak{g}$
 and $\rho$ is the Weyl vector for the simple
 Lie algebra $\mathfrak{g}$ of the rank $r$:
 \be
 \lb{weylv}
 \rho = \frac{1}{2} \sum_{\alpha >0} \alpha =
 \sum_{f=1}^r \lambda_{(f)} \;
 \ee
 (the summations are over positive roots $\alpha$
 and fundamental weights $\lambda_{(f)}$).
  The choice of the root space metric
 for calculating the scalar product in the right-hand side of (\ref{2casim}) must be consistent with the definitions (\ref{defMK})
 and (\ref{Qcas})
 which give $c_{(2)}^{(\rm ad)}  = 1$. In fact, we can
 choose an arbitrary root space metric, but then
 we should normalize the values (\ref{2casim}) in a special way
   \be
 \lb{2cas1}
 c_{(2)}^{(\Lambda)} : =
 \frac{C_{(2)}(\Lambda)}{C_{(2)}(\Lambda_{(\rm ad)})} \; ,
\ee
($\Lambda_{(\rm ad)}$ is the highest weight of
the adjoint representation of $\mathfrak{g}$)
such that
 the value of the
  quadratic Casimir operator is automatically fixed
  in the adjoint representation
  as $c_{(2)}^{(\rm ad)}  = 1$. We stress that
  another convenient choice of the root space metric
  is such that \cite{LanMan}
  \be
  \lb{C2abg}
  C_{(2)}(\Lambda_{(\rm ad)})=2{\sf t}=
  2(\alpha+\beta+\gamma) \; ,
  \ee
   where Vogel parameters
  $\alpha,\beta,\gamma$ are given in Table {\sf \ref{tab1}}.

 \vspace{0.3cm}

 In \cite{IsKriv,IKP,IKMkr,IsaKri2},
 we considered the cases $\ad^{\otimes k}$  when all
 representations $T^{(\Lambda_i)}$ in the left-hand side
 of (\ref{tttt}) are the adjoint representations
 $\Lambda_i = \Lambda_{(\rm ad)}$. In this case,
 the value of the $k$-split Casimir operator is
\begin{equation}
\lb{casik}
\hat{c}^{(\Lambda)}_{(1...k)} =
\frac{1}{2}(c_{(2)}^{(\Lambda)}-k \, c_{(2)}^{(\rm ad)})
=\frac{1}{2}c_{(2)}^{(\Lambda)}- \frac{k}{2} \; ,
\end{equation}
where $c_{(2)}^{(\Lambda)}$ are defined in (\ref{2cas1}),
and we substituted $c_{(2)}^{(\rm ad)} = 1$.

\vspace{0.2cm}

  \noindent
 {\bf Examples 4.} The tensor representations of
 $\mathfrak{sl}_N$, $\mathfrak{so}_N$
 and $\mathfrak{sp}_N$ are
 characterized by the Young diagrams $[\lambda_1,...\lambda_m]$, and
 the highest weights $\Lambda_{[\lambda_1,...\lambda_m]}$ are equal to
 (see e.g. sect. 3.4.2 in \cite{IsRub2})
  \be
 \lb{weight}
 \begin{array}{l}
  {\rm for} \; \mathfrak{sl}_N: \;\;\;\;
 \Lambda_{[\lambda_1,...\lambda_{N-1}]} =
 \sum\limits_{i=1}^{N-1}
 \bigl(\lambda_i - \frac{|\lambda|}{N}\bigr) e^{(i)}
 - \frac{|\lambda|}{N}\,  e^{(N)} ,
 \;\;\;\; |\lambda| :=
 \sum\limits_{i=1}^{N-1} \lambda_i
 \;\;\;\;\; (\lambda_i|_{_{i>m}}=0) \, ; \\ [0.4cm]
 {\rm for} \; \mathfrak{so}_N \; {\rm and} \;\mathfrak{sp}_{N=2r}: \;\;\;\;
 \Lambda_{[\lambda_1,...\lambda_m]} =
 \sum\limits_{i=1}^{m} \lambda_i e^{(i)} \; ,
 \;\;\;\; (m \leq [\frac{N}{2}]) \; ,
 \end{array}
 \ee
 where $e^{(i)}$ are basis vectors in the root spaces.
 In the representation of $\mathfrak{g}$ with highest
weight $\Lambda$, the quadratic Casimir operator $C$ has
the eigenvalue (\ref{2casim}), where $\rho$ is
the Weyl vector (\ref{weylv}):
 \be
 \lb{delta2}
   {\rm for} \;\; \mathfrak{sl}_N: \;\;\;\;\;\;
 \rho = \sum\limits_{i=1}^{N} \Bigl(\frac{N+1}{2}-i\Bigr)e^{(i)}\; ;
 \ee
 \be
 \lb{delta1}
 \begin{array}{c}
 {\rm for} \;\; \mathfrak{so}_N: \;\;\;
 \rho = \sum\limits_{i=1}^{[\frac{N}{2}]} (\frac{N}{2}-i)e^{(i)} \; ;
 \;\;\;\;\; {\rm for} \;\; \mathfrak{sp}_{N=2r}: \;\;\;
 \rho = \sum\limits_{i=1}^{\frac{N}{2}} (\frac{N}{2}-i+1)e^{(i)}\; .
 \end{array}
 \ee
 Finally, taking the orthonormal metric
 $(e^{(i)},e^{(j)})=\delta^{ij}$
 in the root spaces of $\mathfrak{sl}_N$ and $\mathfrak{so}_N$,
  we obtain  the eigenvalues for the quadratic Casimir
   operators (\ref{2casim})  in the $\mathfrak{sl}_N$ case
   (see e.g. \cite{GrossT}):
   \be
 \lb{c2sln}
 \begin{array}{c}
 C_{(2)}([\lambda_1,...,\lambda_{N-1}]) =
 \sum\limits_{\ell=1}^{N-1} \lambda_\ell^2 -
 2 \sum\limits_{\ell=1}^{N-1} \ell  \lambda_\ell
 +(N+1)|\lambda| - \frac{|\lambda|^2}{N}  = \\ [0.2cm]
 = -\sum\limits_{\ell=1}^{k} a_\ell^2 +
 2 \sum\limits_{\ell=1}^{k} \ell  a_\ell
 +(N-1)|\lambda| - \frac{|\lambda|^2}{N}  \; ,
 \end{array}
\ee
where $\lambda^{\sf T}=[a_1,a_2,...,a_k]$,
$|\lambda^{\sf T}|= |\lambda|$, and in
 $\mathfrak{so}_N$ case:
  \be
 \lb{c2son}
 \mathfrak{so}_N:\;\;\;
 C_{(2)}([\lambda_1,...,\lambda_m]) =
 \sum_{\ell=1}^m (\lambda_\ell^2 +  \lambda_\ell(N-2\ell) ) =
 \sum_{\ell=1}^m (\lambda_\ell^2 -
 2 \ell  \lambda_\ell) + N |\lambda| \; ,
 \ee
 where $m \leq [\frac{N}{2}]$. In the case of $\mathfrak{sp}_N$ we have:
  \be
 \lb{c2spn}
 \mathfrak{sp}_N:\;\;  C_{(2)}([\lambda_1,...,\lambda_m]) =
 \frac{1}{2} \sum_{\ell=1}^m (\lambda_\ell^2 +  \lambda_\ell(N+2-2\ell) ) =
 \frac{1}{2}  \sum_{\ell=1}^m (\lambda_\ell^2 -
 2 \ell  \lambda_\ell) + \frac{1}{2}(N+2) |\lambda|  ,
 \ee
 where $m \leq \frac{N}{2}$ and, in the root space of $\mathfrak{sp}_N$,
 according to the normalization (\ref{C2abg}),
 we choose the orthogonal metric as
 $(e^{(j)},e^{(\ell)})=\frac{1}{2} \delta^{j \ell}$.

  Recall that, for the Lie algebra $\mathfrak{g}$, the dimension of
 the representation $T^{(\Lambda)}$
 with the highest weight $\Lambda$
 is given by the Weyl formula
 \be
 \lb{weyldim}
 \dim T^{(\Lambda)} =
 \prod_{\alpha > 0} \frac{(\Lambda + \rho,\alpha)}{(\rho,\alpha)} \; ,
 \ee
where the product is performed over all
 positive roots $\alpha$ of $\mathfrak{g}$.

\subsubsection{Universal description of Lie algebras
 using the split Casimir operator technique}

Note that the eigenvalues (\ref{gfad}) can take the same value
for different irreducible representations $T^{(\Lambda)}$
in the decomposition (\ref{tttt}). We denote by $a_i$
the eigenvalues $\hat{c}_{1...k}^{(\Lambda)}$
that take mismatched values
(the index $i=1,2,...,M$ numerates these eigenvalues).
Then
 the $k$-split Casimir operator (\ref{char03}) in the representation
 (\ref{tttt}) satisfies the characteristic identity
 \be
 \lb{char05}
\prod_{i=1}^M (\hat{C}_{\Lambda_1,...,\Lambda_k} - a_i) = 0 \, ,
\ee
which leads to the definition of projectors
\be
  \lb{char06}
P_{a_j} = \prod_{i \neq j}
\frac{(\hat{C}_{\Lambda_1,...,\Lambda_k} - a_i)}{
(a_j -  a_i)} \; .
 \ee
The images of these projectors are invariant subspaces
$V_{a_j} \subset V_{\Lambda_1} \otimes \cdots \otimes V_{\Lambda_k}$
with dimensions
\be
  \lb{char07}
\dim V_{a_j}  = {\rm Tr}_{1,2,...,k} (P_{a_j})  \; .
 \ee
 Then, the trace of the $m$-th power
 of the $k$-split Casimir operator in the representation
 (\ref{tttt}) is defined by the formula
 \be
 \lb{char08}
{\rm Tr}_{12...k}(\hat{C}_{\Lambda_1,...,\Lambda_k})^m  =
 {\rm Tr}_{12...k} \Bigl( \sum_{j=1}^M \; a_j^m \; P_{a_j} \Bigr) =
 \sum_{j=1}^M \; a_j^m \; \dim V_{a_j} \; .
\ee

 As we mentioned above, the invariant subspaces $V_{a_j}$
 can be reducible since
the quadratic Casimir operator (\ref{Qcas}) can take
the same eigenvalue (\ref{gfad})
in different irreducible representations $T^{(\Lambda)}$.
 When
we consider the decomposition (\ref{tttt}) of the representation
$\ad^{\otimes k}$, the eigenvalues $a_j = \hat{c}_{1...k}^{(\Lambda)}$
are written in the form (\ref{casik}) and are determined only
by the value (\ref{2cas1}) of the quadratic Casimir
operator in the representations $T^{(\Lambda)}$. In this
case we call the direct sum of all representations $T^{(\Lambda)}$,
 that act in the subspace $V_{a_j}$,
as {\it Casimir representation} (associated
with the eigenvalue $a_j$), and call the spaces $V_{a_j}$ ($j=1,2,...,M$)
as {\it Casimir subspaces}.

Our method \cite{IsPr1,IsKriv,IKP,IsPr2,IKMkr}
of constructing the universal formulas
for simple Lie algebras $\mathfrak{g}$ is based on the
possibility to write eigenvalues $a_j = \hat{c}_{1...k}^{(\Lambda)}$
 in the universal form
$a_j(\alpha,\beta,\gamma)$ in terms of the Vogel parameters.
In all known examples (see, e.g., Table 2 in \cite{IKMkr}),
the eigenvalues (\ref{gfad}) are linear functions of
$\hat{\alpha},\hat{\beta},\hat{\gamma}$.
When considering the expansion of $\mathfrak{g}^{\otimes k}$,
it is sufficient (in view of (\ref{casik}) and (\ref{2cas1}))
to find only universal expressions for the eigenvalues (\ref{2casim})
of the quadratic Casimir operators.
The set of all representations $T^{(\Lambda)}$ (for all simple Lie algebras
 $\mathfrak{g}$) associated with
the same universal eigenvalue
$a_j(\alpha,\beta,\gamma)= \hat{c}_{1...k}^{(\Lambda)}$
is called the {\it universal multiplet}.

\vspace{0.1cm}

In the case of the Lie algebras
$\mathfrak{g}=\mathfrak{sl}_N,\mathfrak{so}_N$, an efficient method (based on several conjectures) was developed in \cite{RMcr1} and \cite{RMcr2} to obtain universal expressions
for the Casimir eigenvalues (\ref{2casim}) (for the representations  $\Lambda$ arising in the decomposition of $\mathfrak{g}^{\otimes k}$).
For representations appearing in the decomposition of $\mathfrak{g}^{\otimes 5}$, it was proved in \cite{RMcr2} that such expressions are valid for all simple Lie algebras $\mathfrak{g}$.

 \newtheorem{not5}[not2]{Remark\itshape}
\begin{not5}\label{not5}
 The dual $\mathfrak{sl}_N$
representations, denoted as $(\mu,\lambda)$ and $(\lambda,\mu)$,
 have the same value of the quadratic Casimir operator, and thus they
  are combined into a single Casimir representation
  for which we use notation
 $(\mu,\lambda)_{dual} = (\mu,\lambda)+(\lambda,\mu)$, if
 $\mu \neq \lambda$.

 The universal multiplets that include
 Casimir $\mathfrak{sl}_N$-representations 
 arising  in decompositions (\ref{ad2sl1}),
 (\ref{ad3sl}), (\ref{ad4sl}) of $\ad^{\otimes k}|_{k=2,3,4}$
 and also include related Casimir $\mathfrak{so}_N$-representa\-tions
 arising in decompositions
 (\ref{soad23}), (\ref{ad4}) of $\ad^{\otimes k}|_{k=2,3,4}$
  are summarized in \cite{IKP}, \cite{IKMkr}.
 Here, in Tables \ref{tab11}-\ref{tab13},
 we present a subset of these universal multiplets (only for the cases
 $\ad^{\otimes 2}$ and $\ad^{\otimes 3}$) that combine
 the Casimir $\mathfrak{sl}_N$ and $\mathfrak{so}_N$
 representations expressed in terms of the Young diagrams.
  We denote
the singlet representations by $[\emptyset]$, and
the dash indicates a representation whose dimension is zero.

  \begin{table}[h]
  \centering
  \caption{ } \label{tab11}
	 \vspace{0.2cm}
	\begin{tabular}{|c|c|c|c|c|c|c|}
		\hline
		$\;\;$ & $X_0$ & $X_1=Y_1$ & $X_2$&  $Y_2$ &
		$Y_2'$ & $Y_2''$ \\
		\hline
		$\mathfrak{sl}_N$
&\footnotesize $([\emptyset],[\emptyset])$
  &\footnotesize $([1],[1])$   &\footnotesize
  $([2],[1^2])_{dual}$ &
		\footnotesize $([2],[2])$ &\footnotesize  $([1^2],[1^2])$
  &\footnotesize  $([1],[1])$	 \\
		\hline
		$\mathfrak{so}_N$  &
\footnotesize $[\emptyset]$ &
		\footnotesize $[1^2]$&\footnotesize $[2,1^2]$ &
		\footnotesize $[2^2]$ &\footnotesize $[1^4]$
    &\footnotesize $[2]$ \\
		\hline
	\end{tabular}

  \centering
  \caption{ } \label{tab12}
	 \vspace{0.2cm}
	\begin{tabular}{|c|c|c|c|c|c|c|}
		\hline
		$\;\;$ & $X_3$ & $X_3'$ & $X_3''$&  $Y_3$ &
		$Y_3'$ & $Y_3''$ \\
		\hline
		$\mathfrak{sl}_N$ &\footnotesize $([2,1],[2,1])$
  &\footnotesize  $([3],[1^3])_{dual}$   &\footnotesize  --
  &\footnotesize  $([3],[3])$
&\footnotesize  $([1^3],[1^3])$
   &\footnotesize  $([\emptyset],[\emptyset])$
		 \\
		\hline
		$\mathfrak{so}_N$  &\footnotesize $[2^3]+[3,1^3]$ &
		\footnotesize $[3,1^3]$&\footnotesize --
      & \footnotesize $[3^2]$
    &\footnotesize $[1^6]$ &\footnotesize -- \\
		\hline
	\end{tabular}

  \centering
  \caption{ } \label{tab13}
	 \vspace{0.2cm}
	\begin{tabular}{|c|c|c|c|c|c|c|}
		\hline
		$\;\;$ &   $B$ &
		$B'$ & $B''$
		& $C$ &
		$C'$ & $C''$  \\
		\hline
		$\mathfrak{sl}_N$  &
		\footnotesize $([1^2],[1^2])$ &\footnotesize  $([2],[2])$
  &\footnotesize  $([2,1],[2,1])$
		&\footnotesize  $([3],[2,1])_{dual}$
&\footnotesize  $([2,1],[1^3])_{dual}$
   &\footnotesize  -- \\
		\hline
		$\mathfrak{so}_N$  &
		\footnotesize -- &\footnotesize $[3,1]$
    &\footnotesize $[2^2,1^2]$ & \footnotesize $[3,2,1]$
    &\footnotesize $[2,1^4]$ &\footnotesize -- \\
		\hline
	\end{tabular}
\end{table}

\end{not5}

\noindent
{\bf Examples 5.} Consider $\mathfrak{sl}_N$ representations
$Y_n = ([n], \, [n])$. Recall that $Y_1 = ([1], \, [1])$
is $\ad$-representation of $\mathfrak{sl}_N$.
According to the rules (\ref{rule01})
the tensor product $Y_n \times Y_1$ is decomposed as follows
\be
\lb{yny01}
\begin{array}{c}
Y_n \times Y_1 = ([n], \, [n])\times ([1], \, [1]) =
([n+1], \, [n+1])
+ \bigl([n,1],[n+1]\bigr)_{dual} + \\ [0.2cm]
+ ([n,1],[n,1])  +
 \bigl([n],[n-1,1]\bigr)_{dual}
 + 2 \cdot ([n], \, [n]) + ([n-1],[n-1]) \; .
 \end{array}
\ee
For the cases of the Lie algebras
$\mathfrak{so}_N$ and $\mathfrak{sp}_N$
the decompositions of
$Y_n \times Y_1$ are written as
\be
\lb{yny02}
\begin{array}{c}
\mathfrak{so}_N: \;\;\;\;\; Y_n \times Y_1 =
[n^2] \times [1^2] = [(n+1)^2] +[n+1,n,1] + [n^2,1^2]
+ \\ [0.2cm]
 + [n,n-1,1] + [n^2] +[n+1,n-1] + [(n-1)^2] \; ,
  \end{array}
\ee
\be
\lb{yny02sp}
\begin{array}{c}
\mathfrak{sp}_{N=2r}: \;\;\;  Y_n \times Y_1 = [2n] \times [2] = [2n+2] + [2n+1,1]
 + [2n,2] + \\ [0.2cm]
 + [2n-1,1] +[2n]  +[2n-2]\; .
  \end{array}
\ee
 Introduce the operation of summation \cite{Macd}
 of two Young diagrams
 $\lambda=[\lambda_1,\lambda_2,...]$ and
 $\mu=[\mu_1,\mu_2,...]$
 \be
\lb{rule3}
 (\lambda+\mu)_i = \lambda_i+\mu_i  \; .
 \ee
Comparing decompositions (\ref{yny01}) and
(\ref{yny02sp}) we notice that the right-hand side of
 (\ref{yny02sp})
 can be obtained from the right-hand side of (\ref{yny01})
 if there exists a relationship between Young diagrams
 for $\mathfrak{sl}_N$ and $\mathfrak{sp}_N$
 ($N$ is even) Casimir
 representations that is established by the operation
  \be
\lb{rule2}
(\lambda,\mu)_{dual} \;\; \rightarrow \;\; (\lambda+\mu) \; , \;\;\;\;
(\lambda,\lambda) \;\; \rightarrow \;\; (\lambda+\lambda) \; .
\ee
Here, in the left-hand sides, we have
 $\mathfrak{sl}_N$ (Casimir) representations
written in the composite form (via two partitions
$\lambda$ and $\mu$) and in
the right-hand sides we have
the $\mathfrak{sp}_N$-representations
associated to the Young diagrams
 which are obtained by the summation (\ref{rule3})
of the partitions $\lambda$ and $\mu$.
There is only one exception to rule (\ref{rule2}), which
occurs when comparing
decompositions (\ref{yny01}) and (\ref{yny02sp}),
namely one of the pair of the
$\mathfrak{sl}_N$-representations $([n],[n])$ in
 the decomposition (\ref{yny01})
maps to nothing, i.e. to the $\mathfrak{sp}_N$-representation $[-]$,
 which has zero dimension.
 Our conjecture is that
 the representations of $\mathfrak{sl}_N$ and
 $\mathfrak{sp}_N$ (which are
  related by the rule (\ref{rule2}))
 should pack into one universal multiplet.
 In all known cases, the relation (\ref{rule2}) between (Casimir) representations of $\mathfrak{sl}_N$ and $\mathfrak{sp}_N$ is almost always fulfilled, with only few exceptions (as with
 the mapping $([n],[n]) \to [-]$ in the example discussed above). We defer the description of these exceptions
 to our future publication.

We will call operation (\ref{rule3}), which relates the
representations of $\mathfrak{sl}_N$ and $\mathfrak{sp}_N$, the horizontal sum of Young diagrams by analogy with the vertical sum
of Young diagrams, which, as was discovered in \cite{RMcr2}, relates the representations of $\mathfrak{sl}_N$ and $\mathfrak{so}_N$. In \cite{RMcr2}, the vertical sum operation was applied to find universal form
for eigenvalues of quadratic Casimir operators in the representations arising in the decomposition
of $\mathfrak{g}^{\otimes 5}$.


 \section{Beyond Vogel universality}
 \setcounter{equation}0

\subsection{Irreps $Y_{n}$,
  $Y'_{n}$ and universal decomposition of $[1] \otimes Y_n$
  and $[1] \otimes Y_n'$ for simple Lie algebras of classical series}

  Let $Y_{n}$  be the Cartan $n$-power of the adjoint representation
  $Y_1=\ad$ of the simple Lie algebra $\mathfrak{g}$.
  Denote the highest weight of the adjoint representation
 as $\lambda_{\ad}$, then the highest weight of
 the representation $Y_{n}$ is $(n \, \lambda_{\ad})$.

  \subsubsection{The
  $\mathfrak{sl}_N$-irreps $Y_{n}$,
  $Y'_{n}$ and decompositions of $[1] \otimes Y_n$ and $[1] \otimes Y_n'$}

The Young diagram of the $\ad$-representation of $\mathfrak{sl}_N$
is $[2,1^{N-2}]$. Therefore, the Young diagram for the
$Y_n$-representation is $[2n,n^{N-2}]$ and, according to
the notation adopted in section {\bf \ref{class}}
(see Proposition {\bf \ref{test2}} and
Remark {\sf \ref{not2}}), we associate this representation with a pair of diagrams $([n],[n])$.
The dimensions of the $\mathfrak{sl}_N$-irreps
 $Y_{n}=[2n,n^{N-2}]=([n],[n])$ and
 $Y'_{n}=[2^n,1^{N-2n}]=([1^n],[1^n])$ are evaluated
 with the help of (\ref{abc7}), (\ref{dimYk}):
 \be
 \lb{dimYn}
 \dim Y_{n} = \left( \frac{(N+n-2)!}{n! (N-2)!} \right)^2
 \frac{(N+2n-1)}{(N-1)} \; ,
 \ee
 \be
 \lb{dimY'n}
 \dim Y'_{n} = \left( \frac{(N+1)!}{n! (N-n+1)!} \right)^2
 \frac{(N-2n+1)}{(N+1)} \; .
 \ee
 We note (see (\ref{dim2T})) that $\dim Y_{n}$ is obtained from
 $\dim Y_{n}'$ by formal exchange $N \to -N$, which is
 in agreement with the exchange of the Vogel parameters
 $\hat{\alpha} \leftrightarrow \hat{\beta}$ for the $\mathfrak{sl}_N$ algebra
 (see Table {\sf \ref{tab1}}).

 In the $\mathfrak{sl}_N$ case, the decompositions of the tensor product
 of the defining representation and representations $Y_n,Y'_n$  are
 \be
 \lb{slfy2}
 \begin{array}{c}
  [1] \otimes Y_n  =
 [2n+1,n^{N-2}] + [2n,n+1,n^{N-3}] + [2n-1,(n-1)^{N-2}]
 \;\;\;\; \Leftrightarrow \\ [0.2cm]
 ([1],[\emptyset]) \otimes ([n],[n])
 =([n+1],[n]) + ([n,1],[n]) + ([n],[n-1]) \; ,
  \end{array}
 \ee
 \be
 \lb{slfypr2}
 \begin{array}{c}
 [1] \otimes Y_n'  =
 [2^{n+1},1^{N-2n-1}] +  [3,2^{n-1},1^{N-2n}] + [2^n,1^{N-2n+1}]
 \;\;\;\; \Leftrightarrow
 \\ [0.2cm]
 ([1],[\emptyset]) \otimes ([1^n],[1^n])
 =([1^{n+1}],[1^n]) + ([2,1^{n-1}],[1^n]) + ([1^n],[1^{n-1}])\; .
 \end{array}
\ee
The dimensions of the representations
which appear in the decompositions (\ref{slfy2})
and (\ref{slfypr2}) are equal to
\be
\lb{dimYn123}
\begin{array}{c}
\dim ([2n+1,n^{N-2}]) =
\dfrac{(N+2n)(N+n-1)}{(N+2n-1) (n+1)} \dim Y_n \; , \\ [0.3cm]
\dim (2n,n+1,n^{N-3}) =   \dfrac{(N+n-1)n(N-2)}{
  (n+1)(N+n-2)} \dim Y_n \; , \\ [0.3cm]
\dim ([2n-1,(n-1)^{N-2}]) =
\dfrac{(N+2n-2)n}{(N+n-2)(N+2n-1)} \dim Y_n \; ,
\end{array}
\ee
\be
\lb{dimYnpr123}
\begin{array}{c}
\dim ([2^{n+1},1^{N-2n-1}]) =
\dfrac{(N-2n)(N-n+1)}{(N-2n+1) (n+1)} \dim Y_n' \; , \\ [0.3cm]
\dim (3,2^{n-1},1^{N-2n}) =   \dfrac{(N-n+1)n(N+2)}{
  (n+1)(N-n+2)} \dim Y_n' \; ,  \\ [0.3cm]
\dim ([2^n,1^{N-2n+1}]) =
\dfrac{(N-2n+2)n}{(N-n+2)(N-2n+1)} \dim Y_n' \; .
\end{array}
\ee
Taking into account the decompositions (\ref{slfy2})
and  (\ref{slfypr2}),
we obtain obvious identities for dimensions (\ref{dimYn}),
(\ref{dimYn123}) and
(\ref{dimYn}), (\ref{dimYnpr123})
{\small \be
\lb{dimidi}
\begin{array}{c}
N \cdot \dim (Y_n) =
\dim ([2n+1,n^{N-2}]) + \dim ([2n,n+1,n^{N-3}])
 + \dim ([2n-1,(n-1)^{N-2}]) \; , \\ [0.2cm]
N \cdot \dim (Y_n') =
\dim ([2^{n+1},1^{N-2n-1}]) +  \dim ([3,2^{n-1},1^{N-2n}]) +
\dim ([2^n,1^{N-2n+1}])\; .
  \end{array}
\ee  }


 By means of (\ref{c2sln}), we evaluate the
 eigenvalues of the quadratic
 Casimir operators for the $\ad$ representation
 and for the representations appearing in (\ref{slfy2})
 {\small \be
 \lb{c2yn1}
 \begin{array}{c}
C_2([1]) = N - 1/N  \; , \;\;\;\;
C_{(2)}([2,1^{N-2}]) =C_{(2)}({\ad}) =2N \; ,   \\ [0.2cm]
C_2(Y_n) = 2n(n-1+N) \; , \;\;\;\;
 C_2([2n+1,n^{N-2}]) = \dfrac{(2Nn^2+2nN^2-1+N^2)}{N} \; , \\ [0.2cm]
 C_2([2n,n+1,n^{N-3}]) =
 \dfrac{(2Nn^2-2nN-2N+2nN^2-1+N^2)}{N} \; , \\ [0.2cm]
  C_2([2n-1,(n-1)^{N-2}])=\dfrac{(2Nn^2-4nN+2N+2nN^2-N^2-1)}{N}
 \end{array}
 \ee }
  We
 normalize values (\ref{c2yn1}) as in (\ref{2cas1}), and
  due to (\ref{gfad}), we obtain the eigenvalues
 of the 2-split Casimir operator $\hat{C}_{\Box \otimes Y_n}$
 in the representation
 $([1]\otimes Y_n)$ for the $\mathfrak{sl}_N$ case:
 \be
 \lb{slcYn}
 \begin{array}{c}
 \hat{c}_2^{[2n+1,n^{N-2}]} = \frac{1}{2} (c_{(2)}^{[2n+1,n^{N-2}]}
 - c_{(2)}^{[1]} - c_{(2)}^{Y_n} ) = \frac{n}{2N} \, ,
 \\ [0.2cm]
 \hat{c}_{(2)}^{[2n,n+1,n^{N-3}]} =
 \frac{1}{2} (c_{(2)}^{[2n,n+1,n^{N-3}]}
 - c_{(2)}^{[1]} - c_{(2)}^{Y_n})  = - \frac{1}{2N} \, ,
 \\ [0.2cm]
 \hat{c}_{(2)}^{[2n-1,(n-1)^{N-2}]} =
 \frac{1}{2} (c_{(2)}^{[2n-1,(n-1)^{N-2}]}
 - c_{(2)}^{[1]} - c_{(2)}^{Y_n})  = -\frac{(n-1+N)}{2N} \; .
 \end{array}
 \ee
 Thus, the characteristic identity is
 \be
 \lb{slhYn}
 \begin{array}{c}
   (\hat{C}_{\Box \otimes Y_n} + \frac{1}{2N})
  (\hat{C}_{\Box \otimes Y_n} - \frac{n}{2N})
  (\hat{C}_{\Box \otimes Y_n} + \frac{(n-1+N)}{2N}) = 0 \; ,
  \end{array}
 \ee
 where we denote the $\mathfrak{sl}_N$
 defining representation $[1]$ as $\Box$.

 \vspace{0.2cm}

  \newtheorem{not7}[not2]{Remark\itshape}
\begin{not7}\label{not7}
 By means of (\ref{c2sln})
 we obtain the eigenvalues of the Casimir operators for
 the representations in the decomposition
 (\ref{slfypr2}) of $[1] \times Y_n'$:
  \be
 \lb{slypr2}
 \begin{array}{c}
 C_2(Y_n')  =  C_2([2^{n},1^{N-2n}]) = 2n(N-n+1) \; , \\ [0.2cm]
 C_2([3,2^{n-1},1^{N-2n}])= 2+2n+N-2n^2+2Nn-1/N  \; , \\ [0.2cm]
  C_2([2^{n+1},1^{N-2n-1}])=  N-2n^2+2Nn-1/N \; , \\ [0.2cm]
  C_2([2^n,1^{N-2n+1}])= 4n-N-2-2n^2+2Nn-1/N \; .
 \end{array}
\ee
 Then, we obtain (according to (\ref{gfad})) the eigenvalues
 of the 2-split Casimir operator in the representation
 $[1]\otimes Y_n'=\Box \otimes Y_n'$ for the $\mathfrak{sl}_N$ case:
 \be
 \lb{slcynp}
 \begin{array}{c}
 \hat{c}_2^{[3,2^{n-1},1^{N-2n}]}  = \frac{1}{2N} \, ,
 \;\;\;\;
 \hat{c}_{(2)}^{[2^{n+1},1^{N-2n-1}]}  = - \frac{n}{2N} \, ,
\;\;\;\;
 \hat{c}_{(2)}^{[2^n,1^{N-2n+1}]}  = -\frac{(N-n+1)}{2N} \; ,
 \end{array}
 \ee
 and the characteristic identity is
 \be
 \lb{slhynp}
 \begin{array}{c}
   (\hat{C}_{\Box \otimes Y_n'} - \frac{1}{2N})
  (\hat{C}_{\Box \otimes Y_n'} + \frac{n}{2N})
  (\hat{C}_{\Box \otimes Y_n'} + \frac{(N-n+1)}{2N}) = 0 \; .
  \end{array}
 \ee
\end{not7}

  \subsubsection{The
  $\mathfrak{so}_N$ and $\mathfrak{sp}_N$ irreps $Y_{n}$,
  $Y'_{n}$ and their tensor products
  $[1] \otimes Y_n$ and $[1] \otimes Y_n'$}

  The dimensions of the $\mathfrak{so}_N$-irreps
  $Y_{n}=[n^2]$, $Y'_{n}=[1^{2n}]$ are
  evaluated by means of (\ref{dimak})
  and (\ref{dimYk})
  (Vogel parameters are given in Table {\sf \ref{tab1}}):
   \be
 \lb{dimYso}
 \dim Y_{n} =  \frac{(N+n-5)!\, (N+n-4)!\, (N+2n-2)!}{n!\,
 (n+1)!\, (N-4)!\, (N-2)!\, (N+2n-5)!} \; ,
 \ee
 \be
 \lb{dimY'so}
 \dim Y'_{n} =  \frac{N!}{(2n)! \, (N-2n)!}  \; ,
 \ee

 In the $\mathfrak{so}_N$ case, the decompositions of the tensor product
 of the defining $[1]=\Box$ and representations $Y_n,Y'_n$ are
 \be
 \lb{sofyn}
  [1] \otimes Y_n  =
 [n+1,n] + [n^2,1] + [n,n-1] \; ,
 \ee
 \be
 \lb{sofynp}
 [1] \otimes Y_n'  =
 [1^{2n+1}] + [2,1^{2n-1}] + [1^{2n-1}] \; ,
 \ee
 The dimensions of the $\mathfrak{so}_N$ representations
that appear in the decompositions (\ref{sofyn})
and (\ref{sofynp}) are equal to
\be
\lb{dimso123}
\begin{array}{c}
\dim ([n+1,n]) =
2 \dfrac{(N+2n)(N+n-3)}{(N+2n-3) (n+2)} \dim Y_n \; , \\ [0.3cm]
\dim ([n^2,1]) =   \dfrac{n(N-4)(N+n-3)}{
  (n+2)(N+n-5)} \dim Y_n \; , \\ [0.3cm]
\dim ([n,n-1]) = 
2 \dfrac{n(N+2n-6)}{(N+2n-3) (N+n-5)} \dim Y_n \; .
\end{array}
\ee
\be
\lb{dimpso123}
\begin{array}{c}
\dim ([1^{2n+1}]) =
 \dfrac{N!}{(N-2n-1)! (2n+1)!} = \dfrac{N-2n}{2n+1} \, \dim Y_n' \; ,
 \\ [0.4cm]
\dim ([2,1^{2n-1}]) =   \dfrac{2n(N-2n)(N+2)N!}{
  (2n+1)!(N-2n+1)!} =
  \dfrac{2n(N-2n)(N+2)}{(2n+1)(N-2n+1)} \, \dim Y_n' \; ,
  \\ [0.4cm]
\dim ([1^{2n-1}]) =
\dfrac{N!}{(N-2n+1)! (2n-1)!}
= \dfrac{2n}{N-2n+1} \, \dim Y_n' \; .
\end{array}
\ee

  By using (\ref{c2son}) we deduce the values
  of the quadratic Casimir operators
  for the defining, $\ad$ and $Y_n$ representations, and
  for representations arising in (\ref{sofyn})
\be
 \lb{sofyn1}
 \begin{array}{c}
 C_{(2)}([1]) = N - 1 \, , \;\;\;
 C_{(2)}([1^{2}]) = C_{(2)}({\ad}) =2(N-2) \, , \\ [0.3cm]
 C_{(2)}(Y_n) = C_{(2)}([n^2]) =2n(n-3+N) \, , \\ [0.3cm]
 C_{(2)}([n+1,n]) = 2n^2-4n-1+N(2n+1) \, , \\ [0.3cm]
 C_{(2)}([n^2,1]) = 2n^2-6n-5+N(2n+1) \, , \\ [0.3cm]
 C_{(2)}([n,n-1]) = 2n^2-8n+5+N(2n-1) \, .
 \end{array}
 \ee
 Then, taking into account the value
 $C_{(2)}(\ad)$, we normalize  all values
 (\ref{sofyn1}) by means of (\ref{2cas1})
   and finally, according to (\ref{gfad}), we obtain the eigenvalues
 of the 2-split Casimir operator in the representations arising in
 the decomposition of
 $([1]\otimes Y_2(\alpha))$ for $\mathfrak{so}_N$ case:
 \be
 \lb{socYn}
 \begin{array}{c}
 \hat{c}_{(2)}^{[n^2,1]} = \frac{1}{2} (c_{(2)}^{[n^2,1]}
 - c_{(2)}^{[1]} - c_{(2)}^{Y_n} ) = - \frac{1}{N-2} \, ,
 \\ [0.2cm]
 \hat{c}_{(2)}^{[n+1,n]} =
 \frac{1}{2} (c_{(2)}^{[n+1,n]}
 - c_{(2)}^{[1]} - c_{(2)}^{Y_n})  = \frac{n}{2(N-2)} \, ,
 \\ [0.2cm]
 \hat{c}_{(2)}^{[n,n-1]} =
 \frac{1}{2} (c_{(2)}^{[n,n-1]}
 - c_{(2)}^{[1]} - c_{(2)}^{Y_n}) = -\frac{N+n-3}{2(N-2)} \; .
 \end{array}
 \ee
 Thus, the characteristic identity for $\hat{C}_{\Box \otimes Y_n}$
 is
 \be
 \lb{sohYn}
 \begin{array}{c}
   (\hat{C}_{\Box \otimes Y_n} + \frac{1}{N-2})
  (\hat{C}_{\Box \otimes Y_n} - \frac{n}{2(N-2)})
  (\hat{C}_{\Box \otimes Y_n} + \frac{N+n-3}{2(N-2)}) = 0 \; ,
  \end{array}
 \ee
 where we denote the $\mathfrak{so}_N$
 defining representation $[1]$ as $\Box$.


 \newtheorem{not8}[not2]{Remark\itshape}
\begin{not8}\label{not8}
 For the $\mathfrak{so}_N$-representations arising
 in the decomposition of $[1] \times Y_n'$ (\ref{sofynp})
 we respectively obtain the values
  of the quadratic Casimir operators
 \be
 \lb{sofynp1}
 \begin{array}{c}
 C_{(2)}(Y'_n) = C_{(2)}([1^{2n}]) =2n(N-2n) \, , \\ [0.3cm]
 C_{(2)}([1^{2n+1}]) = (2n+1)(N-2n-1) \, , \\ [0.3cm]
 C_{(2)}([2,1^{2n-1}]) = (2n+1)(N-2n+1) \, , \\ [0.3cm]
 C_{(2)}([1^{2n-1}]) = (2n-1)(N-2n+1) \, .
 \end{array}
 \ee
  and, according to (\ref{gfad}), we deduce the eigenvalues
 of the 2-split Casimir operator in the representation
 $[1]\otimes Y_n'=\Box \otimes Y_n'$
 (for the $\mathfrak{so}_N$ case):
 \be
 \lb{socynp}
 \begin{array}{c}
 \hat{c}_2^{[1^{2n+1}]}  = -\frac{n}{N-2} \, ,
 \;\;\;\;
 \hat{c}_{(2)}^{[2,1^{2n-1}]}  = \frac{1}{2(N-2)} \, ,
  \;\;\;\;
 \hat{c}_{(2)}^{[1^{2n-1}]}  = -\frac{(N-2n)}{2(N-2)}\; .
 \end{array}
 \ee
 Thus, the characteristic identity is
 \be
 \lb{sohynp}
 \begin{array}{c}
   (\hat{C}_{\Box \otimes Y_n'} +\frac{n}{N-2} )
  (\hat{C}_{\Box \otimes Y_n'} - \frac{1}{2(N-2)} )
  (\hat{C}_{\Box \otimes Y_n'} + \frac{(N-2n)}{2(N-2)}) = 0 \; .
  \end{array}
 \ee
 \end{not8}

  \newtheorem{not8b}[not2]{Remark\itshape}
\begin{not8b}\label{not8b}
  The dimensions of the $\mathfrak{sp}_N$-irreps
  $Y_{n}=[2n]$ and $Y'_{n}=[2^{n}]$ are
  evaluated by means of (\ref{dimYk})
  (Vogel parameters for $\mathfrak{sp}_N$
  are given in Table {\sf \ref{tab1}}):
   \be
 \lb{dimYsp}
 \dim Y_{n} =  \frac{\Gamma(2n+N)}{\Gamma(N)\, \Gamma(2n+1)} \; ,
 \ee
 \be
 \lb{dimY'sp}
 \dim Y'_{n} =  \frac{(N+2-2n)(N+3-2n)(N+4-2n)}{(N-n+4)(n+1)(N+2)(N+3)}
 \Bigl( \frac{\Gamma(N+4)}{\Gamma(n+1) \Gamma(N-n+4)} \Bigr)^2\; .
 \ee
 The decompositions of the tensor products $[1] \otimes Y_n$
 and $[1] \otimes Y_n'$ are
 \be
 \lb{spfyn}
 \begin{array}{c}
  [1] \otimes Y_n  =
 [2n+1] + [2n,1] + [2n-1] \; , \\ [0.2cm]
 [1] \otimes Y_n'  =
 [3,2^{n-1}] + [2^{n},1] + [2^{n-1},1] \; ,
 \end{array}
 \ee
 and by using (\ref{c2spn}) we deduce the values of the
 quadratic Casimir operators in the representations
 of $\mathfrak{sp}_N$ arising in
 both sides of (\ref{spfyn}):
 \be
 \lb{spfyn1}
 \begin{array}{c}
 C_{(2)}([1]) = \frac{1}{2}(N + 1) \, , \;\;\;
 C_{(2)}([2]) = C_{(2)}(Y_1) =C_{(2)}({\ad}) =(N+2) \, , \\ [0.2cm]
 C_{(2)}(Y_n) = C_{(2)}([2n]) =n(2n+N)  , \;\;
C_{(2)}(Y_n') = C_{(2)}([2^n]) =n(N-n+3) \, ,
  \\ [0.3cm]
 C_{(2)}([2n+1]) = \frac{1}{2}(2n+1)(N+2n+1) , \\ [0.2cm]
 C_{(2)}([2n,1]) = \frac{1}{2}(2n+1)(N+2n-1) \, , \\ [0.2cm]
 C_{(2)}([2n-1]) = \frac{1}{2}(2n-1)(N+2n-1) \, , \\ [0.2cm]
  C_{(2)}([3,2^{n-1}]) =\frac{1}{2}(5+6n-2n^2+ (2n+1)N)\, , \\ [0.2cm]
 C_{(2)}([2^n,1]) = \frac{1}{2}(1+4n-2n^2+ (2n+1)N) \, , \\ [0.2cm]
  C_{(2)}([2^{n-1},1]) =\frac{1}{2}(-5+8n-2n^2+ (2n-1)N ) \, .
 \end{array}
 \ee
 By making the same calculations as in the case of
 $\mathfrak{so}_N$ and using formulas (\ref{gfad}),
 we obtain the eigenvalues of the
 2-split Casimir operator in the representations arisen in
 the decompositions
 $([1]\otimes Y_n(\alpha))$ and $([1]\otimes Y_n(\beta))$
 for $\mathfrak{sp}_N$ case:
 \be
 \lb{spcYn}
 \hat{c}_{(2)}^{[2n+1]} =  \dfrac{n}{N+2} \, , \;\;\;\;
 \hat{c}_{(2)}^{[2n,1]} =  -\dfrac{1}{2(N+2)} \, , \;\;\;\;
 \hat{c}_{(2)}^{[2n-1]} =  -\dfrac{N+2n}{2(N+2)} \, ,
 \ee
  \be
 \lb{spcYnpr}
 \hat{c}_{(2)}^{[3,2^{n-1}]} =  \dfrac{1}{N+2} \, , \;\;\;\;
 \hat{c}_{(2)}^{[2^{n},1]} =  -\dfrac{n}{2(N+2)} \, , \;\;\;\;
 \hat{c}_{(2)}^{[2^{n-1},1]} =  -\dfrac{N-n+3}{2(N+2)} \, .
 \ee
 Thus, the characteristic identities for $\mathfrak{sp}_N$ operators
 $\hat{C}_{\Box \otimes Y_n}$ and $\hat{C}_{\Box \otimes Y_n'}$ are
 \be
 \lb{sphYn}
 \begin{array}{c}
   (\hat{C}_{\Box \otimes Y_n} + \frac{1}{2(N+2)})
  (\hat{C}_{\Box \otimes Y_n} - \frac{n}{N+2})
  (\hat{C}_{\Box \otimes Y_n} + \frac{N+2n}{2(N+2)}) = 0 \; ,
  \end{array}
 \ee
 \be
 \lb{sphYnp}
 \begin{array}{c}
   (\hat{C}_{\Box \otimes Y_n'} - \frac{1}{(N+2)})
  (\hat{C}_{\Box \otimes Y_n'} + \frac{n}{2(N+2)})
  (\hat{C}_{\Box \otimes Y_n'} + \frac{N-n+3}{2(N+2)}) = 0 \; .
  \end{array}
 \ee
 We note that identities (\ref{sohYn}) and (\ref{sohynp})
 for operators $\hat{C}_{\Box \otimes Y_n}$ and
 $\hat{C}_{\Box \otimes Y_n'}$ in the $\mathfrak{so}_{N}$ case are related, respectively, to identities (\ref{sphYnp}) and
 (\ref{sphYn}) for operators $\hat{C}_{\Box \otimes Y_n'}$
  and $\hat{C}_{\Box \otimes Y_n}$ in the $\mathfrak{sp}_N$ case
  by the transformation $N \to -N$, which is known as
  the Cvitanovi\'{c}-Mkrtchyan duality trnsformation,
  which relates characters of the $\mathfrak{so}_N$ and
  $\mathfrak{sp}_N$ algebras.
\end{not8b}

\subsubsection{Universal characteristic identities
for $\hat{C}_{\Box \otimes Y_n}$ and $\hat{C}_{\Box \otimes Y_n'}$
 for Lie algebras of classical series}

 Characteristic identities
  (\ref{slhYn}), (\ref{sohYn}) and (\ref{sphYn}) (for Lie algebras of classical series)
  are written for the split Casimir operator taken
  in the representation $\Box \otimes Y_n$ in the unified form
  \be
  \lb{charn0}
 \Bigl(\hat{C}_{\Box \otimes Y_n} +
 \frac{1}{2}+\frac{\hat{\alpha}}{2}(1-n)\Bigr)
  (\hat{C}_{\Box \otimes Y_n} + n \frac{\hat{\alpha}}{2})
  (\hat{C}_{\Box \otimes Y_n} + \frac{\hat{\beta}}{2}) = 0 \; ,
  \ee
    where $\hat{\alpha}$ and  $\hat{\beta}$ are the
    homogeneous Vogel
  parameters, which we define in (\ref{hompar}) and
  list in Table {\sf \ref{tab1}}.
  In this Table, we also add the parameter $\dim \Box$, which is
  the dimension of the defining representation $\Box$,
 give dimensions of the simple Lie algebras $\mathfrak{g}$
and add the normalized values $c_{(2)}^{(\Box)}$ of the quadratic Casimir
 operator in the defining representation $\Box$.

  \begin{table}[h!] 
  \centering
 \caption{\sf Vogel parameters and other characteristics of the
 simple Lie alpgbras} 
	\label{tab1}  \vspace{0.2cm}
	\begin{tabular}{|c|c|c|c|c|c|c|c|c|}
		\hline
  $\;\;$ & $\mathfrak{sl}(N)$ & $\mathfrak{so}(N)$&
 $\mathfrak{sp}(N)$ &
		$\mathfrak{g}_2$ & $\mathfrak{f}_4$
		& $\mathfrak{e}_6$ &  $\mathfrak{e}_7$  &
		$\mathfrak{e}_8$   \\
		\hline
		$\alpha$ &\footnotesize $-2$ &\footnotesize $ -2$
 &\footnotesize  $-2$ &
		\footnotesize $-2$ &\footnotesize  $-2$ &\footnotesize  $-2$
		&\footnotesize  $-2$  &\footnotesize  $-2$  \\
		\hline
		$\beta$ &\footnotesize $2$ &\footnotesize $ 4$
     &\footnotesize  $1$ &
		\footnotesize $10/3$ &\footnotesize  $5$ &\footnotesize  $6$
		&\footnotesize  $8$  &\footnotesize  $12$  \\
		\hline
		$\gamma$ &\footnotesize $N$ &\footnotesize $N-4$
     &\footnotesize  $(N+4)/2$ &
		\footnotesize $8/3$ &\footnotesize  $6$ &\footnotesize  $8$
		&\footnotesize  $12$  &\footnotesize  $20$  \\
		\hline
\footnotesize ${\sf t} = C_{(2)}(\ad)/2$ &\footnotesize $N$ &
\footnotesize $N-2$
     &\footnotesize  $(N+2)/2$ &
		\footnotesize $4$ &\footnotesize  $9$ &\footnotesize  $12$
		&\footnotesize  $18$  &\footnotesize  $30$  \\
		\hline
\footnotesize $\hat{\alpha}=\alpha/(2{\sf t})$ &\footnotesize $-1/N$
     &\footnotesize $ -1/(N-2)$   &\footnotesize  $-2/(N+2)$ &
		\footnotesize $-1/4$ &\footnotesize  $-1/9$ &\footnotesize  $-1/12$
		&\footnotesize  $-1/18$  &\footnotesize  $-1/30$  \\
		\hline
\footnotesize $\hat{\beta}=\beta/(2{\sf t})$  &\footnotesize $1/N$ &
		\footnotesize $2/(N-2)$&\footnotesize $1/(N+2)$ &
		\footnotesize $5/12$ &\footnotesize $5/18$ &\footnotesize $1/4$
     &  \footnotesize $2/9$ &  \footnotesize $1/5$ \\
		\hline
  \footnotesize $\hat{\gamma}=\gamma/(2{\sf t})$  &
  \footnotesize $1/2$ & \footnotesize
  $\frac{N-4}{2(N-2)}$ &\footnotesize $\frac{N+4}{2(N+2)}$&
  \footnotesize $1/3$ &\footnotesize $1/3$
  &\footnotesize $1/3$ & \footnotesize $1/3$  & \footnotesize $1/3$
  \\   \hline
  $\dim\Box $  &\footnotesize $N$ & \footnotesize
  $N$ &\footnotesize $N$&
  \footnotesize $7$ &\footnotesize $26$ &\footnotesize $27$
  &  \footnotesize $56$ &  \footnotesize $248$ \\   \hline
  $\dim\mathfrak{g} $  &\footnotesize $N^2-1$ & \footnotesize
  $\frac{N(N-1)}{2}$ &\footnotesize $\frac{N(N+1)}{2}$&
  \footnotesize $14$ &\footnotesize $52$ &\footnotesize $78$ &
  \footnotesize $133$  &  \footnotesize $248$ \\   \hline
  $c_{(2)}^{(\Box)}$
  &\footnotesize $\frac{N^2-1}{2N^2}$ & \footnotesize
  $\frac{(N-1)}{2(N-2)}$ &\footnotesize $\frac{(N+1)}{2(N+2)}$&
  \footnotesize $1/2$ &\footnotesize $2/3$ &
  \footnotesize $13/18$ & \footnotesize $19/24$  &  \footnotesize $1$
  \\   \hline
  $c_{(2)}^{(\Box)}\cdot \dim\Box$
  &\footnotesize $\frac{\dim \mathfrak{g}}{2N}$ & \footnotesize
  $\frac{\dim \mathfrak{g}}{(N-2)}$ &\footnotesize $\frac{\dim \mathfrak{g}}{(N+2)}$&
  \footnotesize $\frac{\dim \mathfrak{g}}{4}$ &
  \footnotesize $\frac{\dim \mathfrak{g}}{3}$ &
  \footnotesize $\frac{\dim \mathfrak{g}}{4}$ &
  \footnotesize $\frac{\dim \mathfrak{g}}{3}$  &
  \footnotesize $\dim \mathfrak{g}$
  \\   \hline
	\end{tabular}
\end{table}
\noindent

  Characteristic identities (\ref{slhynp}), (\ref{sohynp})
  and (\ref{sphYnp}) are written in the universal form
  (for $\mathfrak{sl}_N$, $\mathfrak{so}_N$
  and $\mathfrak{sp}_N$ algebras) via the Vogel parameters as
 \be
 \lb{charnp0}
 \begin{array}{c}
   \Bigl(\hat{C}_{\Box \otimes Y_n'} +
 \frac{1}{2}+\frac{\hat{\beta}}{2}(1-n)\Bigr)
  (\hat{C}_{\Box \otimes Y_n'} + n \frac{\hat{\beta}}{2})
  (\hat{C}_{\Box \otimes Y_n'} + \frac{\hat{\alpha}}{2}) = 0 \; .
  \end{array}
 \ee
 As expected, this characteristic identity
 is obtained from (\ref{charn0}) by exchanging
 the parameters $\alpha \leftrightarrow \beta$
 which relates the $Y_n=Y_n(\alpha)$
 and $Y_n'=Y_n(\beta)$ representations.

 \subsection{Universal decompositions for $\Box \otimes Y_n$
  in the case of exceptional Lie algebras}

\begin{figure}[h!]

\unitlength=3.7mm
\begin{picture}(17,4)(-4,0)

\put(0.5,1.6){${\sf G}_2$}

\put(8.5,2){\circle{0.5}}
\put(8.65,1.8){\line(1,0){2.15}}
\put(8.75,2){\line(1,0){2}}
\put(8.65,2.2){\line(1,0){2.15}}
 \put(9.1,1.65){\Large $<$}
\put(11,2){\circle{0.5}}

\put(8.3,2.5){\scriptsize 1}
\put(11,2.5){\scriptsize 2}

\end{picture}


\unitlength=3.7mm
\begin{picture}(17,3)(-4,0)

\put(0.5,1.8){${\sf F}_4$}

\put(7.5,2){\circle{0.5}}
\put(7.5,2.5){\scriptsize 1}
\put(7.75,2){\line(1,0){2}}
\put(10,2){\circle{0.5}}
\put(10,2.5){\scriptsize 2}
\put(10.22,1.9){\line(1,0){2}}
\put(10.22,2.1){\line(1,0){2}}
 \put(10.8,1.65){\Large $>$}
\put(12.5,2){\circle{0.5}}
\put(12.5,2.5){\scriptsize 3}
\put(12.75,2){\line(1,0){2}}
\put(15,2){\circle{0.5}}
\put(15,2.5){\scriptsize 4}

\end{picture}

\unitlength=3.6mm
\begin{picture}(17,3)(-4,0)

\put(0.5,1.8){${\sf E}_6$}

\put(5.5,2){\circle{0.5}}
\put(5.75,2){\line(1,0){2}}
\put(8,2){\circle{0.5}}
\put(8.25,2){\line(1,0){2}}
\put(10.5,2){\circle{0.5}}
\put(10.75,2){\line(1,0){2}}
\put(13,2){\circle{0.5}}
\put(13.25,2){\line(1,0){2}}
\put(15.5,2){\circle{0.5}}

\put(10.5,3.5){\circle{0.5}}
\put(10.5,2.25){\line(0,1){1}}

\put(5.5,1){\scriptsize  1}
\put(8,1){\scriptsize 2}
\put(10.5,1){\scriptsize 3}
\put(13,1){\scriptsize 4}
\put(15.5,1){\scriptsize 6}
\put(9.5,3.5){\scriptsize 5}

\end{picture}

\unitlength=3.6mm
\begin{picture}(17,3.5)(-4,0)

\put(0.5,1.8){${\sf E}_7$}

\put(5.5,2){\circle{0.5}}
\put(5.75,2){\line(1,0){2}}
\put(8,2){\circle{0.5}}
\put(8.25,2){\line(1,0){2}}
\put(10.5,2){\circle{0.5}}
\put(10.75,2){\line(1,0){2}}

\put(13,2){\circle{0.5}}
\put(13.25,2){\line(1,0){2}}
\put(15.5,2){\circle{0.5}}
\put(15.75,2){\line(1,0){2}}
\put(18,2){\circle{0.5}}

\put(13,3.5){\circle{0.5}}
\put(13,2.25){\line(0,1){1}}

\put(5.5,1){\scriptsize  1}
\put(8,1){\scriptsize 2}
\put(10.5,1){\scriptsize 3}
\put(13,1){\scriptsize 4}
\put(15.5,1){\scriptsize 5}
\put(18,1){\scriptsize 7}
\put(12,3.5){\scriptsize 6}

\end{picture}

\unitlength=3.6mm
\begin{picture}(22,3.5)(-4,0)

\put(0.5,1.3){${\sf E}_8$}

\put(5.5,1.5){\circle{0.5}}
\put(5.5,0.5){\scriptsize  1}
\put(5.75,1.5){\line(1,0){2}}
\put(8,1.5){\circle{0.5}}
\put(8,0.5){\scriptsize 2}
\put(8.25,1.5){\line(1,0){2}}
\put(10.5,1.5){\circle{0.5}}
\put(10.5,0.5){\scriptsize 3}
\put(10.75,1.5){\line(1,0){2}}
\put(13,1.5){\circle{0.5}}
\put(13,0.5){\scriptsize 4}
\put(13.25,1.5){\line(1,0){2}}

\put(15.5,1.5){\circle{0.5}}
\put(15.5,0.5){\scriptsize 5}
\put(15.75,1.5){\line(1,0){2}}
\put(18,1.5){\circle{0.5}}
\put(18,0.5){\scriptsize 6}
\put(20.5,1.5){\circle{0.5}}
\put(20.5,0.5){\scriptsize 8}
\put(18.25,1.5){\line(1,0){2}}

\put(15.5,3){\circle{0.5}}
\put(14.5,3){\scriptsize 7}
\put(15.5,1.75){\line(0,1){1}}

\end{picture}

\caption{\label{XXX1}  {\sf Dynkin diagrams for the exceptional
 Lie algebras.} }
\end{figure}
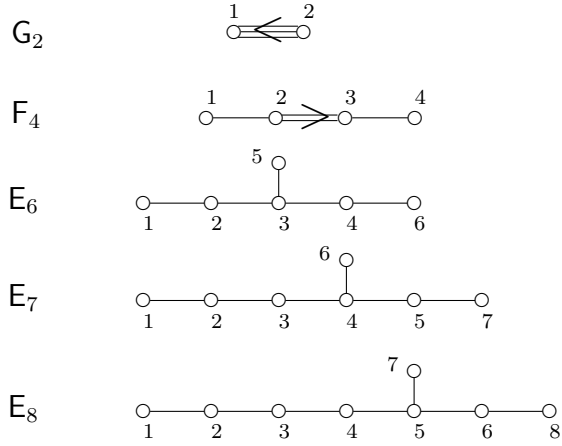

  For all exceptional Lie algebras
  $\mathfrak{g} = \{ \mathfrak{g}_2, \mathfrak{f}_4,
  \mathfrak{e}_6, \mathfrak{e}_7, \mathfrak{e}_8\}$ we denote
  the fundamental representation of $\mathfrak{g}$
  with minimal dimension by $\Box$ and
  the Cartan powers of $n$ adjoint representations
  of $\mathfrak{g}$ by $Y_n$.
  We use the numbering of the simple roots and fundamental weights
  adopted in \cite{IsRub2} according to the numbering
  of nodes in Dynkin diagrams shown
  in Figure {\sf \ref{XXX1}}.

    \subsubsection{The case of the $\mathfrak{g}_2$
    algebra. Description of irreps $Y_{n}$
  and decomposition of $\Box \otimes Y_n$\label{secg2}}

  It is convenient (see e.g. \cite{IsRub2})
  to consider the root space of the
  algebra $\mathfrak{g}_2$ as a two-dimensional
   plane  embedded in a 3-dimensional space $\mathbb{R}^3$, such that
  the 3-dimensional vectors $(x_1,x_2,x_3)$ satisfy the relation
  $x_1+x_2+x_3=0$.  The fundamental weights
  $\lambda_{(1)},\lambda_{(2)}$ (we also use the notation
  $\lambda_{[[d]]}$, where $d$ denotes the dimension
 of the representation $[[d]]$), the Weyl vector $\rho$
 (see \cite{Burb} and \cite{IsRub2} and references therein) and
 the highest weight
 vectors for the representations $Y_n$ are
  \be
 \lb{g2yn1}
 \begin{array}{c}
  \lambda_{[[7]]} = \lambda_{(1)} = (0,-1,1) \; , \;\;\;\;
  \lambda_{[[14]]} = \lambda_{\ad} =\lambda_{(2)} = (-1,-1,2) \; ,
    \\ [0.2cm]
  \lambda_{Y_n} = n \cdot \lambda_{(2)} = (-n,-n,2n) \; , \;\;\;\;
  \rho = \sum\limits_{i=1}^2 \lambda_{(i)} = (-1,-2,3) \; .
 \end{array}
 \ee
 We denote the representation with the highest weight
 $ \lambda=(\lambda_1,\lambda_2,\lambda_3)
 \bigr|_{\lambda_3=-\lambda_1-\lambda_2}$
 as $T_\lambda$,
 and one can find from (\ref{weyldim}) that
 \be
 \lb{dimg2}
 \dim T_\lambda
 = - \frac{1}{5!} \prod_{1 \leq i<j \leq 3} (\lambda'_i-\lambda'_j)
 (\lambda'_i+\lambda'_j)\; ,
 \ee
 where $\lambda'_1=\lambda_1-1$, $\lambda'_2=\lambda_2-2$ and
 $\lambda'_3= - \lambda'_1 - \lambda'_2 = \lambda_3+3$.

   The decomposition of the tensor product
 of the defining representation $T_{\lambda_{(1)}}=\Box$ and
 the representation $Y_n=T_{n  \lambda_{(2)}}$ is
  \be
 \lb{g2fyn}
 T_{\lambda_{(1)}} \otimes T_{n  \lambda_{(2)}}
 = T_{n \lambda_{(2)}+\lambda_{(1)}} +
 T_{(n-1) \lambda_{(2)}+\lambda_{(1)}}  +
 T_{(n-1) \lambda_{(2)}+ 2 \lambda_{(1)}}  \; .
 \ee
 With the help of (\ref{dimg2}) we obtain
 \be
 \lb{dimg2a}
 \begin{array}{c}
 \dim T_{\Lambda_1^{(n)}} =
 \dfrac{4(n+3)(3n+7)}{(2n+3)(3n+4)} \dim Y_n \, , \;\;\;
  \dim T_{\Lambda_2^{(n)}} =
 \dfrac{4n(3n+2)}{(2n+3)(3n+5)} \dim Y_n \, , \\ [0.3cm]
  \dim T_{\Lambda_3^{(n)}} =
 \dfrac{27n(n+3)}{(3n+4)(3n+5)} \dim Y_n \, ,
 \end{array}
 \ee
 where $\dim Y_n \equiv \dim T_{_{n  \lambda_{(2)}}}$
 is given by formula (\ref{dimYk}), and we denote
  \be
 \lb{g2Ln}
 \Lambda_1^{(n)}= n \lambda_{(2)}+ \lambda_{(1)} \; , \;\;\;\;
 \Lambda_2^{(n)}= (n-1) \lambda_{(2)}+ \lambda_{(1)}\; , \;\;\;\;
  \Lambda_3^{(n)}= (n-1) \lambda_{(2)}+ 2\lambda_{(1)}\; .
 \ee
 According to the normalization (\ref{C2abg}),
 we choose the root space metric of
 $\mathfrak{g}_2$ as
 $(e^{(i)},e^{(j)})=\frac{1}{3}\delta^{ij}$, and
 applying formula (\ref{2casim}) we deduce
 \be
 \lb{g2Yn2}
 \begin{array}{c}
 C_{(2)}(\lambda_{(1)})= 4 \, , \;\;\;
 C_{(2)}(\lambda_{(2)})= C_{(2)}(\ad)= 8 \, , \;\;\;
 C_{(2)}(n \lambda_{(2)})= 2n(n+3) \, ,  \\ [0.2cm]
  C_{(2)}(\Lambda_1^{(n)})= 2(n^2 +4n+2) \, , \;\;\;
 C_{(2)}( \Lambda_2^{(n)})=  2(n^2 +2n-1) \, ,
 \\ [0.2cm]
 C_{(2)}(\Lambda_3^{(n)})= 2n^2 +6n+4/3 \, .
 \end{array}
 \ee
 Finally, by using (\ref{2cas1}) and (\ref{gfad}) we obtain
 (for the algebra $\mathfrak{g}_2$) the
 eigenvalues of the split Casimir operator
 in the representation $\Box \otimes Y_n$
 \be
 \lb{g2Yn3}
 \hat{c}_{(2)}^{\Lambda_1^{(n)}} = \frac{n}{8} \, , \;\;\;
 \hat{c}_{(2)}^{\Lambda_2^{(n)}} = -\frac{n+3}{8}
 \, , \;\;\; \hat{c}_{(2)}^{\Lambda_3^{(n)}}= -\frac{1}{6} \, ,
\ee
and the characteristic identity in this case is written as
\be
 \lb{g2Yn4}
 \Bigl( \hat{C}_{\Box \otimes Y_n} - \frac{n}{8}\Bigr)
 \Bigl( \hat{C}_{\Box \otimes Y_n} + \frac{n+3}{8}\Bigr)
 \Bigl( \hat{C}_{\Box \otimes Y_n} + \frac{1}{6}\Bigr) = 0  \; .
 \ee

  \subsubsection{The case of the $\mathfrak{f}_4$
    algebra. Irreps $Y_{n}$
  and decomposition of $\Box \otimes Y_n$\label{secf4}}

  The root space of the
  algebra $\mathfrak{f}_4$ is the four-dimensional
   Euclidean space $\mathbb{R}^4$.  The fundamental weights
   $\lambda_{(1)},\lambda_{(2)},\lambda_{(3)},\lambda_{(4)}
    \in \mathbb{R}^4$, the Weyl vector $\rho$
    (see \cite{Burb} and \cite{IsRub2}), and the highest
    weight vector for the
 special representation $Y_n$, which we use
 below, are
  \be
 \lb{f4yn1}
 \begin{array}{c}
 \lambda_{[[52]]} = \lambda_{\ad} =
 \lambda_{(1)} = (1,0,0,1) \, , \;\;\;
  \lambda_{[[1274]]} = \lambda_{(2)} = (1,1,0,2)
     \, ,  \\ [0.2cm]
   \lambda_{[[273]]} = \lambda_{(3)} =
      \frac{1}{2}(1,1,1,3) \, , \;\;
      \lambda_{[[26]]} = \lambda_{(4)} = (0,0,0,1) \, , \;\;\;
      \\ [0.2cm]
  \lambda_{Y_n} = n \lambda_{(1)} = (n,0,0,n) \, ,  \;\;\;\;
    \rho = \sum\limits_{i=1}^4 \lambda_{(i)} =
    \frac{1}{2}\,(5,3,1,11) \; .
      \end{array}
 \ee
 Here the number $d$ in the notation $\lambda_{[[d]]}$
 denotes  the dimension of the representation.

  The decomposition of the tensor product
 of the defining representation $\Box=T_{\lambda_{(4)}}$ and
 the representation $Y_n=T_{n  \lambda_{(1)}}$ is
  \be
 \lb{f4fyn}
 T_{\lambda_{(4)}} \otimes T_{n  \lambda_{(1)}}
 = T_{n \lambda_{(1)}+\lambda_{(4)}} +
 T_{(n-1) \lambda_{(1)}+\lambda_{(4)}}  +
 T_{(n-1) \lambda_{(1)}+ \lambda_{(3)}}  \; .
 \ee
 With the help of the Weyl formula (\ref{weyldim}) we obtain
 \be
 \lb{dimf4a}
 \begin{array}{c}
 \dim T_{\Lambda_1^{(n)}} =
 \dfrac{3(n+8)(2n+13)}{(n+3)(n+4)} \dim Y_n \, , \;\;\;
  \dim T_{\Lambda_2^{(n)}} =
 \dfrac{3n(2n+3)}{(n+4)(n+5)} \dim Y_n \, , \\ [0.3cm]
  \dim T_{\Lambda_3^{(n)}} =
 \dfrac{14n(n+8)}{(n+3)(n+5)} \dim Y_n \, ,
 \end{array}
 \ee
  where $\dim Y_n \equiv \dim T_{_{n  \lambda_{(1)}}}$
 is given by formula (\ref{dimYk}), and we denote
 \be
 \lb{f4Ln}
 \Lambda_1^{(n)} = n \lambda_{(1)} + \lambda_{(4)} \; , \;\;\;\; \Lambda_2^{(n)} = (n-1) \lambda_{(1)} + \lambda_{(4)}
 \; , \;\;\;\;  \Lambda_3^{(n)} =
 (n-1) \lambda_{(1)} + \lambda_{(3)} \; .
 \ee
  According to (\ref{2casim}), we find the values
 of the quadratic Casimir operators in special representations
 of $\mathfrak{f}_4$ that interests us:
 \be
 \lb{f4Yn2}
 \begin{array}{c}
 C_{(2)}(\lambda_{(4)})= 12 \, , \;\;\;
 C_{(2)}(\lambda_{(1)})= C_{(2)}(\ad)= 18 \, , \;\;\;
 C_{(2)}(n \lambda_{(1)})= 2n(n+8) \, ,  \\ [0.2cm]
  C_{(2)}(\Lambda_1^{(n)})= 2(n^2+9n+6) \, , \;\;\;
 C_{(2)}( \Lambda_2^{(n)})=  2(n^2 +7n-2) \, ,
 \\ [0.2cm]
 C_{(2)}(\Lambda_3^{(n)})= 2(n^2 +8n+3) \, .
 \end{array}
 \ee
  Finally, by using (\ref{2cas1}) and (\ref{gfad}), we obtain
 the eigenvalues of the split Casimir operator
 in the representation $\Box \otimes Y_n$ for
 the $\mathfrak{f}_4$ algebra
 \be
 \lb{f4Yn3}
 \hat{c}_{(2)}^{\Lambda_1^{(n)}} = \frac{n}{18} \, , \;\;\;
 \hat{c}_{(2)}^{\Lambda_2^{(n)}} = -\frac{n+8}{18}
 \, , \;\;\; \hat{c}_{(2)}^{\Lambda_3^{(n)}}= -\frac{1}{6} \, ,
\ee
 and the characteristic identity for the split Casimir
 operator $\hat{C}_{\Box \otimes Y_n}$ is
\be
 \lb{f4Yn4}
 \Bigl( \hat{C}_{\Box \otimes Y_n} - \frac{n}{18}\Bigr)
 \Bigl( \hat{C}_{\Box \otimes Y_n} + \frac{n+8}{18}\Bigr)
 \Bigl( \hat{C}_{\Box \otimes Y_n} + \frac{1}{6}\Bigr) = 0 \; .
 \ee

  \subsubsection{The case of the $\mathfrak{e}_6$
    algebra. Irreps $Y_{n}$
  and decomposition of $\Box \otimes Y_n$\label{sece6}}

  The 6-dimensional weight space of the Lie algebra
  $\mathfrak{e}_6$ is embedded into the
 8-dimensional space $\mathbb{R}^8$ as a hyperplane orthogonal to the
 vectors $e^{(1)}+e^{(8)}$ and $e^{(1)}+e^{(2)}+2e^{(8)}$
 (see Problem 3.1.2 in \cite{IsRub2}).
    The fundamental weights $\lambda_{(i)}$, $i=1,...,6$,
    the Weyl vector $\rho$ and the highest weight vectors $\lambda_{Y_n}$ of the representations $Y_n$ are
   {\small \be
 \lb{e6yn1}
 \begin{array}{c}
 \lambda_{[[{\sf 27}]]} = \lambda_{(1)} =
 (-\frac{1}{3},-\frac{1}{3},1,0,0,0,0,\frac{1}{3}) \; , \;\;\;\;
  \lambda_{[[{\sf 351}]]} = \lambda_{(2)} =
 (-\frac{2}{3},-\frac{2}{3},1,1,0,0,0,\frac{2}{3})
 \; ,  \\ [0.2cm]
 \lambda_{(3)} =
 (-1,-1,1,1,1,0,0,1) \; , \;\;\;\;
  \lambda_{[[\overline{\sf 351}]]} = \lambda_{(4)} =
    (-\frac{5}{6},-\frac{5}{6},
    \frac{1}{2},\frac{1}{2},\frac{1}{2},
    \frac{1}{2},-\frac{1}{2},\frac{5}{6}) \; , \\ [0.2cm]
      \lambda_{[[78]]} = \lambda_{\ad} = \lambda_{(5)} =
  \frac{1}{2}(-1,-1,1,1,1,1,1,1) \, ,   \\ [0.2cm]
  \lambda_{[[\overline{\sf 27}]]} = \lambda_{(6)} =
  (-\frac{2}{3},-\frac{2}{3},0,0,0,0,0,\frac{2}{3}), \;\;\;
   \lambda_{Y_n} = n \cdot \lambda_{(5)} =
  \frac{n}{2}(-1,-1,1,1,1,1,1,1)
    ,   \\ [0.2cm]
    \rho = \sum\limits_{i=1}^6 \lambda_{(i)} =
   (-4,-4,4,3,2,1,0,4) \;
      \end{array}
 \ee}
 We use the definitions
  of fundamental weights and the Weyl vector $\rho$ for
  the algebra $\mathfrak{e}_6$ given in \cite{IsRub2}
  (see also \cite{Burb},
 where the numbering of simple roots and fundamental
 weights differs from that adopted here\footnote{Our numbering of the fundamental weights is given in Fig. \ref{XXX1}.}).
  The decomposition of the tensor product
 of the defining representation $\Box=T_{\lambda_{(1)}}$ and
 the representation $Y_n=T_{n  \lambda_{(5)}}$ is
  \be
 \lb{e6fyn}
 T_{\lambda_{(1)}} \otimes T_{n  \lambda_{(5)}}
 = T_{n \lambda_{(5)}+\lambda_{(1)}} +
 T_{(n-1) \lambda_{(5)}+\lambda_{(1)}}  +
 T_{(n-1) \lambda_{(5)}+ \lambda_{(4)}}  \; .
 \ee
  With the help of the Weyl formula (\ref{weyldim}) we obtain
 \be
 \lb{dime6a}
 \begin{array}{c}
 \dim T_{\Lambda_1^{(n)}} =
 \dfrac{12(n+9)(n+11)}{(n+4)(2n+11)} \dim Y_n \, , \;\;\;
  \dim T_{\Lambda_2^{(n)}} =
 \dfrac{12n(n+2)\dim Y_n}{(n+7)(2n+11)}  \, , \\ [0.3cm]
  \dim T_{\Lambda_3^{(n)}} =
 \dfrac{15n(n+11)}{(n+4)(n+7)} \dim Y_n \, ,
 \end{array}
 \ee
  where $\dim Y_n \equiv \dim T_{_{n  \lambda_{(5)}}}$
 is given by the general formula (\ref{dimYk}) with the
 choice of the Vogel parameters as in
 Table {\sf \ref{tab1}}, and we denote
 \be
 \lb{e6Ln}
 \Lambda_1^{(n)} = n \lambda_{(5)} + \lambda_{(1)} \; , \;\;\;\; \Lambda_2^{(n)} = (n-1) \lambda_{(5)} + \lambda_{(1)}
 \; , \;\;\;\;  \Lambda_3^{(n)} =
 (n-1) \lambda_{(5)} + \lambda_{(4)} \; .
 \ee
 In the same way as before,
  we find (by making use of (\ref{2casim})) the values
 of the quadratic Casimir operators for the
 special representations of $\mathfrak{e}_6$
 of interest to us:
 \be
 \lb{e6Yn2}
 \begin{array}{c}
 C_{(2)}(\lambda_{(1)})= \frac{52}{3} \, , \;\;\;
 C_{(2)}(\lambda_{(5)})= C_{(2)}(\ad)= 24 \, , \;\;\;
 C_{(2)}(n \lambda_{(5)})= 2n(n+11) \, ,  \\ [0.2cm]
  C_{(2)}(\Lambda_1^{(n)})= 2(n^2+12n+\frac{26}{3}) \, , \;\;\;
 C_{(2)}( \Lambda_2^{(n)})=  2(n^2 +10n-\frac{7}{3}) \, ,
 \\ [0.2cm]
 C_{(2)}(\Lambda_3^{(n)})= 2(n^2 +11n+\frac{14}{3}) \, .
 \end{array}
 \ee
  Finally, we apply (\ref{2cas1}), (\ref{gfad}) and obtain
 the eigenvalues of the split Casimir operator in the representations
 appearing in the decomposition of $\Box \otimes Y_n$ for
 the $\mathfrak{e}_6$ algebra
 \be
 \lb{e6Yn3}
 \hat{c}_{(2)}^{\Lambda_1^{(n)}} = \frac{n}{24} \, , \;\;\;
 \hat{c}_{(2)}^{\Lambda_2^{(n)}} = -\frac{n+11}{24}
 \, , \;\;\; \hat{c}_{(2)}^{\Lambda_3^{(n)}}= -\frac{1}{6} \, .
\ee
 Thus, the characteristic identity for split Casimir
 operator $\hat{C}_{\Box \otimes Y_n}$
 (for the Lie algebra $\mathfrak{e}_6$) is
\be
 \lb{e6Yn4}
 \Bigl( \hat{C}_{\Box \otimes Y_n} - \frac{n}{24}\Bigr)
 \Bigl( \hat{C}_{\Box \otimes Y_n} + \frac{n+11}{24}\Bigr)
 \Bigl( \hat{C}_{\Box \otimes Y_n} + \frac{1}{6}\Bigr) = 0
 \ee

  \subsubsection{The case of the $\mathfrak{e}_7$
    algebra. Irreps $Y_{n}$
  and decomposition of $\Box \otimes Y_n$\label{sece7}}

  The seven dimensional weight space of the Lie algebra
  $\mathfrak{e}_7$ is embedded into the
 8-dimensional space $\mathbb{R}^8$ as a hyperplane orthogonal to the
 vector $e^{(1)}+e^{(8)}$ (see, e.g., section 3.1.2 in \cite{IsRub2}).
    The fundamental weights,
    highest weight vectors of the
 representations $Y_n$ (Cartan $n$-power
 of the adjoint representation), and the Weyl vector $\rho$, are defined as
 (see Problem 3.4.18 in \cite{IsRub2}, and \cite{Burb},
 where the numbering of simple roots and fundamental
 weights differs from that adopted here and given in Fig. \ref{XXX1})
   {\small \be
 \lb{e7yn1}
 \begin{array}{c}
  \lambda_{[[56]]} = \lambda_{(1)} =
     (-\frac{1}{2},1,0,0,0,0,0,\frac{1}{2}) \; , \;\;\;\;
    \lambda_{(2)}  = (-1,1,1,0,0,0,0,1) \; ,  \\ [0.2cm]
  \lambda_{(3)}  =
     (-\frac{3}{2},1,1,1,0,0,0,\frac{3}{2}) \; , \;\;\;\;
     \lambda_{(4)}  = (-2,1,1,1,1,0,0,2) \; , \\ [0.2cm]
     \lambda_{(5)}  = \frac{1}{2}
    (-3,1,1,1,1,1,-1,3) \; , \;\;\;\;\;
     \lambda_{[[{\sf 912}]]} = \lambda_{(6)} =
    \frac{1}{2}(-2,1,1,1,1,1,1,2) \; , \\ [0.2cm]
      \lambda_{[[{\sf 133}]]} = \lambda_{\ad} = \lambda_{(7)} =
  (-1,0,0,0,0,0,0,1) \; , \;\;\;\;
     \lambda_{Y_n} = n\lambda_{(7)} =
    (-n,0,0,0,0,0,0,n) \; , \\ [0.2cm]
    \rho = \sum\limits_{i=1}^7 \lambda_{(i)} =
   (-\frac{17}{2},5,4,3,2,1,0,\frac{17}{2}) \; .
      \end{array}
 \ee}
  The decomposition of the tensor product
 of the defining representation $\Box=T_{\lambda_{(1)}}$ and
 the representation $Y_n=T_{n  \lambda_{(7)}}$ is
  \be
 \lb{e7fyn}
 T_{\lambda_{(1)}} \otimes T_{n  \lambda_{(7)}}
 = T_{n \lambda_{(7)}+\lambda_{(1)}} +
 T_{(n-1) \lambda_{(7)}+\lambda_{(1)}}  +
 T_{(n-1) \lambda_{(7)}+ \lambda_{(6)}}  \; .
 \ee
  With the help of the Weyl formula (\ref{weyldim}) we obtain
 \be
 \lb{dime7a}
 \begin{array}{c}
 \dim T_{\Lambda_1^{(n)}} =
 \dfrac{24(n+14)(n+17)}{(n+6)(2n+17)} \dim Y_n \, , \;\;\;
  \dim T_{\Lambda_2^{(n)}} =
 \dfrac{24n(n+3)}{(n+11)(2n+17)} \dim Y_n \, , \\ [0.3cm]
  \dim T_{\Lambda_3^{(n)}} =
 \dfrac{32n(n+17)}{(n+6)(n+11)} \dim Y_n \, ,
 \end{array}
 \ee
  where $\dim Y_n \equiv \dim T_{_{n  \lambda_{(7)}}}$
 is defined in (\ref{dimYk}) (with $\mathfrak{e}_7$
 Vogel parameters) and we denote
 \be
 \lb{e7Ln}
 \Lambda_1^{(n)} = n \lambda_{(7)}+ \lambda_{(1)}\; , \;\;\;\;
 \Lambda_2^{(n)} =
 (n-1) \lambda_{(7)}+ \lambda_{(1)}\; , \;\;\;\; \Lambda_3^{(n)}=
 (n-1) \lambda_{(7)}+ \lambda_{(6)} \; .
 \ee
By using (\ref{2casim}) we deduce the values of the Casimir operators
for the $\mathfrak{e}_7$ representations arising in the
decomposition (\ref{e7fyn}):
 \be
 \lb{e7Yn2}
 \begin{array}{c}
 C_{(2)}(\lambda_{(1)})= \frac{57}{2} \, , \;\;\;
 C_{(2)}(\lambda_{(7)})= C_{(2)}(\ad)= 36 \, , \;\;\;
 C_{(2)}(n \lambda_{(7)})= 2n(n+17) \, ,  \\ [0.2cm]
  C_{(2)}(\Lambda_1^{(n)})= 2n^2+36n+\frac{57}{2} \, , \;\;\;
 C_{(2)}( \Lambda_2^{(n)})=  2n^2 +32n-\frac{11}{2}) \, ,
 \\ [0.2cm]
 C_{(2)}(\Lambda_3^{(n)})= (2n+33)(n+\frac{1}{2}) \, .
 \end{array}
 \ee
 Finally, we use (\ref{2cas1}), (\ref{gfad}) and obtain
 the eigenvalues of the split Casimir operator
 in the representations arising in the decomposition
 of $\Box \otimes Y_n$ for the $\mathfrak{e}_7$ algebra
 \be
 \lb{e7Yn3}
 \hat{c}_{(2)}^{\Lambda_1^{(n)}} = \frac{n}{36} \, , \;\;\;
 \hat{c}_{(2)}^{\Lambda_2^{(n)}} = -\frac{n+17}{36}
 \, , \;\;\; \hat{c}_{(2)}^{\Lambda_3^{(n)}}= -\frac{1}{6} \, .
\ee
The characteristic identity for the split Casimir
 operator $\hat{C}_{\Box \otimes Y_n}$ is
\be
 \lb{e7Yn4}
 \Bigl( \hat{C}_{\Box \otimes Y_n} - \frac{n}{36}\Bigr)
 \Bigl( \hat{C}_{\Box \otimes Y_n} + \frac{n+17}{36}\Bigr)
 \Bigl( \hat{C}_{\Box \otimes Y_n} + \frac{1}{6}\Bigr) = 0
 \ee

  \subsubsection{The Lie algebra $\mathfrak{e}_8$. Irreps $Y_{n}$
  and decomposition of $\Box \otimes Y_n$}

    The root space of the Lie algebra
  $\mathfrak{e}_8$ is the
 8-dimensional Euclidean space $\mathbb{R}^8$.
    The fundamental weights, the Weyl vector $\rho$,
 and the highest weight vector of the
 representation $Y_n$ are defined as
   {\small \be
 \lb{e8yn1}
 \begin{array}{c}
  \lambda_{[[248]]} = \lambda_{(1)} = \lambda_{\ad} =
     (1,0,0,0,0,0,0,1) \; , \;\;\;\;
    \lambda_{[[{\sf 30380}]]} = \lambda_{(2)}  = (1,1,0,0,0,0,0,2) \; ,  \\ [0.2cm]
  \lambda_{(3)}  =
     (1,1,1,0,0,0,0,3) \; , \;\;\;\;
     \lambda_{(4)}  = (1,1,1,1,0,0,0,4) \; , \\ [0.2cm]
     \lambda_{(5)}  = \frac{1}{2}
    (1,1,1,1,1,0,0,5) \; , \;\;\;\;\;
      \lambda_{(6)} =
    \frac{1}{2}(1,1,1,1,1,1,-1,7) \; , \\ [0.2cm]
       \lambda_{(7)} =  \frac{1}{2}(1,1,1,1,1,1,1,5) \; , \;\;\;\;
  \lambda_{[[{\sf 3875}]]} = \lambda_{(8)} =(0,0,0,0,0,0,0,2) \; , \\ [0.2cm]
     \lambda_{Y_n} = n\lambda_{(1)} =
    (n,0,0,0,0,0,0,n) \; , \;\;\; 
    \rho = \sum\limits_{i=1}^7 \lambda_{(i)} =
   (6,5,4,3,2,1,0,23) \; .
      \end{array}
 \ee}
We use the definitions of the fundamental weights and
 the Weyl vector $\rho$, which are given in \cite{IsRub2}
 (see Problem 3.4.18) and \cite{Burb},
 where the numbering of simple roots and fundamental
 weights differs from that adopted here\footnote{The correspondence is
 $\lambda_{(1)} =\omega_{(8)}$, $\lambda_{(2)} =\omega_{(7)}$,
 $\lambda_{(3)} =\omega_{(6)}$, $\lambda_{(4)} =\omega_{(5)}$, $\lambda_{(5)} =\omega_{(4)}$,
 $\lambda_{(6)} =\omega_{(3)}$, $\lambda_{(7)} =\omega_{(2)}$,
 $\lambda_{(8)} =\omega_{(1)}$, where $\omega_{(i)}$
 are the fundamental weights defined in \cite{Burb}. Our numeration
 coincides with numeration in \cite{IsRub2}; see Fig. \ref{XXX1}.}.

  The decomposition of the tensor product
 of the defining representation $\Box=T_{\lambda_{(1)}}=\ad=Y_1$ and
 the representation $Y_n=T_{n\, \lambda_{(1)}}$ (the Cartan power of
 the adjoint representation) for the $\mathfrak{e}_8$ case is
 \be
 \lb{e8fa}
 \begin{array}{l}
 T_{_{\lambda_{(1)}}} \otimes T_{_{n\, \lambda_{(1)}}} = \\ [0.2cm]
= T_{_{(n+1)\, \lambda_{(1)}}} +
  T_{_{(n-1) \, \lambda_{(1)}+\lambda_{(2)}}}
 + T_{_{(n-1) \, \lambda_{(1)}+\lambda_{(8)}}}
  + T_{_{n\, \lambda_{(1)}}} + T_{_{(n-2) \, \lambda_{(1)}+\lambda_{(2)}}}
  + T_{_{(n-1) \, \lambda_{(1)}}} \; ,
   \end{array}
 \ee

 Comparing (\ref{e8fa})
 with decompositions (\ref{g2fyn}), (\ref{f4fyn}),
 (\ref{e6fyn}), (\ref{e7fyn}) for $\Box \times Y_n$
 for other exceptional Lie algebras $\mathfrak{g}_2,
 \mathfrak{f}_4,\mathfrak{e}_6,\mathfrak{e}_7$,
 we come to the conclusion that we cannot
 observe universality for $\mathfrak{e}_8$ case
 in the context of considering the decomposition
 of $\Box \otimes Y_n$. However,
 due to the fact that for  $\mathfrak{e}_8$
 the defining representation $\Box$
 coincides with the adjoint representation $Y_1$, we have
  for the decomposition (\ref{e8fa}) (in the
  $\mathfrak{e}_8$ case) the standard Vogel universal
  prescription in terms of the Vogel parameters specific for the
  tensor products $Y_1 \times Y_n$ (see eqs. (\ref{yny01}),
  (\ref{yny02}) and (\ref{yny02sp}) for the $\mathfrak{sl}_N$, $\mathfrak{so}_N$
  and $\mathfrak{sp}_N$ cases)
 arising in the decomposition of the
  tensor products $\ad^{\otimes (n+1)}$ of the adjoint representations.

   \subsubsection{Irreps $Y_{n}'$
  and decompositions of $\Box \otimes Y_n'$
  for exceptional Lie algebras. General remarks}

 It is remarkable that all characteristic identities
 (\ref{g2Yn4}), (\ref{f4Yn4}), (\ref{e6Yn4}) and (\ref{e7Yn4}),
 for the exceptional
 Lie algebras $\mathfrak{g}_2$, $\mathfrak{f}_4$,
 $\mathfrak{e}_6$, $\mathfrak{e}_7$
 can be written in a uniform way by using the Vogel parameters
  \be
  \lb{charn1}
 \Bigl(\hat{C}_{\Box \otimes Y_n} +
 \frac{1}{2}+\frac{\hat{\alpha}}{2}(1-n)\Bigr)
  \Bigl(\hat{C}_{\Box \otimes Y_n} + n \frac{\hat{\alpha}}{2}\Bigr)
  \Bigl(\hat{C}_{\Box \otimes Y_n} + \frac{\hat{\gamma}}{2}\Bigr) = 0 \; .
  \ee
  Then we can unify identity (\ref{charn0}) for Lie algebras
  of the classical series with identity (\ref{charn1})
  and write an identity that will be valid for all
  simple Lie algebras (except the $\mathfrak{e}_8$ algebra)
  \be
  \lb{charn2}
 \Bigl(\hat{C}_{\Box \otimes Y_n} +
 \frac{1}{2}(1+\alpha(1-n))\Bigr)
  \Bigl(\hat{C}_{\Box \otimes Y_n} + n \frac{\hat{\alpha}}{2}\Bigr)
  \Bigl(\hat{C}_{\Box \otimes Y_n} + \frac{\hat{\beta}}{2}\Bigr)
  \Bigl(\hat{C}_{\Box \otimes Y_n} + \frac{\hat{\gamma}}{2}\Bigr) = 0 \; .
  \ee

 Unfortunately, the uniform identity (\ref{charnp0}) for
 the split Casimir
  operator $\hat{C}_{\Box \otimes Y_n'}$
  in the representation $\Box \otimes Y_n'$
  for Lie algebras
  of the classical series cannot to be generalized
   to the case of all
  simple Lie algebras (including exceptional Lie algebras),
  as in the case of
  characteristic identity
  (\ref{charn2}) for the operator
  $\hat{C}_{\Box \otimes Y_n}$  in the representation
  $\Box \otimes Y_n$. The natural conjecture
  for the characteristic identity that could be obtained from
   (\ref{charn2}) by the exchange
   $\alpha \leftrightarrow \beta$, as we did to obtain
    (\ref{charnp0}) from (\ref{charn0}), does not work.

Indeed, we have the following decompositions of the
tensor product $\Box \otimes Y_2'$ of the defining $\Box$ and
 $Y_2' = Y_2(\beta)$ representations  for the case of
the exceptional Lie algebras
 \be
 \lb{figna}
\begin{array}{l}
\mathfrak{g}_2: \;\;\;
 [[7]] \otimes [[27]] =
 [[77]] + [[64]] + [[14]] + [[7]] + [[27]]  \; , \\ [0.2cm]
 \mathfrak{f}_4: \;\;\; [[26]] \otimes [[324]] =
 [[2652]] + [[4096]] + [[1053]] + [[273]] + [[324]] + [[26]] \; ,  \\ [0.2cm]
 \mathfrak{e}_6: \;\;\;   [[27]] \otimes [[650]] =
 [[7722]] + [[7371]] + [[1728]] + [[351']] + [[351]] + [[27]] \; ,  \\ [0.2cm]
 \mathfrak{e}_7: \;\;\;   [[56]] \otimes [[1539]] =
 [[51072]] + [[27664]] + [[6480]] + [[912]] + [[56]]  \; ,  \\ [0.2cm]
 \mathfrak{e}_8: \;\;\; [[248]] \otimes  [[3875]] =
 [[779247]] + [[147250]] + [[30380]] + [[3875]] + [[248]] \; .
\end{array}
 \ee
 Here the case of $\mathfrak{e}_8$ corresponds to the case of
 the standard Vogel universality observed in the decompositions of
 $\ad \otimes  Y_2' \subset \ad^{\otimes 3}$.
As we see, in the decompositions (\ref{figna}) of the representations
$\Box \otimes Y_2'$, a different number of terms arises.
For example, in the cases of
 $\mathfrak{f}_4$ and $\mathfrak{e}_6$ algebras, the decomposition has six terms, while for the cases of
 $\mathfrak{g}_2$ and $\mathfrak{e}_7$ algebras, the decomposition has only
 five terms. Thus, the corresponding characteristic identity must have at least six factors, which is difficult to reconcile with the universal
 identity (\ref{charnp0}) for a series of the classical Lie
 algebras, where there is only three factors.
 Therefore, writing the universal characteristic identity for
 the split Casamir operator $\hat{C}_{\Box \otimes Y_n'}$
 for all simple Lie algebras (except $\mathfrak{e}_8$) is apparently impossible.

 \newtheorem{not9}[not2]{Remark\itshape}
\begin{not9}\label{not9}
We summarize the results obtained in (\ref{c2yn1}),
(\ref{slypr2}),  (\ref{sofyn1}),(\ref{sofynp1}),
(\ref{g2Yn2}), (\ref{f4Yn2}), (\ref{e6Yn2}), (\ref{e7Yn2})
and give a general universal formulae for the normalized
eigenvalues (\ref{2cas1}) of the quadratic Casimir
 operators in the representations
$Y_n$ and $Y_n'$  for all simple Lie algebras
(see Theorem 1.1 in \cite{LanMan})
 \be
 \lb{c2yn}
 c_{(2)}^{Y_n} = \frac{C_{(2)}(Y_n)}{C_{(2)}(\ad)} =
 n(1-\hat{\alpha} n+\hat{\alpha}) \; , \;\;\;\;
 c_{(2)}^{Y_n'} = \frac{C_{(2)}(Y_n')}{C_{(2)}(\ad)} =
 n(1-\hat{\beta} n+\hat{\beta}) \; .
 \ee
\end{not9}

  \section{Universal dimensions
   for subrepresentations in
   decompositions of $\Box \otimes Y_n$
   and $\Box \otimes Y_n'$\label{univdim}}
   \setcounter{equation}0

 \subsection{Projectors on the Casimir
 subspaces and traces for
 $\hat{C}^m_{\Box \otimes Y_n}$ and $\hat{C}^m_{\Box \otimes Y_n'}$}

   \subsubsection{Traces of
  powers of 2-split Casimir and
  higher Casimir operators}

Consider the images of the $2$-split Casimir operator
$\hat{C}_{(12)} = {\sf g}^{ab} X_a \otimes X_b$ in a representation
 $T^{(1)} \otimes T^{(2)}$:
\be
\lb{trcas03}
 \hat{C}_{T^{(1)} \otimes T^{(2)}} =
  {\sf g}^{ab} T^{(1)}(X_a) \otimes T^{(2)}(X_b) \; ,
\ee
where $T^{(1)}$ and $T^{(2)}$ are two irreps of the Lie algebra
$\mathfrak{g}$.

First,  we rewrite general formulas (\ref{char05}) --
(\ref{char08}) for
 the case of the 2-split Casimir operator (\ref{trcas03}).
 The characteristic identity (\ref{char05})
 is written as
\be
  \lb{charn5}
\prod_{j=1}^M (\hat{C}_{T^{(1)}\otimes T^{(2)}} - a_j) = 0 \, ,
\ee
where $a_j$ are the mismatched eigenvalues of
$\hat{C}_{T^{(1)}\otimes T^{(2)}}$ defined in
(\ref{gfad}) (for $k=2$). The corresponding projectors (\ref{char06}),
dimension formula (\ref{char07})
and the formula (\ref{char08}) for the trace of the $m$-th power
 of the operator $\hat{C}_{T^{(1)}\otimes T^{(2)}}$ are
\be
  \lb{charn6}
P_{a_\ell} = \prod_{j\neq \ell}
\frac{(\hat{C}_{T^{(1)}\otimes T^{(2)}} - a_j)}{(a_\ell - a_j)}
\; , \;\;\;\; \dim V_{a_\ell} = {\rm Tr}_{12}(P_{a_\ell}) \; ,
 \ee
 \be
 \lb{cask01}
{\rm Tr}_{12}(\hat{C}_{T^{(1)}\otimes T^{(2)}})^m  =
 {\rm Tr}_{12} \sum_{\ell=1}^M \; a_\ell^m \; P_{a_\ell} =
 \sum_{\ell=1}^M \; a_\ell^m \; \dim V_{a_\ell} \; .
\ee

Then, using (\ref{casnorm1}) and
 (\ref{casnorm2}), we deduce
\be
\lb{trcas04b}
{\rm Tr}_{1} \bigl( \hat{C}_{T^{(1)}\otimes T^{(2)}}\bigr) = 0 =
{\rm Tr}_{2} \bigl( \hat{C}_{T^{(1)}\otimes T^{(2)}} \bigr) \; ,
\ee
\be
\lb{trcas04}
\begin{array}{c}
{\rm Tr}_{12} \bigl( \hat{C}_{T^{(1)}\otimes T^{(2)}}\bigr)^2  =
 {\sf g}^{a_1 b_1} {\sf g}^{a_2 b_2}
 {\rm Tr}_1 \bigl(T^{(1)}(X_{a_1}) \, T^{(1)}(X_{a_2}\bigr)
 {\rm Tr}_2 \bigl(T^{(2)}(X_{b_1})  T^{(2)}(X_{b_2})\bigr) =  \\ [0.2cm]
 = d^{^{(T^{(1)})}} \, c_{(2)}^{(T^{(2)})} \, \dim T^{(2)} =
 d^{^{(T^{(2)})}} \, c_{(2)}^{(T^{(1)})} \, \dim T^{(1)}
 = d^{^{(T^{(1)})}} d^{^{(T^{(2)})}}\,  \dim \mathfrak{g} \; ,
 \end{array}
\ee
In particular, for any irrep $T^{(1)}=T$ with
the highest weight $\lambda$
and adjoint representation $T^{(2)}=\ad$, we obtain
\be
\lb{trcas01}
{\rm Tr}_{12} \bigl( \hat{C}_{T \otimes \ad } \bigr)^2 =
 c_{(2)}^{(T)} \, \dim T =  d^{^{(T)}}\,  \dim \mathfrak{g}
 \;\;\; \Rightarrow \;\;\; d^{^{(T)}} =
 \frac{c_{(2)}^{(T)} \, \dim T}{\dim \mathfrak{g} }  \; ,
\ee
where we take into account $d^{^{(\ad)}}=1$
and $c_{(2)}^{(T)} \equiv c_{(2)}^{(\lambda)} =
C_{(2)}(\lambda)/C_{(2)}(\ad)$ (the definition of
$C_{(2)}(\lambda)$ is given
in (\ref{2casim})). Substitution of (\ref{trcas01})
into (\ref{trcas04}) yields formula
 \be
 \lb{trcas05}
{\rm Tr}_{12} \bigl( \hat{C}_{T^{(1)} \otimes T^{(2)}} \bigr)^2 =
\frac{c_{(2)}^{(T^{(1)})} \, c_{(2)}^{(T^{(2)})} \, \dim T^{(1)}
\, \dim T^{(2)}}{\dim \mathfrak{g}} \; ,
 \ee
 which we will use below.

 The formula (\ref{trcas01}) can be generalized
 (with the help of (\ref{cask01})
 and Schur's Lemma) as follows
  \be
  \lb{khcas01}
  \begin{array}{c}
 {\rm Tr}_{2}(\hat{C}_{T \otimes \ad})^m  =
 T_{a_1} \cdots T_{a_m} \, {\sf g}^{a_1...a_m}
 = c_{(m)}^{(T)} \, I_{_T} , \;\;\;\;\;\;
 c_{(m)}^{T} = \frac{1}{\dim T} \,
 \sum\limits_{\ell=1}^M \; a_\ell^m \; \dim V_{a_\ell} \; , \\ [0.3cm]
 {\sf g}^{a_1 ... a_m} := {\sf g}^{a_1 b_1} \cdots {\sf g}^{a_m b_m}
{\rm Tr}\bigl({\rm ad}(X_{b_1}) \cdots {\rm ad}(X_{b_m})\bigr) \; ,
 \end{array}
  \ee
  where $I_T$ and $c_{(m)}^{T}$ are respectively unit operator and
  eigenvalue of the higher $m$-Casimir
 in the space of the representation $T$, and $a_\ell$ are eigenvalues
 of $\hat{C}_{T \otimes \ad}$ in the Casimir spaces $V_{a_\ell}$.

  \newtheorem{pro4}[pro2]{Proposition}
 \begin{pro4}\label{pro4}
 Let $T^{(1)}$ and $T^{(2)}$ be two representations of
 the simple Lie algebra $\mathfrak{g}$ and representation
 $T^{(2)}$ (or $T^{(1)}$) is equivalent
 to the skew symmetric representation such that
 \be
 \lb{skewrep}
 (T^{(2)}(X_a))^{\sf T} = - A\, T^{(2)}(X_a)\, A^{-1} \; ,
 \;\;\;\;\; (\forall X_a \in \mathfrak{g}) \; ,
 \ee
 where $A$ is some constant matrix. Then, for normalization
 accepted in (\ref{defMK}) and (\ref{trcas03}), we have
 identity
 \be
 \lb{c2c3}
 {\rm Tr}_{12} \bigl( \hat{C}_{T^{(1)} \otimes T^{(2)}} \bigr)^3 =
 - \frac{1}{4} \,
 {\rm Tr}_{12} \bigl( \hat{C}_{T^{(1)} \otimes T^{(2)}} \bigr)^2 \; .
 \ee
\end{pro4}
{\bf Proof.} Introduce the concise notation
$T^{(i)}_{a} = T^{(i)}(X_{a})$. First of all,
in view of (\ref{skewrep}), we find
 \be
 \lb{t2skew}
d^{(2)}_{b_1 b_2 b_3} :=
{\rm Tr} \bigl(T^{(2)}_{b_1}T^{(2)}_{b_2}T^{(2)}_{b_3}\bigr) =
{\rm Tr} \bigl(T^{(2)}_{b_1}T^{(2)}_{b_2}T^{(2)}_{b_3}\bigr)^{\sf T} =
- {\rm Tr} \bigl(T^{(2)}_{b_3}T^{(2)}_{b_2}T^{(2)}_{b_1}\bigr) \; ,
 \ee
 and this means that the tensor $d^{(2)}_{b_1 b_2 b_3}$ is completely
 anti-symmetric with respect to all permutations of $(b_1, b_2, b_3)$.
 Then, for the left hand side of (\ref{c2c3})
we deduce
 {\small $$
 \begin{array}{c}
{\rm Tr}_{12} \bigl( \hat{C}_{T^{(1)} \otimes T^{(2)}} \bigr)^3 =
{\sf g}^{a_1 b_1} {\sf g}^{a_2 b_2}{\sf g}^{a_3 b_3} \,
{\rm Tr}_{1} \bigl(T^{(1)}_{a_1}T^{(1)}_{a_2}T^{(1)}_{a_3}\bigr) \,
{\rm Tr}_{2} \bigl(T^{(2)}_{b_1}T^{(2)}_{b_2}T^{(2)}_{b_3}\bigr) =
\\ [0.2cm]
 = \dfrac{1}{4}
  {\sf g}^{a_1 b_1} {\sf g}^{a_2 b_2}{\sf g}^{a_3 b_3} \,
 {\rm Tr}_{1} \bigl(
 [T^{(1)}_{a_1}T^{(1)}_{a_2}]_{_{-}} \, T^{(1)}_{a_3} \bigr) \,
{\rm Tr}_{2} \bigl([T^{(2)}_{b_1}T^{(2)}_{b_2}]_{_{-}} \,
T^{(2)}_{b_3}\bigr) =
\\ [0.2cm]
 = \dfrac{1}{4}
  {\sf g}^{a_1 b_1} {\sf g}^{a_2 b_2}{\sf g}^{a_3 b_3} \,
 X_{a_1 a_2}^{d_1} \,  X_{b_1 b_2}^{d_2} {\rm Tr}_{1} \bigl(
 T^{(1)}_{d_1}\, T^{(1)}_{a_3} \bigr) \,
{\rm Tr}_{2} \bigl( T^{(2)}_{d_2} \,
T^{(2)}_{b_3}\bigr) = \\ [0.2cm]
 = - \, \dfrac{1}{4} {\sf g}^{d_1 d_2}{\sf g}^{a_3 b_3} \,
 {\rm Tr}_{1} \bigl( T^{(1)}_{d_1}\, T^{(1)}_{a_3} \bigr) \,
{\rm Tr}_{2} \bigl( T^{(2)}_{d_2} \,
T^{(2)}_{b_3}\bigr) =
- \dfrac{1}{4} \,
 {\rm Tr}_{12} \bigl( \hat{C}_{T^{(1)} \otimes T^{(2)}} \bigr)^2 \; ,
 \end{array}
 $$ }
 where in the second equality we took into account
  that tensor
 ${\rm Tr}_{2}\bigl(T^{(2)}_{a_1}\, T^{(2)}_{a_2} \,
 T^{(2)}_{a_3} \bigr)$ is completely anti-symmetric. \hfill \qed

\vspace{0.1cm}

  \noindent
 {\bf Corollary.} The condition (\ref{skewrep}) is fulfilled
 for all Cartan powers $Y_n$ of the
 adjoint representation of the simple Lie algebras $\mathfrak{g}$.
 Indeed, for the adjoint representation $\ad=Y_1$
 the condition (\ref{skewrep}) is valid
 \be
 \lb{skewrep2}
 X_{af}^{c} = - \, {\sf g}_{f b}\, X_{ad}^{b} \, {\sf g}^{d c} \; .
 \ee
 The generators $X_a$
 of the Lie algebra $\mathfrak{g}$ in the representation $Y_n$
 are
 $$
 (T^{(Y_n)}(X_a))^{(b_1...b_n)}_{(c_1...c_n)} \equiv
 X^{(b_1...b_n)}_{a(c_1...c_n)} =
 X_{a(c_1} ^{(b_1} \delta^{b_2}_{c_2} \cdots \delta^{b_n)}_{c_n)}
 $$
 where parentheses in $(b_1...b_n)$ indicate
 the symmetrization over
 all indices $b_1,...,b_n$ and in view of
 (\ref{skewrep2}) we also obtain for $T^{(Y_n)}(X_a)$ the condition
(\ref{skewrep})
 $$
  X^{(d_1...d_n)}_{a(f_1...f_n)} = - \,
 {\sf g}_{f_1 b_1} \cdots {\sf g}_{f_n b_n}\,  X^{(b_1...b_n)}_{a(c_1...c_n)}
  {\sf g}^{c_1 d_1} \cdots {\sf g}^{c_n d_n} \; .
 $$
 Thus, for any irreducible representation $T$
 of the simple Lie algebra $\mathfrak{g}$ we have identity
 (\ref{c2c3})
  \be
 \lb{c2c3yn}
 {\rm Tr}_{12} \bigl( \hat{C}_{T \otimes Y_n} \bigr)^3 =
 - \frac{1}{4} \,
 {\rm Tr}_{12} \bigl( \hat{C}_{T \otimes Y_n} \bigr)^2 \; ,
 \ee
 for 2-split Casimir operators (\ref{trcas03})
 with normalization defined in (\ref{defMK}), (\ref{Qcas}).

 \subsubsection{The case of the $\mathfrak{sl}_N$ algebra}

By using decompositions (\ref{slfy2}) and (\ref{slfypr2}),
eigenvalues (\ref{slcYn}),  and (\ref{slcynp}),
we deduce, by means of (\ref{cask01}), the traces of powers of the
split Casimir operator in the
representations $\Box \otimes Y_n$ and $\Box \otimes Y_n'$
in the case of
$\mathfrak{sl}_N$:
\be
\lb{trksl0}
\begin{array}{c}
{\rm Tr}_{12}\hat{C}_{\Box \otimes Y_n}^k =
(\frac{n}{2N})^k \cdot \dim ([2n+1,n^{N-2}])
+ (-\frac{1}{2N})^k \cdot \dim (2n,n+1,n^{N-3}) + \\ [0.2cm]
+(-\frac{n-1+N}{2N})^k \cdot \dim ([2n-1,(n-1)^{N-2}]) \; ,
\end{array}
\ee
\be
\lb{ptrksl0}
\begin{array}{c}
{\rm Tr}_{12}\hat{C}_{\Box \otimes Y_n'}^k =
(-\frac{n}{2N})^k \cdot \dim ([2^{n+1},1^{N-2n-1}])
+ (\frac{1}{2N})^k \cdot \dim (3,2^{n-1},1^{N-2n}) + \\ [0.2cm]
+(-\frac{N-n+1}{2N})^k \cdot \dim ([2^n,1^{N-2n+1}]) \; .
\end{array}
\ee
In particular for $k=1,2,3$ we obtain:
\be
\lb{trC2}
{\rm Tr} \hat{C}_{\Box \otimes Y_n} =0 , \;\;\;
{\rm Tr} \hat{C}_{\Box \otimes Y_n}^2 =
\frac{n (N+n-1)}{2 N^2} \dim (Y_n) , \;\;\;
{\rm Tr} \hat{C}_{\Box \otimes Y_n}^3 =
-\frac{1}{4} {\rm Tr} \hat{C}_{\Box \otimes Y_n}^2 \; ,
\ee
\be
\lb{trpC2}
{\rm Tr} \hat{C}_{\Box \otimes Y_n'} =0 , \;\;\;
{\rm Tr} \hat{C}_{\Box \otimes Y_n'}^2 =
\frac{n (N-n+1)}{2 N^2} \dim (Y_n') , \;\;\;
{\rm Tr} \hat{C}_{\Box \otimes Y_n'}^3 =
-\frac{1}{4} {\rm Tr} \hat{C}_{\Box \otimes Y_n'}^2 \; .
\ee
As expected, the last identities in (\ref{trC2})
and (\ref{trpC2}) coincides with
 (\ref{c2c3}).


  \subsubsection{The case of the $\mathfrak{so}_N$ algebra}

In the same way,  in the case of
$\mathfrak{so}_N$, by using dimensions (\ref{dimso123}),
(\ref{dimpso123}),
eigenvalues (\ref{socYn}), (\ref{socynp})  and formula
(\ref{cask01}),  we obtain the traces for powers of the
split Casimir operator in the
representation $\Box \otimes Y_n$ and
$\Box \otimes Y_n'$:
\be
\lb{trkso0}
\begin{array}{c}
{\rm Tr}_{12}\hat{C}_{\Box \otimes Y_n}^k =
(\frac{n}{2(N-2)})^k \cdot \dim ([n+1,n])
+ (-\frac{1}{N-2})^k \cdot \dim (n^2,1) + \\ [0.2cm]
+(-\frac{N+n-3}{2(N-2)})^k \cdot \dim ([n,n-1]) \; ,
\end{array}
\ee
\be
\lb{trkso1}
\begin{array}{c}
{\rm Tr}_{12}\hat{C}_{\Box \otimes Y_n'}^k =
(-\frac{n}{(N-2)})^k \cdot \dim ([1^{2n+1}])
+ (\frac{1}{2(N-2)})^k \cdot \dim ([2,1^{2n-1}]) + \\ [0.2cm]
+(-\frac{N-2n}{2(N-2)})^k \cdot \dim ([1^{2n-1}]) \; .
\end{array}
\ee
For $k=1,2,3$ we deduce in the $\mathfrak{so}_N$ case:
\be
\lb{Ck23}
{\rm Tr} \hat{C}_{\Box \otimes Y_n} = 0 , \;\;\;
{\rm Tr} \hat{C}_{\Box \otimes Y_n}^2 =
\frac{(N+n-3)n}{(N-2)^2} \cdot \dim Y_n
\, , \;\;\;\;
{\rm Tr} \hat{C}_{\Box \otimes Y_n}^3 =
-\frac{1}{4} {\rm Tr} \hat{C}_{\Box \otimes Y_n}^2 \; ,
\ee
\be
\lb{Ckpr23}
{\rm Tr} \hat{C}_{\Box \otimes Y_n'} =0  , \;\;\;\;
{\rm Tr} \hat{C}_{\Box \otimes Y_n'}^2 =
\frac{(N-2n)n}{(N-2)^2} \cdot \dim Y_n'
\, , \;\;\;\;
{\rm Tr} \hat{C}_{\Box \otimes Y_n'}^3 =
-\frac{1}{4} {\rm Tr} \hat{C}_{\Box \otimes Y_n'}^2 \; .
\ee
The last identities in (\ref{Ck23}) and
(\ref{Ckpr23}) coincide with (\ref{c2c3}).


 \subsubsection{The case of the exceptional Lie algebras}

For the Lie algebras $\mathfrak{g}_2$, $\mathfrak{f}_4$,
$\mathfrak{e}_6$, $\mathfrak{e}_7$ we have (see, respectively,
subsections {\bf \ref{secg2}}, {\bf \ref{secf4}},
{\bf \ref{sece6}} and {\bf \ref{sece7}},)
\be
\lb{trkgfe}
\begin{array}{c}
\mathfrak{g}_2: \; {\rm Tr}_{12}\hat{C}_{\Box \otimes Y_n}^k =
(\frac{n}{8})^k \cdot
\dim (T_{\Lambda_1^{(n)}})
+ (-\frac{n+3}{8})^k \cdot
\dim (T_{\Lambda_2^{(n)}})
+(-\frac{1}{6})^k \cdot
\dim (T_{\Lambda_3^{(n)}}) \; , \\ [0.3cm]
\mathfrak{f}_4: \; {\rm Tr}_{12}\hat{C}_{\Box \otimes Y_n}^k =
(\frac{n}{18})^k \cdot
\dim (T_{\Lambda_1^{(n)}})
+ (-\frac{n+8}{18})^k \cdot
\dim (T_{\Lambda_2^{(n)}})
+(-\frac{1}{6})^k \cdot
\dim (T_{\Lambda_3^{(n)}}) \; , \\ [0.3cm]
\mathfrak{e}_6: \; {\rm Tr}_{12}\hat{C}_{\Box \otimes Y_n}^k =
(\frac{n}{24})^k \cdot
\dim (T_{\Lambda_1^{(n)}})
+ (-\frac{n+11}{24})^k \cdot
\dim (T_{\Lambda_2^{(n)}})
+(-\frac{1}{6})^k \cdot
\dim (T_{\Lambda_3^{(n)}}) \; , \\ [0.3cm]
\mathfrak{e}_7: \; {\rm Tr}_{12}\hat{C}_{\Box \otimes Y_n}^k =
(\frac{n}{36})^k \cdot
\dim (T_{\Lambda_1^{(n)}})
+ (-\frac{n+17}{36})^k \cdot
\dim (T_{\Lambda_2^{(n)}})
+(-\frac{1}{6})^k \cdot
\dim (T_{\Lambda_3^{(n)}}) \; , \\ [0.3cm]
\end{array}
\ee
and, for all these cases and normalization
accepted in (\ref{defMK}), (\ref{Qcas}), we obtain
identity (\ref{c2c3})
 \be
\lb{Ckgfe23}
{\rm Tr} \hat{C}_{\Box \otimes Y_n}^3 =
-\frac{1}{4} {\rm Tr} \hat{C}_{\Box \otimes Y_n}^2 \; .
\ee

 \subsection{Universal dimensions
   for Casimir representations in the
   decompositions of $\Box \otimes Y_n$}

   The characteristic identity (\ref{charn2}) yields
  four projectors (\ref{charn6}) which are written as
  {\small \be
 \lb{proj01}
 \begin{array}{c}
 P^{(n)}_{a_1} = \dfrac{ (2 \, \hat{C}_{\Box \otimes Y_n} + n \hat{\alpha})
  \bigl(2 \, \hat{C}_{\Box \otimes Y_n} + \hat{\beta}\bigr)
  \bigl(2 \, \hat{C}_{\Box \otimes Y_n} + \hat{\gamma}\bigr)}{
  (2 n \hat{\alpha}-\hat{\alpha}-1)
  (\hat{\alpha}n+ \hat{\beta} -\hat{\alpha}-1)
  (\hat{\alpha}n+ \hat{\gamma} -\hat{\alpha}-1)} \; ,
 \\ [0.4cm]
 P^{(n)}_{a_2} =\dfrac{ \Bigl(\hat{C}_{\Box \otimes Y_n} +
 \frac{1}{2}+\frac{\hat{\alpha}}{2}(1-n)\Bigr)
  \bigl(2 \, \hat{C}_{\Box \otimes Y_n} + \hat{\beta}\bigr)
  \bigl(2 \, \hat{C}_{\Box \otimes Y_n} + \hat{\gamma}\bigr)}{
  \Bigl(\frac{1}{2}+\frac{\hat{\alpha}}{2} -  n \hat{\alpha}\Bigr)
  (\hat{\beta}-  n \hat{\alpha})
  (\hat{\gamma} -  n \hat{\alpha})} \; , \\ [0.2cm]
 P^{(n)}_{a_3} = \dfrac{ \Bigl(\hat{C}_{\Box \otimes Y_n} +
 \frac{1}{2}+\frac{\hat{\alpha}}{2}(1-n)\Bigr)
  \bigl(2 \hat{C}_{\Box \otimes Y_n} + n \hat{\alpha}\bigr)
   \bigl(2  \hat{C}_{\Box \otimes Y_n} + \hat{\gamma}\bigr)}{
  \Bigl(-\frac{\hat{\beta}}{2} +
 \frac{1}{2}+\frac{\hat{\alpha}}{2}(1-n)\Bigr)
  (\hat{\beta} - n \hat{\alpha})
  (\hat{\beta}- \hat{\gamma})} \; , \\ [0.2cm]
   P^{(n)}_{a_4} = \dfrac{ \Bigl(\hat{C}_{\Box \otimes Y_n} +
 \frac{1}{2}+\frac{\hat{\alpha}}{2}(1-n)\Bigr)
  \bigl(2 \hat{C}_{\Box \otimes Y_n} + n \hat{\alpha}\bigr)
   \bigl(2 \hat{C}_{\Box \otimes Y_n} + \hat{\beta}\bigr)}{
  \Bigl(-\frac{\hat{\gamma}}{2} +
 \frac{1}{2}+\frac{\hat{\alpha}}{2}(1-n)\Bigr)
  \bigl(\hat{\gamma} - n \hat{\alpha}\bigr)
  \bigl(\hat{\gamma}- \hat{\beta}\bigr)} \; ,
 \end{array}
 \ee}
 where the eigenvalues are
 \be
 \lb{eigenv}
  a_1 = - \frac{1}{2}(1+\hat{\alpha}(1-n)) \; , \;\;\;\;
  a_2 = -  n \frac{\hat{\alpha}}{2} \; , \;\;\;\;
  a_3 = - \frac{\hat{\beta}}{2} \; , \;\;\;\;
  a_4 = -  \frac{\hat{\gamma}}{2} \; .
 \ee
 According to the formulas (\ref{g2Ln}), (\ref{f4Ln}),
 (\ref{e6Ln}) and  (\ref{e7Ln}),
 we denote the representations, which correspond to these eigenvalues,
 respectively, as $\Lambda^{(n)}_2(a_1)$,
 $\Lambda^{(n)}_1(a_2)$, $\Lambda^{(n)}_4(a_3)$, $\Lambda^{(n)}_3(a_4)$
 (see Table {\sf \ref{tab6}}), and we have the
 universal decomposition
  \be
  \lb{decomp}
 \Box \otimes Y_n =  \Lambda^{(n)}_1 + \Lambda^{(n)}_2 +
 \Lambda^{(n)}_3 + \Lambda^{(n)}_4 \; .
  \ee
  The standard method for calculating the dimensions of
  eigenspaces
  $V_{a_\ell} \subset  V_{\Box} \otimes V_{Y_n}$
  (extracted by the projectors $P^{(n)}_{a_\ell}$)
  is to calculate the traces of these projectors
  (\ref{charn6}).
  This can be done by using the values of the traces
  of the powers of the split Casimir operator
 \be
 \lb{TRCK}
 \begin{array}{c}
 {\rm Tr}_{12} {\bf 1} = \dim (\Box) \, \dim(Y_n) \; , \;\;\;\;\;\;
 {\rm Tr}_{12} \hat{C}_{\Box \otimes Y_n} = 0 \, , \\ [0.3cm]
 {\rm Tr}_{12} (\hat{C}_{\Box \otimes Y_n})^2 =
 \dfrac{c_{(2)}^{(\Box)} \, c_{(2)}^{Y_n} \, \dim (\Box) \,
 \dim(Y_n)}{\dim \mathfrak{g}} \; , \;\;\;\;\;\;
 {\rm Tr}_{12} (\hat{C}_{\Box \otimes Y_n})^3
 =  -  \dfrac{1}{4} \, {\rm Tr}_{12} \hat{C}_{\Box \otimes Y_n}^2 \; ,
 \end{array}
 \ee
 where the universal expressions (in terms of the Vogel parameters)
 are known for $\dim(Y_n)$, $\dim (\mathfrak{g})$ and $c_{(2)}^{Y_n}$
 (see, respectively, equations (\ref{dimYk}), (\ref{dimY1}) and (\ref{c2yn})), while the
 values of $\dim (\Box)$ and $c_{(2)}^{(\Box)}$
 are given in Table {\sf \ref{tab1}}.
 Another equivalent method of finding $\dim V_{a_\ell}$ is to
 solve the linear system of 4 equations which follow from
  (\ref{cask01}) and (\ref{TRCK})
  \be
 \lb{linsys}
 \sum_{\ell =1}^4 A_{k\ell} \dim_\ell = {\rm Tr}_{12} (\hat{C}_{\Box \otimes Y_n})^{k-1}
 \;\;\;\;\; (k=1,2,3,4) \; ,
 \ee
 where $\dim_\ell:= \dim V_{a_\ell}$,
 the components ${\rm Tr}_{12} (\hat{C}_{\Box \otimes Y_n})^{k-1}$
 are defined in (\ref{TRCK}) and $A_{k\ell}=a_\ell^{k-1}$ is the Vandermonde matrix of the eigenvalues (\ref{eigenv}).
 Both methods give the same result:
{\small \be
 \lb{dimYn1}
 \dim V_{a_1}=\frac{\Bigl( 4 \hat{\alpha}
  (\hat{\alpha} n-\hat{\alpha}-1)(n-1) c_{(2)}^{(\Box)}
 - \hat{\alpha} \hat{\beta}\hat{\gamma} \dim \mathfrak{g}\Bigr) n\dim(Y_n)\dim(\Box)}{
 (\hat{\alpha} n-\hat{\alpha}+\hat{\gamma}-1)(\hat{\alpha} n-\hat{\alpha}+\hat{\beta}-1)
 (1+\hat{\alpha}-2\hat{\alpha} n)\dim \mathfrak{g}},
 \ee}
 {\small \be
 \lb{dimYn2}
 \dim V_{a_2}=\frac{(1-\hat{\alpha} n+\hat{\alpha})
 \Bigl( 4 n (1-\hat{\alpha} n) c_{(2)}^{(\Box)}
  + \hat{\beta}\hat{\gamma} \dim \mathfrak{g}\Bigr) \dim(Y_n)\dim(\Box)}{
 (\hat{\gamma}-\hat{\alpha} n)(\hat{\beta}-\hat{\alpha} n)
 (1+\hat{\alpha}-2\hat{\alpha} n)\dim \mathfrak{g}},
 \ee}
  {\small \be
 \lb{dimYn3}
 \dim V_{a_3}=\frac{(1-\hat{\alpha} n+\hat{\alpha})
 \Bigl( 4 (1-\hat{\beta}) c_{(2)}^{(\Box)}
  + \hat{\alpha}\hat{\gamma} \dim \mathfrak{g}\Bigr) n\dim(Y_n)\dim(\Box)}{
 (\hat{\gamma}-\hat{\beta})(\hat{\alpha} n-\hat{\beta})
 (1+\hat{\alpha}-\hat{\alpha} n-\hat{\beta})\dim \mathfrak{g}},
 \ee}
  {\small \be
 \lb{dimYn4}
 \dim V_{a_4}=\frac{(1-\hat{\alpha} n+\hat{\alpha})
 \Bigl( 4 (1-\hat{\gamma}) c_{(2)}^{(\Box)}
  + \hat{\alpha}\hat{\beta} \dim \mathfrak{g}\Bigr) n\dim(Y_n)\dim(\Box)}{
 (\hat{\beta}-\hat{\gamma})(\hat{\alpha} n-\hat{\gamma})
 (1+\hat{\alpha}-\hat{\alpha} n-\hat{\gamma})\dim \mathfrak{g}},
 \ee}
 Expressions for $\dim V_{a_3}$ and $\dim V_{a_4}$ are related
 to each other by
 the exchange $\hat{\beta} \leftrightarrow \hat{\gamma}$.
 One can directly check that we have
 $\dim V_{a_4}=0$ and $\dim V_{a_3}=0$
 respectively for the Lie algebras of the classical
 and exceptional series of the simple Lie algebras. This fact
 obviously follows
 from identities (\ref{charn0}) and (\ref{charn1}).
 This also defines the universal expressions for the Casimir operator
  values $c_{(2)}^{(\Box)}$ in the case
  of the Lie algebras of the classical and exceptional series:
   \be
 \lb{c2box}
 \begin{array}{c}
  c_{(2)}^{(\Box)} =
  \dfrac{\hat{\alpha}\hat{\beta} \dim \mathfrak{g}}{4 (\hat{\gamma}-1)}
  \;\;\;\; ({\rm for} \;
  \mathfrak{sl}, \mathfrak{so}, \mathfrak{sp}) \; ,
 \;\;\;\;\;\;\;
  c_{(2)}^{(\Box)}=
 \dfrac{\hat{\alpha}\hat{\gamma} \dim \mathfrak{g}}{
  4 (\hat{\beta}-1) } \;\;\;\;
  ({\rm for} \; \mathfrak{g}_2, \mathfrak{f}_4,
 \mathfrak{e}_6, \mathfrak{e}_7)\; , 
 \end{array}
 \ee
 $$
 c_{(2)}^{(\Box)}= \dfrac{\hat{\alpha}\dim \mathfrak{g}}{4}
  \Bigl(
  \dfrac{ \epsilon \; \hat{\beta} }{ (\hat{\gamma}-1)} +
   \dfrac{(1-\epsilon) \; \hat{\gamma} }{(\hat{\beta}-1) } \Bigr)
    \; ,
 $$
 where $\epsilon =1$ for the case of $\mathfrak{sl}$,
 $\mathfrak{so}$, $\mathfrak{sp}$ and $\epsilon =0$
 for the case of $\mathfrak{g}_2$, $\mathfrak{f}_4$,
 $\mathfrak{e}_6$, $\mathfrak{e}_7$.
The eigenvalues (\ref{c2box}) coincide with the values of
$c_{(2)}^{(\Box)}$ listed in Table {\sf \ref{tab1}}.

\begin{table}[h] 
  \centering
 \caption{ } 
	\label{tab6}  \vspace{0.2cm}
	\begin{tabular}{|c|c|c|c|c||c|}
		\hline
  $\;\;$ & $\Lambda_1^{(n)}$ & $\Lambda_2^{(n)}$&
  $\Lambda_3^{(n)}$ & $\Lambda_4^{(n)}$ & $Y_n$   \\   \hline
    $\mathfrak{sl}_N$  &\footnotesize $([n+1],[n])$ & \footnotesize
  $([n],[n-1])$ &\footnotesize $0$&
  \footnotesize $([n,1],[n])$  & \footnotesize $([n],[n])$ \\   \hline
     $\mathfrak{so}_N$  &\footnotesize $[n+1,n]$ & \footnotesize
  $[n,n-1]$ &\footnotesize $0$&  \footnotesize $[n^2,1]$
  &  \footnotesize $[n^2]$ \\   \hline
  $\mathfrak{sp}_{N=2r}$  &\footnotesize $[2n+1]$ & \footnotesize
  $[2n-1]$ &\footnotesize $0$&  \footnotesize $[2n,1]$
  &  \footnotesize $[2n]$ \\   \hline
	$\mathfrak{g}_2$ &\footnotesize $n\lambda_{(2)}+\lambda_{(1)}$
   &\footnotesize $(n-1)\lambda_{(2)}+\lambda_{(1)}$
    &\footnotesize  $(n-1)\lambda_{(2)}+2\lambda_{(1)}$
    &  \footnotesize $0$ &\footnotesize $n\lambda_{(2)}$  \\   \hline
  $\mathfrak{f}_4$  &\footnotesize $n\lambda_{(1)}+\lambda_{(4)}$ &
  \footnotesize $(n-1)\lambda_{(1)}+\lambda_{(4)}$&
  \footnotesize $(n-1)\lambda_{(1)}+\lambda_{(3)}$ &
	\footnotesize $0$  &\footnotesize $n\lambda_{(1)}$  \\ \hline
  $\mathfrak{e}_6$  &\footnotesize $n\lambda_{(5)}+\lambda_{(1)}$
  & \footnotesize $(n-1)\lambda_{(5)}+\lambda_{(1)}$
  &\footnotesize $(n-1)\lambda_{(5)}+\lambda_{(4)}$&
  \footnotesize $0$  &\footnotesize $n\lambda_{(5)}$   \\  \hline
  $\mathfrak{e}_7$  &\footnotesize $n\lambda_{(7)}+\lambda_{(1)}$
  & \footnotesize $(n-1)\lambda_{(7)}+\lambda_{(1)}$
  &\footnotesize $(n-1)\lambda_{(7)}+\lambda_{(6)}$&
  \footnotesize $0$ &\footnotesize $n\lambda_{(7)}$ \\
  \hline \\ [-0.45cm] \hline
  {\small $\hat{c}_{(2)}$ } &
   \footnotesize $a_2=- n \hat{\alpha}/2$ & \footnotesize
  $a_1=- (1+\hat{\alpha}(1-n))/2$ &
  \footnotesize $a_4=- \hat{\gamma}/2$&
  \footnotesize $a_3=- \hat{\beta}/2$ & -- \\   \hline
	\end{tabular}
\end{table}

Finally, at the end of this subsection,
we present in Table {\sf \ref{tab6}}
the structure of the universal multiplets
 $\Lambda_i^{(n)}|_{i=1,2,3,4}$
arising in the decomposition (\ref{decomp}) of
the tensor product $\Box \otimes Y_n$
for all simple Lie algebras (except $\mathfrak{e}_8$).
The numbers ''$0$'' indicate (in Table {\sf \ref{tab6}})
that the corresponding representations
have dimensions equal to zero. From the data in
Table {\sf \ref{tab6}} one can see that
 the rule (\ref{rule2}), which connect
 the Young diagrams for the $\mathfrak{sl}_N$ and
 $\mathfrak{sp}_N$ representations, is perfectly fulfilled
 for the universal multiplets $\Lambda^{(n)}_i|_{i=1,2,4}$
 and $Y_n$.

 \subsection{
  Decomposition of
   $\Box \otimes Y_1 = \Box \otimes \ad$ and
   universal colour factors in gauge theories}

  In this subsection, we consider the
  special case $n=1$ for the decomposition
   of $\Box \otimes Y_n$, i.e., we consider the decomposition of
   $\Box \otimes \ad$. This case is important for
   applications (e.g., for the evaluation of universal
   colour factors for an infinite set of special
   Feynman ladder diagrams) of gauge theories with matter fields,
   where gauge fields and matter fields are respectively
   realized in the adjoint ($\ad$) and
   defining ($\Box$) representations of the gauge
  groups. Remarkably, for $n=1$, the characteristic
  identity (\ref{charn2})
  is symmetric under
  all permutations of the Vogel parameters
  $\alpha,\beta,\gamma$:
  \be
  \lb{char1}
 \Bigl(\hat{C}_{\Box \otimes \ad} +
 \frac{1}{2}\Bigr)
  \Bigl(\hat{C}_{\Box \otimes \ad} + \frac{\hat{\alpha}}{2}\Bigr)
  \Bigl(\hat{C}_{\Box \otimes \ad} + \frac{\hat{\beta}}{2}\Bigr)
  \Bigl(\hat{C}_{\Box \otimes \ad} + \frac{\hat{\gamma}}{2}\Bigr) = 0 \; .
  \ee
   One can visualize the split Casimir operator
   $\hat{C}_{\Box \otimes \ad}$
 in the representation $\Box \otimes \ad$ as
 the following Feynman diagram:


     \unitlength=6mm
 \begin{picture}(20,4)(-2,0)

\put(1.5,1.8){\small
$(\hat{C}_{\Box \otimes \ad})^{\alpha_1 a_2}_{\;\; \beta_1 b_2}\;  = \;
 T^{\alpha_1}_{\;\; \beta_1}(X_d) \; X^{a_2}_{c\, b_2} \,
 {\sf g}^{c\, d} \;\; = \;\; $}

 {\linethickness{0.3mm}
 \put(13.5,1){\vector(1,0){5}}
 \put(13.5,1){\vector(1,0){1.5}}
 }

\put(16.3,2.4){\scriptsize $d$}
\put(15.4,1.2){\scriptsize $d$}

\put(12.8,3.2){\footnotesize $a_2$}
 \put(18.5,3.2){\footnotesize $b_2$}

 \put(12.8,0.8){\footnotesize $\alpha_1$}
 \put(18.7,0.7){\footnotesize $\beta_1$}

\thicklines

\put(16,2.82){\oval(0.2,0.2)[l]}
\put(16,2.59){\oval(0.2,0.2)[r]}
\put(16,2.35){\oval(0.2,0.2)[l]}
\put(16,2.11){\oval(0.2,0.2)[r]}
\put(16,1.87){\oval(0.2,0.2)[l]}
\put(16,1.63){\oval(0.2,0.2)[r]}
\put(16,1.39){\oval(0.2,0.2)[l]}
\put(16,1.15){\oval(0.2,0.2)[r]}

  \put(13.54,3){\oval(0.22,0.21)[t]}
  \put(13.77,3){\oval(0.22,0.21)[b]}
 \put(14,3){\oval(0.22,0.21)[t]}
 \put(14.23,3){\oval(0.22,0.21)[b]}
 \put(14.46,3){\oval(0.22,0.21)[t]}
 \put(14.69,3){\oval(0.22,0.21)[b]}
 \put(14.92,3){\oval(0.22,0.21)[t]}
 \put(15.15,3){\oval(0.22,0.21)[b]}
 \put(15.38,3){\oval(0.22,0.21)[t]}
 \put(15.61,3){\oval(0.22,0.21)[b]}
 \put(15.84,3){\oval(0.22,0.21)[t]}
 \put(16.07,3){\oval(0.22,0.21)[b]}

 \put(16.3,3){\oval(0.22,0.21)[t]}
 \put(16.53,3){\oval(0.22,0.21)[b]}
 \put(16.76,3){\oval(0.22,0.21)[t]}
 \put(16.99,3){\oval(0.22,0.21)[b]}
 \put(17.22,3){\oval(0.22,0.21)[t]}
 \put(17.45,3){\oval(0.22,0.21)[b]}
 \put(17.68,3){\oval(0.22,0.21)[t]}
 \put(17.91,3){\oval(0.22,0.21)[b]}
 \put(18.14,3){\oval(0.22,0.21)[t]}
 \put(18.37,3){\oval(0.22,0.21)[b]}

 \put(21.5,2){(4.26{\bf b})}



\end{picture}

\noindent
where the waved lines correspond to
the gauge fields realized in the $\ad$-representation
and the solid line corresponds to
the matter fields in the defining representation $T=\Box$.
In other words, the split Casimir
 operator $\hat{C}_{\Box \otimes \ad}$ is equal to the colour factor
for the Feynman diagram, which describes the exchange by a gauge
particle between propagating gauge and matter particles.

  Now we note that the first and second formula in (\ref{c2box})
  are related by the permutation
  $\hat{\beta} \leftrightarrow \hat{\gamma}$. By taking into
  account the symmetry under all permutations of the Vogel
  parameters $\alpha,\beta,\gamma$, we can permute the parameters
  $\hat{\beta} \leftrightarrow \hat{\gamma}$
  for the Lie algebras
  of the classical series so that the content of Tables
  {\sf \ref{tab1}, \ref{tab6}} (for $n=1$) is changed,
  as it is indicated in
  Table {\sf \ref{tab7}}. In order not to confuse the new choice
  of parameters $\hat{\beta}$ and $\hat{\gamma}$ with the old one,
  we introduce the notation $\hat{\beta}' =\hat{\gamma}$,
  $\hat{\gamma}\,' =\hat{\beta}$ for the Lie algebras of the classical
  series and $\hat{\beta}' =\hat{\beta}$,
  $\hat{\gamma}\,' =\hat{\gamma}$ for the exceptional Lie algebras.
  In this case, we have
  $\hat{\gamma}\,'=1/3$ for the Lie algebras
  $\mathfrak{sl}_3$ and $\mathfrak{so}_8$
  as for all exceptional Lie algebras, and the second
  formula for the Casimir operator value in (\ref{c2box}):
   \be
  \lb{c2box0}
   c_{(2)}^{(\Box)}=
 \dfrac{\hat{\alpha}\hat{\gamma}^{\,\prime} \dim \mathfrak{g}}{
  4 (\hat{\beta}'-1) }  \; ,
  \ee
  becomes valid (universal) for all simple Lie algebras, except $\mathfrak{e}_8$.

  \begin{table}[h] 
  \centering
 \caption{ } 
	\label{tab7}  \vspace{0.2cm}
	\begin{tabular}{|c|c|c|c||c|c|c|c|}
		\hline
  $\;\;$ & $\hat{\alpha}$ & $\hat{\beta}'$ & $\hat{\gamma}\, '$ &
  $\Lambda_1^{(n=1)}$ & $\Lambda_2^{(n=1)}$&
  $\Lambda_3^{(n=1)}$ & $\Lambda_4^{(n=1)}$   \\   \hline
    $\mathfrak{sl}_N$ &\scriptsize $-\frac{1}{N}$
    &\scriptsize $\frac{1}{2}$
    &\scriptsize $\frac{1}{N}$  &
    \footnotesize $([2],[1])$ & \footnotesize
  $([1],[\emptyset])$ &\footnotesize $([1^2],[1])$&
  \footnotesize  $0$ \\   \hline
     $\mathfrak{so}_N$ &\scriptsize $-\frac{1}{N-2}$ &
  \scriptsize $\frac{N-4}{2N-4}$ & \scriptsize $\frac{2}{N-2}$
   & \footnotesize $[2,1]$ & \footnotesize
  $[1]$ &\footnotesize  $[1^3]$ &  \footnotesize $0$ \\   \hline
   $\mathfrak{sp}_{N=2r}$ &\scriptsize $-\frac{2}{N+2}$ &
  \scriptsize $\frac{N+4}{2N+4}$ & \scriptsize $\frac{1}{N+2}$
   & \footnotesize $[3]$ & \footnotesize
  $[1]$ &\footnotesize  $[2,1]$ &  \footnotesize $0$ \\   \hline
	$\mathfrak{g}_2$ & \scriptsize $-1/4$ &
 \scriptsize $5/12$ & \scriptsize $1/3$  &
 \footnotesize $\lambda_{(2)}+\lambda_{(1)}$
   &\footnotesize $\lambda_{(1)}$
    &\footnotesize  $2\lambda_{(1)}$
    &  \footnotesize $0$  \\   \hline
  $\mathfrak{f}_4$ & \scriptsize  $-1/9$ &
   \scriptsize  $5/18$ &\scriptsize  $1/3$ &
   \footnotesize $\lambda_{(1)}+\lambda_{(4)}$ &
  \footnotesize $\lambda_{(4)}$&
  \footnotesize $\lambda_{(3)}$ &
	\footnotesize $0$   \\ \hline
  $\mathfrak{e}_6$& \scriptsize  $-1/12$ &
   \scriptsize  $1/4$  & \scriptsize  $1/3$  &
   \footnotesize $\lambda_{(5)}+\lambda_{(1)}$
  & \footnotesize $\lambda_{(1)}$
  &\footnotesize $\lambda_{(4)}$&
  \footnotesize $0$    \\  \hline
  $\mathfrak{e}_7$ & \scriptsize  $-1/18$  &
   \scriptsize  $2/9$ & \scriptsize  $1/3$ &
   \footnotesize $\lambda_{(7)}+\lambda_{(1)}$
  & \footnotesize $\lambda_{(1)}$
  &\footnotesize $\lambda_{(6)}$&
  \footnotesize $0$  \\   \hline \\ [-0.45cm] \hline
   {\small $\hat{c}_{(2)}$ } & -- & -- & --  &
   \footnotesize $a_2=- \hat{\alpha}/2$ & \footnotesize
  $a_1=- 1/2$ &
  \footnotesize $a_4=- \hat{\gamma}'/2$&
  \footnotesize $a_3=- \hat{\beta}'/2$  \\   \hline
	\end{tabular}
\end{table}

In the case $n=1$, after substitution (\ref{c2box0}),
 formulas (\ref{dimYn1}) -- (\ref{dimYn4}) for
 $\dim V_{a_i}$
are drastically simplified:
 \be
 \lb{dimVn1}
\begin{array}{c}
\dim V_{a_1}^{(1)} = \dim \Lambda^{(n=1)}_2 = \dim \Box \; , \;\;\;\;\;\;
\dim V_{a_3}^{(1)} = \dim \Lambda^{(n=1)}_4 = 0 \; , \\ [0.2cm]
\dim V_{a_2}^{(1)} = \dim \Lambda^{(n=1)}_1 =
\dfrac{(1-\hat{\alpha}-\hat{\beta})(\hat{\gamma}-1)}{
\hat{\alpha}\hat{\beta} (\hat{\gamma}-\hat{\alpha})}
\dim \Box \; , \\ [0.2cm]
\dim V_{a_4}^{(1)} = \dim \Lambda^{(n=1)}_3 =
\dfrac{(1-\hat{\beta}-\hat{\gamma})(\hat{\alpha}-1)}{
\hat{\gamma}\hat{\beta} (\hat{\alpha}-\hat{\gamma})}
\dim \Box \; ,
\end{array}
\ee
where $a_i$ are the eigenvalues
 of the quadratic split Casimir operator
in the representations
$\Lambda_k^{(1)}$ (see Table {\sf \ref{tab7}}):
 \be
 \lb{eigenv1}
a_1=\hat{c}_{(2)}^{(\Lambda_2^{(1)})} = -1/2 \, ,
\;\;\; a_2=\hat{c}_{(2)}^{(\Lambda_1^{(1)})}=-\hat{\alpha}/2\, ,  \;\;\;
a_3=\hat{c}_{(2)}^{(\Lambda_4^{(1)})}=-\hat{\beta}'/2\, ,  \;\;\; a_4=\hat{c}_{(2)}^{(\Lambda_3^{(1)})}=-\hat{\gamma}'/2 \, .
 \ee

Finally, we show how one can obtain universal expressions
for colour (group) factors for a set of Feynman diagrams.
We consider the simplest case of the $L$-loop ladder
diagrams with the closed solid (matter) line

     \unitlength=6mm
 \begin{picture}(20,5)(1,-1)

\put(4,1.8){\small
$(\hat{C}_{\Box \otimes \ad}^{\, ^L}
)^{\alpha \, a_2}_{\;\;\; \alpha \, b_2}
\;\;  = \;\; $}

 { 
 \put(10.7,1){\vector(1,0){8}}
 \put(10.7,1){\vector(1,0){1.5}}
 \put(10.7,1){\vector(1,0){4}}
 \put(18.7,-0.4){\vector(-1,0){8}}
}

 \put(10.7,0.3){\oval(1.4,1.4)[l]}
\put(18.7,0.3){\oval(1.4,1.4)[r]}

\put(14.9,1.9){$\dots$}

\put(12.6,2.4){\scriptsize $d_0$}
\put(11.7 ,1.2){\scriptsize $d_0$}

\put(14,2.4){\scriptsize $d_1$}
\put(13.1 ,1.2){\scriptsize $d_1$}

\put(16.8,2.4){\scriptsize $d_L$}
\put(15.9,1.2){\scriptsize $d_L$}

\put(9.9,3.2){\footnotesize $a_2$}
 \put(18.5,3.2){\footnotesize $b_2$}


\thicklines

\put(12.4,2.82){\oval(0.2,0.2)[l]}
\put(12.4,2.59){\oval(0.2,0.2)[r]}
\put(12.4,2.35){\oval(0.2,0.2)[l]}
\put(12.4,2.11){\oval(0.2,0.2)[r]}
\put(12.4,1.87){\oval(0.2,0.2)[l]}
\put(12.4,1.63){\oval(0.2,0.2)[r]}
\put(12.4,1.39){\oval(0.2,0.2)[l]}
\put(12.4,1.15){\oval(0.2,0.2)[r]}

\put(13.8,2.82){\oval(0.2,0.2)[l]}
\put(13.8,2.59){\oval(0.2,0.2)[r]}
\put(13.8,2.35){\oval(0.2,0.2)[l]}
\put(13.8,2.11){\oval(0.2,0.2)[r]}
\put(13.8,1.87){\oval(0.2,0.2)[l]}
\put(13.8,1.63){\oval(0.2,0.2)[r]}
\put(13.8,1.39){\oval(0.2,0.2)[l]}
\put(13.8,1.15){\oval(0.2,0.2)[r]}

\put(16.6,2.82){\oval(0.2,0.2)[l]}
\put(16.6,2.59){\oval(0.2,0.2)[r]}
\put(16.6,2.35){\oval(0.2,0.2)[l]}
\put(16.6,2.11){\oval(0.2,0.2)[r]}
\put(16.6,1.87){\oval(0.2,0.2)[l]}
\put(16.6,1.63){\oval(0.2,0.2)[r]}
\put(16.6,1.39){\oval(0.2,0.2)[l]}
\put(16.6,1.15){\oval(0.2,0.2)[r]}

    \put(10.78,3){\oval(0.22,0.21)[t]}
  \put(11.01,3){\oval(0.22,0.21)[b]}
   \put(11.24,3){\oval(0.22,0.21)[t]}
  \put(11.47,3){\oval(0.22,0.21)[b]}
   \put(11.7,3){\oval(0.22,0.21)[t]}
  \put(11.93,3){\oval(0.22,0.21)[b]}
    \put(12.16,3){\oval(0.22,0.21)[t]}
  \put(12.39,3){\oval(0.22,0.21)[b]}
   \put(12.62,3){\oval(0.22,0.21)[t]}
  \put(12.85,3){\oval(0.22,0.21)[b]}
   \put(13.08,3){\oval(0.22,0.21)[t]}
  \put(13.31,3){\oval(0.22,0.21)[b]}
  \put(13.54,3){\oval(0.22,0.21)[t]}
  \put(13.77,3){\oval(0.22,0.21)[b]}
 \put(14,3){\oval(0.22,0.21)[t]}
 \put(14.23,3){\oval(0.22,0.21)[b]}
 \put(14.46,3){\oval(0.22,0.21)[t]}
 \put(14.69,3){\oval(0.22,0.21)[b]}
 \put(14.92,3){\oval(0.22,0.21)[t]}
 \put(15.15,3){\oval(0.22,0.21)[b]}
 \put(15.38,3){\oval(0.22,0.21)[t]}
 \put(15.61,3){\oval(0.22,0.21)[b]}
 \put(15.84,3){\oval(0.22,0.21)[t]}
 \put(16.07,3){\oval(0.22,0.21)[b]}

 \put(16.3,3){\oval(0.22,0.21)[t]}
 \put(16.53,3){\oval(0.22,0.21)[b]}
 \put(16.76,3){\oval(0.22,0.21)[t]}
 \put(16.99,3){\oval(0.22,0.21)[b]}
 \put(17.22,3){\oval(0.22,0.21)[t]}
 \put(17.45,3){\oval(0.22,0.21)[b]}
 \put(17.68,3){\oval(0.22,0.21)[t]}
 \put(17.91,3){\oval(0.22,0.21)[b]}
 \put(18.14,3){\oval(0.22,0.21)[t]}
 \put(18.37,3){\oval(0.22,0.21)[b]}

 \put(24,1.5){(4.30{\bf b})}


\end{picture}

\noindent
where we used the graphical representation (4.26{\bf b})
of the split Casimir operator $\hat{C}_{\Box \otimes \ad}$.
By using Schur's Lemma (see also formula (\ref{khcas01})),
we write the left hand-side of
(4.30{\bf b}) as follows:
 \be
 \lb{ladder}
\begin{array}{c}
{\rm Tr}_1(\hat{C}_{\Box \otimes \ad})^{\, L} =
  \dfrac{I_2 \,}{\dim \mathfrak{g}}  \,
{\rm Tr}_{12}(\hat{C}_{\Box \otimes \ad})^{\, L} =
\dfrac{I_2 \,}{\dim \mathfrak{g}}  \,
{\rm Tr}_{12}\Bigl(\hat{C}_{\Box \otimes \ad}^{\, L}\sum\limits_i
P_{a_i}^{(1)}\Bigr) = \\ [0.3cm]
=
\dfrac{I_2 \,}{\dim \mathfrak{g}}  \,
{\rm Tr}_{12}\Bigl(\sum\limits_i a_i^{\, L}
P_{a_i}^{(1)}\Bigr) =
\dfrac{I_2 \,}{\dim \mathfrak{g}}  \,
\sum\limits_{i=1}^4 \bigl( a_i^{\, L} \, \dim
V_{a_i}^{(1)} \bigr) = \\ [0.3cm]
=  I_2 \, \dfrac{\dim \Box}{\dim \mathfrak{g}}  \Bigl(
(-\frac{1}{2})^{^L}+ (-\frac{\hat{\alpha}}{2})^{^L} \, \dfrac{(1-\hat{\alpha}-\hat{\beta}')(\hat{\gamma}'-1)}{
\hat{\alpha}\hat{\beta}' (\hat{\gamma}'-\hat{\alpha})}
+ (-\frac{\hat{\gamma}'}{2})^{^L}\,
\dfrac{(1-\hat{\beta}'-\hat{\gamma}')(\hat{\alpha}-1)}{
\hat{\gamma}'\hat{\beta}' (\hat{\alpha}-\hat{\gamma}')}
  \Bigr)\; ,
\end{array}
 \ee
where $I_2$ is the unit operator acting
in the second space of the
 $\ad$-representation, while
 $\dim V_{a_i}^{(1)}$ and $a_i$
 were defined (as functions of the
 Vogel parameters) in (\ref{dimVn1}) and (\ref{eigenv1}).
 Thus, formula (\ref{ladder}) gives the universal expression
 for the colour factors of the
 Feynman diagrams (4.30{\bf b}) in gauge theories
 with all gauge groups, except the gauge group generated by
  the Lie algebra
 $\mathfrak{e}_8$. In the case of $\mathfrak{e}_8$,
 the defining representation $\Box$ coincides with
 the $\ad$-representation, and the solid lines
 in the diagrams (4.26{\bf b}), (4.30{\bf b})
 can be redrawn as wavy lines. In this case, the split Casimir operator
 $\hat{C}_{\Box \otimes \ad} =
 \hat{C}_{\ad \otimes \ad}$ is investigated in the
 framework of the standard Vogel universality
 (see e.g. \cite{IsKriv}) and the colour factor for
 the diagram (4.30{\bf b})
 is evaluated by the methods considered in \cite{IsaPro3}.

\section{Conclusion}
\setcounter{equation}0

The main results of this paper are the universal answers given in
(\ref{dimYn1})-(\ref{dimYn4}) for the dimensions
 of the representations $\Lambda^{(n)}_i$ listed in Table {\sf \ref{tab6}} and universal formulas
(\ref{c2box}) for the Casimir eigenvalue $c_{(2)}^{(\Box)}$. The representations $\Lambda^{(n)}_i$ are not
present in the decomposition of $\ad^{\otimes k}$. They arise in the decomposition of the tensor products $\Box \otimes Y_n$
of the defining representation $\Box$ and any representation $Y_n$,
for example, the defining representation and the
$\ad$-representation $Y_1$. It should be noted that the universality in (\ref{dimYn1})-(\ref{dimYn4}) is not quite explicit. This means that we need to substitute there the universal formulas for $\dim Y_n$,
$\dim \mathfrak{g}$, and formulas (\ref{c2box}) for the
 eigenvalues of the quadratic Casimir
$c_{(2)}^{(\Box)}$ in the defining representation.
At the end of the paper, we give an example of applying
the obtained results for
 evaluating universal colour (group) factors for an infinite
 set of Feynman diagrams in non-Abelian gauge theories
 with the gauge group generated by any simple Lie algebra
 (except $\mathfrak{e}_8$).

Now we discuss some promising directions of investigations
that follow from the results obtained in this paper.
First of all, it is clear that
there exists a uniform (valid for all simple Lie algebras) decomposition
of the tensor product $Y_1 \otimes Y_n$
(see equations (\ref{yny01}), (\ref{yny02}), (\ref{yny02sp}) and
(\ref{e8fa})). Thus, one can find
universal formulas for dimensions
of an infinite series of
representations arising in the decomposition of
$Y_1 \otimes Y_n$.

By using formulas (\ref{khcas01}) and (\ref{cask01}), one can
find universal expressions for values of
the highest Casimir operators in
the defining $\Box$ and in
 $\Lambda_{i}^{(n)}|_{i=1,2,3,4}$ representations.

 It is not possible
  (without introducing an additional discrete parameter $\epsilon$)
  to find a single universal formula for the values
  $c_{(2)}^{(\Box)}$ of the second Casimir operator in (\ref{c2box}),
  so that this formula will be valid simultaneously
  for the classical and exceptional Lie algebras when choosing the
  Vogel parameters, as in Table {\sf \ref{tab1}}.
  Note that formulas in (\ref{c2box}) are related by a permutation of $\hat{\beta}$ with $\hat{\gamma}$.
  This may indicate that when considering universality beyond of the tensor products $\ad^{\otimes k}$, the symmetry under permutations of the Vogel parameters is broken. Apparently, when we select the Vogel parameters
  in Table {\sf \ref{tab1}}, we need to permute $\hat{\beta}$ with $\hat{\gamma}$, either for the Lie algebras of classical series
  or for the exceptional Lie algebras. In this case, the value $\hat{\beta} \, (\hat{\gamma}) = 1/3$ for exceptional algebras will coincide with the value $\hat{\beta}\, (\hat{\gamma}) = 1/3$  for the $\mathfrak{sl}_3$
   and $\mathfrak{so}_8$ algebras, i.e., on the Vogel's plane
   (see \cite{LanMan}) , not only the point
   $D_4 = \mathfrak{so}_8$ but also the point
   $A_2 = \mathfrak{sl}_3$ will fall on the line
   of the exceptional algebras.

It seems that there is a direct generalization
($q$ and $t$ deformations) to the case of
simply laced Lie algebras
of all the universal formulas for dimensions
and identities of the type (\ref{dimidi}) (following from the
decomposition of tensor products $\Box \otimes Y_n$).
In this connection, we should mention recent papers
\cite{BiMiMo,BiMir,Bish}
(see also Sect. 6.1 in \cite{MorSlep} for a review
and references therein). Indeed,
considering the example of the $\mathfrak{sl}_N$-algebra,
the $q$-deformations of dimensions (\ref{dimYn123}) are
\be
\lb{dimq123}
\begin{array}{c}
\dim_q([2n+1,n^{N-2}]) =
\dfrac{[N+2n]_q \, [N+n-1]_q}{[N+2n-1]_q \, [n+1]_q} \dim_q Y_n \; , \\ [0.3cm]
\dim_q ([2n,n+1,n^{N-3}]) =   \dfrac{[N+n-1]_q \, [n]_q \, [N-2]_q}{
  [n+1]_q \, [N+n-2]_q} \dim_q Y_n \; , \\ [0.3cm]
\dim_q ([2n-1,(n-1)^{N-2}]) =
\dfrac{[N+2n-2]_q \, [n]_q}{[N+n-2]_q \, [N+2n-1]_q} \dim_q Y_n \; ,
\end{array}
\ee
where $[A]_q=\frac{(q^A-q^{-A})}{(q-q^{-1})}$
and the $q$-deformation of the first relation in (\ref{dimidi})
is valid in view of the identity
$$
[N]_q = \dfrac{[N+2n]_q \, [N+n-1]_q}{[N+2n-1]_q \, [n+1]_q}+
\dfrac{[N+n-1]_q \, [n]_q \, [N-2]_q}{[n+1]_q \, [N+n-2]_q} +
\dfrac{[N+2n-2]_q \, [n]_q}{[N+n-2]_q \, [N+2n-1]_q} \; .
$$
Analogously, the $q$-deformation of the second
 relation in (\ref{dimidi})
is valid in view of the remarkable symmetric identity
$$
[N+M]_q [M+1]_q [N+1]_q =[N]_q [N+1]_q
+ [N+M+2]_q [M]_q [N]_q + [M]_q[M+1]_q \; ,
$$
where we have to substitute $M\to 2n$ and $N \to N-2n$.
Note that there is an arbitrariness in the definition
of the $q$-dimensions of representations, e.g., $\dim_q (\lambda)
\to q^{c \cdot |\lambda|} \dim_q (\lambda)$, where
$c$ is arbitrary constant and $|\lambda|$ is the number
of cells of the Young diagram $\lambda$
(see  in this regard formula (4.3.80) in \cite{Isa2},
Subsect. 4.3.6).

\vspace{0.3cm}

 {\bf \large Acknowledgments.}
  I would like to thank  all participants of
  the Dubna workshop ''Universal description of Lie algebras,
  Vogel theory and applications''
  in April of 2025 for stimulating discussions.
   I am especially grateful to S.O.Krivonos, A.D.Mironov,
 R.L.Mkrtchyan and A.A.Provorov
  for useful comments and help. I also thank A.Molev and
  O.Ogievetsky for valuable remarks on the operation
 (\ref{rule2}) of summation of Young diagrams.  This work
 was supported by the RSF grant No. 23-11-00311.

\vspace{0.3cm}


\end{document}